\documentclass[final,3p,times]{elsarticle}

\usepackage{graphicx}
\usepackage{multicol,multirow}
\usepackage{amsmath,amssymb,amsfonts}
\usepackage{mathrsfs}
\usepackage{amsthm}
\usepackage{rotating}
\usepackage{appendix}
\usepackage{ifpdf}
\usepackage[T1]{fontenc}
\usepackage{newtxtext}
\usepackage{newtxmath}
\usepackage{textcomp}
\usepackage{xcolor}
\usepackage{lipsum}
\usepackage[colorlinks,allcolors=blue]{hyperref}
\usepackage[export]{adjustbox}

\usepackage{amssymb}
\usepackage{amsmath}
\usepackage{graphicx}
\usepackage{algorithm2e}
\RestyleAlgo{ruled}
\usepackage{caption}
\usepackage{subcaption}
\usepackage{pdflscape}
\usepackage{tabularx}
\usepackage{booktabs}
\usepackage{multirow}
\DeclareMathOperator*{\argmin}{\arg\!\min}

\SetKwInput{KwData}{Inputs}
\SetKwInput{KwResult}{Outputs}

\usepackage[dvipsnames]{xcolor}
\usepackage{soul}
\usepackage{ulem}
\usepackage{dutchcal}
\usepackage{float}
\usepackage{multirow}
\usepackage{fancyhdr}
\usepackage[pscoord]{eso-pic}% The zero point of the coordinate systemis the lower left corner of the page (the default).
\usepackage{nicematrix} %for tables
\usepackage{afterpage} %for wrapping about the landscape
\usepackage{pdflscape}
\usepackage{natbib}
\usepackage{epstopdf}

\usepackage{lineno}

\usepackage[english]{babel}
\addto\captionsenglish{\renewcommand{\appendixname}{Appendix}}

\journal{xxxx}

\begin{document}

\pagestyle{fancy}
\fancyhf{}
\renewcommand{\headrulewidth}{0pt}
\cfoot{{\textsf{Distribution Statement A. Approved for public release. Distribution is unlimited.}}}
\rfoot{\thepage}

\begin{frontmatter}

\title{Projection-based model-order reduction via graph autoencoders suited for unstructured meshes}

\author[1]{Liam K. Magargal}

\author[1]{Parisa Khodabakhshi}
\ead{PAK322@lehigh.edu}

\author[2]{Steven N. Rodriguez}

\author[3]{Justin W. Jaworski}

\author[4]{John G. Michopoulos}

\affiliation[1]{organization={Department of Mechanical Engineering and Mechanics, Lehigh University},%Department and Organization
            city={Bethlehem},
            state={PA},
            country={United States}}

\affiliation[2]{organization={Computational Multiphysics Systems Laboratory, U. S. Naval Research Laboratory},%Department and Organization
            city={Washington},
            state={DC},
            country={United States}}

\affiliation[3]{organization={Kevin T. Crofton Department of Aerospace and Ocean Engineering, Virginia Tech},%Department and Organization
            city={Blacksburg},
            state={VA},
            country={United States}}

\affiliation[4]{organization={Principal Scientist for Material Innovation (Retired), U. S. Naval Research Laboratory},%Department and Organization
            city={Washington},
            state={DC},
            country={United States}}


%\keywords[MSC Codes]{\codes[Primary]{CODE1}; \codes[Secondary]{CODE2, CODE3}}

\begin{abstract} 
This paper presents the development of a graph autoencoder architecture capable of performing projection-based model-order reduction (PMOR) using a nonlinear manifold least-squares Petrov-Galerkin (LSPG) projection scheme. The architecture is particularly useful for advection-dominated flows modeled by unstructured meshes, as it provides a robust nonlinear mapping that can be leveraged in a PMOR setting. The presented graph autoencoder is constructed with a two-part process that consists of (1) generating a hierarchy of reduced graphs to emulate the compressive abilities of convolutional neural networks (CNNs) and (2) training a message passing operation at each step in the hierarchy of reduced graphs to emulate the filtering process of a CNN. The resulting framework provides improved flexibility over traditional CNN-based autoencoders because it is readily extendable to unstructured meshes. We provide an analysis of the interpretability of the graph autoencoder's latent state variables, where we find that the Jacobian of the decoder for the proposed graph autoencoder provides interpretable mode shapes akin to traditional proper orthogonal decomposition modes. To highlight the capabilities of the proposed framework, which is named geometric deep least-squares Petrov-Galerkin (GD-LSPG), we benchmark the method on a one-dimensional Burgers' model with a structured mesh and demonstrate the flexibility of GD-LSPG by deploying it on two test cases for two-dimensional Euler equations that use an unstructured mesh. The proposed framework is more flexible than using a traditional CNN-based autoencoder and provides considerable improvement in accuracy for very low-dimensional latent spaces in comparison with traditional affine projections.
\end{abstract}

\end{frontmatter}

%Some Data journals (DAP, DCE) require an `Impact Statement' section. Comment out this section if it is not required.

% Some math journals (FLO) require a table of contents. Comment out this line if no ToC is needed.
%\localtableofcontents
%\linenumbers

%% main text
\section{Introduction}

Methods in computational mechanics aim to simulate complex physical phenomena via numerical methods. Specifically, approximate solutions are sought by spatially and temporally discretizing the governing equations of a physical system \cite{leveque2002fvm,mazumder2015unstructuredfvm}. In many engineering applications, spatial and temporal resolution must be refined to obtain a sufficiently detailed solution, ultimately resulting in extremely high-dimensional discretized systems. Dimensional compression aims to embed such high-dimensional discretized systems into low-dimensional representations, while retaining essential information. 

Projection-based model-order reduction (PMOR) is a class of approximation methods that aims to reduce the computational cost associated with many-query tasks in computational mechanics, such as design optimization, uncertainty quantification, real-time rendering, etc., while preserving sufficient accuracy of the quantities of interest \cite{benner2015Survey}. PMOR achieves cost savings by projecting the original high-dimensional computational model (known in this context as the full-order model, or FOM in short) onto a precomputed low-dimensional latent space, which is computed using data recovered from the FOM in an offline stage. In the online stage, the projected form of the governing equations can be used to compute low-dimensional solutions, thereby reducing the operation count complexity and achieving cost savings.

In this study, our primary interest is on the dimensionality reduction capabilities of PMOR approaches for methods in computational mechanics that employ unstructured meshes. Two such methods are the finite volume method (FVM) and the finite element method (FEM), which are widely used in the realms of science and engineering for their ability to conduct high-fidelity simulations of complex physical phenomena \cite{reddy2005finiteelement,leveque2002fvm,mazumder2015unstructuredfvm}. Spatial discretization of a domain is typically achieved using one of two main mesh types: structured and unstructured meshes. Structured meshes employ a periodic, grid-like structure to discretize the domain. Conversely, unstructured meshes do not require a grid-like structure and allow mesh components to be arbitrarily ordered \cite{Bern2000MeshGeneration,mazumder2015unstructuredfvm}. This departure from a grid-like structure often makes spatial discretization of physical domains more convenient, allowing for the generation of a higher quality mesh with favorable features. The advantage of unstructured meshes over structured meshes becomes more prominent in domains with complex geometries.

Traditionally, PMOR methods rely on projecting the solution onto a low-dimensional latent space via a subspace approximation method, such as proper orthogonal decomposition (POD) \cite{sirovich1987turbulence}, rational interpolation \cite{baur2011interpolatory}, or balanced truncation \cite{moore1981pca,rezaian2023balanced}. 
Although affine latent spaces have been leveraged extensively to achieve cost savings for a wide variety of linear and nonlinear models \cite{lieu2006aircraft,wentland2023scalablePROM,buitanh2008modelReduction}, PMOR procedures employing affine solution manifolds often fail to accurately model advection-dominated solutions. Such solutions often exhibit features such as sharp gradients, moving shocks and boundaries, and bulk motion, which will be smoothed out using affine solution manifolds. It is well known that these models exhibit a slowly-decaying Kolmogorov $n$-width. The Kolmogorov $n$-width serves as a measure for the error introduced by approximating the solution manifold of a partial differential equation (PDE) with a linear trial manifold of dimension $n$ \cite{ahmed2020kolmogorov,franco2022kolmogorov,peherstorfer2020transport,peherstorfer2020kolmogorov,welper2017interpolation}. When the decay of the Kolmogorov $n$-width is slow, the affine latent space used to approximate the solution must be constructed with a high dimension, leading to marginal model reduction. As a result, a great amount of effort has been made to develop reduced-order models (ROMs) for advection-dominated flows such as adaptive reduced basis schemes \cite{drohmann2011adaptivetime,ohlberger2013freezing}, segmentation of the domain into multiple reduced-order bases \cite{asmsallem2012localbases,dihlmann2011adaptivetime,geelen2022localized}, quadratic manifolds \cite{geelen2023operator,barnett2022quadratic}, modified POD bases \cite{welper2017interpolation,nair2019transportedsnapshot}, and neural-network augmented latent spaces \cite{Barnett2023NNAugmentedPROM,fresca2022poddlrom}. None of these methods, however, have direct knowledge of the geometric structure and topology of the spatially discretized domain. Additionally, many are still fundamentally based on affine latent spaces. In this work, our graph autoencoder aims to leverage knowledge of the geometric structure and topology of the spatially discretized domain to construct a robust nonlinear mapping from the high-dimensional space to the low-dimensional latent space.

Achieving cost-savings via PMOR for linear systems often comes directly from dimensional compression alone \cite{benner2017ModelRedAndApprox,antoulas2009approximation}. However, for nonlinear dynamical systems, projection of the nonlinear term often has an associated operation count complexity that scales directly with the dimension of the FOM, which often results in minimal (if any) cost savings. As a remedy, some studies employ hyper-reduction methods, where the nonlinear terms are computed for a selection of sample points and used to update the corresponding low-dimensional states. Such methods include the discrete empirical interpolation method \cite{barrault2004eim,chaturantabut2010deim, drmac2016deimerror}, Gauss-Newton with approximated tensors method \cite{carlberg2011lspg,carlberg2013gnat}, energy conserving sampling and weighing method \cite{farhat2015ecsw}, and projection tree reduced-order modeling (PTROM) for non-local methods with dense adjacency matrices \cite{rodriguez2022ptrom}. Alternatively, operator inference is often used to learn low-dimensional operators of a nonlinear equation of polynomial form from a regression problem \cite{peherstorfer2016data,benner2020operator,mcquarrie2021data}. Furthermore, some methods have coupled the operator inference framework with a lifting transformation suited for nonlinear problems with general nonlinearity by introducing a change of variables to obtain a polynomial form of the model equations \cite{qian2020lift,khodabakhshi2022opinf}.

Recently, machine learning has been adopted to overcome the limitations of traditional model reduction when applied to advection-dominated flows with slowly-decaying Kolmogorov $n$-widths. Historically, autoencoders have been developed to compress and reconstruct input information, such as images \cite{bank2023autoencoders}, but recently autoencoders have been leveraged in engineering applications. Specifically, the model reduction community has used autoencoders to identify a nonlinear mapping between the high-dimensional system and a low-dimensional latent space \cite{hasegawa2020mlrom,wiewel2019latent} from state solution data alone. Once an autoencoder is trained, the mapping is leveraged to perform online time integration using one of two main classes of approaches. The first class involves learning the low-dimensional dynamics without knowledge of the dynamics of the high-dimensional system. Some studies train a neural network to approximate the low-dimensional update at each time step \cite{wiewel2019latent,kim2019deepfluids,maulik2021advectionrom,fresca2021comprehensive,dutta2022advectionaware}, while others aim to obtain a low-dimensional system of semi-discrete ordinary differential equations (ODEs) \cite{fries2022lasdi,he2023glasdi}. The second class, and the approach that we adopt in this paper, aims to project the governing equations of the semi-discrete high-dimensional dynamical system onto the low-dimensional latent space using the autoencoder, thereby embedding the physics into the resulting ROM \cite{hartman2017deepMOR,kashima2016MORbyAE,lee2020deeplspg,lee2021deepconservation,kim2022masked}.

A common method used across both classes of machine learning-based ROMs is the convolutional neural network (CNN), which is used to construct low-dimensional solution manifolds \cite{hasegawa2020mlrom,wiewel2019latent,kim2019deepfluids}. Because CNNs require inputs to be constructed as a grid, the direct application of CNNs to unstructured meshes for the purpose of model reduction is currently untenable. As a result, a common workaround is to interpolate the unstructured mesh solution onto a structured mesh that can be provided to a CNN \cite{fresca2021comprehensive}. In recent years, graph neural networks (GNNs) have been developed to extract information of interest to the user from sets of unstructured and relational data \cite{battaglia2018graph,zhou2020graph}, making them an appropriate method to generate low-dimensional embeddings of models that use unstructured meshes. While dimensionality reduction and compression using GNNs has been achieved by graph U-nets \cite{gao2019unet} and multiscale graph autoencoders \cite{barwey2023multiscale}, such approaches do not have a decoder that maps directly from the latent state vector to the approximate solution. Instead, they require knowledge of the encoder and cannot be directly used in PMOR. Alternatively, graph autoencoders have been used for fully data-driven model reduction \cite{pichi2024graph,gruber2022comparison}, but such approaches do not include any knowledge of the governing equations.

The main focus of this study is to investigate the abilities of graph autoencoders to perform dimensionality reduction upon unstructured mesh solutions. Toward this end, we present a hierarchical graph autoencoder architecture tailored to generate nonlinear mappings to very low-dimensional latent spaces. Once the low-dimensional latent space is obtained, we embed knowledge of the governing equations into the latent space representation to perform time integration by leveraging a nonlinear manifold least-squares Petrov-Galerkin (LSPG) projection. Traditionally, a nonlinear manifold LSPG projection that leverages a CNN-based autoencoder is referred to as deep LSPG (dLSPG) \cite{lee2020deeplspg,lee2021deepconservation}. In this study, instead, we use a hierarchical graph autoencoder architecture, and the resulting method is named geometric deep LSPG (GD-LSPG). The proposed method is capable of performing PMOR when deploying unstructured meshes and is particularly useful in modeling the highly nonlinear behavior found in advection-dominated flows, while achieving significant dimensionality reduction. We analyze GD-LSPG from two different perspectives. First, we assess GD-LSPG's ability to generate low-dimensional solutions to parameter sets not seen during training. Second, we investigate the interpretability of the nonlinear manifold LSPG scheme. From this perspective, we discover highly interpretable mode shapes from the Jacobian of the decoder for the graph autoencoder, which can be directly related to the POD modes for a classical POD-LSPG projection \cite{carlberg2011lspg,carlberg2013gnat}. Furthermore, we find that the Jacobian of the decoder is closely related to saliency maps \cite{simonyan2013deep}, a method commonly used in image classification to identify features in the image that are most indicative of the image's classification. Ultimately, the proposed method is capable of accurately modeling highly nonlinear behavior found in advection-driven problems while providing interpretable mode shapes to the user.

The paper is organized in the following manner. Section 2 describes the background and preliminaries of the GD-LSPG framework, which includes the FOM and its corresponding residual minimization scheme, a general formulation of performing nonlinear dimension reduction via autoencoders, and a brief overview of graph theory. Section 3 presents the graph autoencoder deployed in GD-LSPG. Section 4 presents the nonlinear manifold LSPG projection and analyzes the interpretability of the graph autoencoder's latent state variables by relating the Jacobian of the decoder for the graph autoencoder to the POD modes used in POD-LSPG. Section 5 includes a set of numerical experiments to investigate the capabilities of our proposed method. Specifically, we apply GD-LSPG to the benchmark one-dimensional (1D) Burgers' model using a structured mesh and to two test cases that use an unstructured mesh to solve the two-dimensional (2D) Euler equations. Namely, the first test case models a setup for a Riemann problem, while the second models a bow shock generated by flow past a cylinder. Finally, Section 6 presents conclusions and discusses avenues for future work.

\section{Background and Preliminaries} \label{sec:problemFormulation}
This section lays the foundation for the introduction of the GD-LSPG method by providing a formulation of the FOM and summarizing preliminaries related to autoencoders and graph theory.  
Specifically, Section \ref{ssec:FOM} introduces the first-order PDE and residual-minimizing time integration scheme on which we develop the GD-LSPG method. Section \ref{ssec:ae} provides a general introduction to performing PMOR with an autoencoder, along with some of the current limitations of autoencoders in the literature. Finally, Section \ref{ssec:graphTheory} presents the basics of graph theory to the reader. 

\subsection{Full-order model} \label{ssec:FOM}
Consider a system of $n_q \in \mathbb{N}$ PDEs where $n_q$ depends on the number of state variables. Using a mesh to spatially discretize the physical domain into $N_c \in \mathbb{N}$ points, the semi-discrete system of the FOM is described by a system of time-continuous ODEs:

\begin{equation} \label{eq:ode1}
    \frac{\mathrm{d} \mathbf{x}}{\mathrm{dt}} = \mathbf{f}\left(\mathbf{x}, t; \boldsymbol{\mu}\right), \quad \quad \mathbf{x}\left(0; \boldsymbol{\mu}\right) = \mathbf{x}^0 \left(\boldsymbol{\mu} \right),
\end{equation}
where $\mathbf{x} \in \mathbb{R}^{N}$ is the semi-discrete state vector, $N = n_q N_c$ denotes the dimension of the FOM, $\boldsymbol{\mu} \in \mathcal{D}$ denotes the parameters, and $\mathbf{f}: \mathbb{R}^{N} \times (0, T_f] \times \mathcal{D} \rightarrow \mathbb{R}^{N}$ is the semi-discretized velocity function.

To approximate the time evolution of the state vector, $\mathbf{x}$, from the system of ODEs, we use the general form in \eqref{eq:ode2},

\begin{equation} \label{eq:ode2}
    \mathbf{r}: \left( \boldsymbol{\xi}; \boldsymbol{\mu} \right) \mapsto \alpha_0 \boldsymbol{\xi} + \sum_{i=1}^{{\tau}} \alpha_i \mathbf{x}^{n+1-i} + \beta_0 \mathbf{f}\left(\boldsymbol{\xi}, t_{n+1}; \boldsymbol{\mu} \right) + \sum_{i=1}^{\tau} \beta_i \mathbf{f}\left(\mathbf{x}^{n+1-i}, t_{n+1-i}; \boldsymbol{\mu} \right),
\end{equation}
in which the value of the state vector, $\mathbf{x}^{n+1}$, at time step $(n+1)\in \mathbb{N}$ is determined by minimizing the time-discrete residual $\mathbf{r}: \mathbb{R}^N \times \mathcal{D} \rightarrow \mathbb{R}^N$ given the state vector at a number of previous time steps. In \eqref{eq:ode2}, $\alpha_i \in \mathbb{R}$ and $\beta_i \in \mathbb{R}, i=0,1,\ldots,\tau$, are constants defined by the time integration scheme, $\boldsymbol{\xi} \in \mathbb{R}^N$ is the sought-after solution of the minimization scheme for the state vector at the $(n+1)^{\mathrm{th}}$ time step, the superscript $n+1-i$ denotes the value of the variable at time step $n+1-i \in \mathbb{N}$, where the time step size $\Delta t \in \mathbb{R}_+$ is chosen to be fixed, i.e., $t_i=i\Delta t$, and $\tau \in \mathbb{N}$ is the number of time steps associated with the time integration scheme.  We note that the time integration scheme is implicit in cases where $\beta_0 \neq 0$.
The state vector at the $(n+1)^{\mathrm{th}}$ time step, $\mathbf{x}^{n+1}$, is defined as the solution of the minimization problem,

\begin{equation} \label{eq:ode2argmin}
    \mathbf{x}^{n+1} = \underset{\boldsymbol{\xi} \in \mathbb{R}^N}{\argmin} \Big\vert \Big\vert \mathbf{r} \left( \boldsymbol{\xi}; \boldsymbol{\mu} \right) \Big\vert \Big\vert _2, \quad n=0,\cdots,N_t-1.
\end{equation}
where $N_t\in\mathbb{N}$ denotes the total number of time steps. With an appropriate selection of coefficients $\alpha_i$ and $\beta_i$, the general formulation of \eqref{eq:ode2} will cover linear multistep schemes, where specific examples are provided in Section \ref{sec:experiments}.

\subsection{Nonlinear dimension reduction via autoencoders} \label{ssec:ae}
A wide variety of nonlinear mappings have been adopted in the literature in recent years to obtain a low-dimensional latent space for PMOR on nonlinear problems. The focus of this study is on the use of autoencoders to approximate a mapping between the high-dimensional system and the low-dimensional latent space \cite{barnett2022quadratic,lee2020deeplspg,lee2021deepconservation,chen2022crom,eivazi2022betavae,pan2022nif}. Autoencoders are a class of deep learning architecture in which the basic idea is to perform dimensional compression on a data set down to a latent space with an encoder, $\mathbf{Enc}: \mathbf{x} \mapsto \mathbf{\hat{x}}$ with $\mathbf{Enc}: \mathbb{R}^N \rightarrow \mathbb{R}^M$, and to reconstruct the data set by decoding the latent space with a decoder, $\mathbf{Dec}: \mathbf{\hat{x}} \mapsto \tilde{\mathbf{x}}$ with $\mathbf{Dec}: \mathbb{R}^M \rightarrow \mathbb{R}^N$, where $M\ll N$. The former is a nonlinear mapping from the high-dimensional state vector, $\mathbf{x}$, to the low-dimensional latent representation, $\hat{\mathbf{x}}$, and the latter is a nonlinear mapping from the low-dimensional embedding to the reconstructed high-dimensional state vector, $\tilde{\mathbf{x}}$.

The encoder and decoder are constructed by a series of layers in which each layer applies a set of predefined functions to the output of the previous layer. The nonlinearity associated with the mapping is introduced through an appropriate selection of functions.
General forms of the encoder and decoder, consisting of $n_h\in\mathbb{N}$ and $n_g\in\mathbb{N}$ layers, respectively, are,
\begin{align}
    &\mathbf{Enc}: (\mathbf{x}; \theta) \mapsto \mathbf{h}_{n_h}(\hspace{1mm} \boldsymbol{\cdot} \hspace{1mm} ; \boldsymbol{\Theta}_{n_h}) \circ \mathbf{h}_{n_h-1}(\hspace{1mm} \boldsymbol{\cdot} \hspace{1mm} ; \boldsymbol{\Theta}_{n_h-1}) \circ \ldots \circ \mathbf{h}_{2}(\hspace{1mm} \boldsymbol{\cdot} \hspace{1mm} ; \boldsymbol{\Theta}_{2})\circ \mathbf{h}_{1}(\mathbf{x} ; \boldsymbol{\Theta}_{1}), \label{eq:general_enc}\\
    &\mathbf{Dec}: (\mathbf{\hat{x}}; \omega) \mapsto \mathbf{g}_{n_g}(\hspace{1mm} \boldsymbol{\cdot} \hspace{1mm}; \boldsymbol{{\Omega}}_{n_g}) \circ \mathbf{g}_{n_g-1}(\hspace{1mm} \boldsymbol{\cdot} \hspace{1mm}; \boldsymbol{{\Omega}}_{n_g-1}) \circ \ldots \circ \mathbf{g}_{2}(\hspace{1mm} \boldsymbol{\cdot} \hspace{1mm}; \boldsymbol{{\Omega}}_{2})\circ \mathbf{g}_{1}(\mathbf{\hat{x}}; \boldsymbol{{\Omega}}_{1}), \label{eq:general_dec}
\end{align}
where $\mathbf{h}_i(\hspace{1mm} \boldsymbol{\cdot} \hspace{1mm} ; \boldsymbol{\Theta}_i), i=1,\ldots,n_h$ and $\mathbf{g}_i(\hspace{1mm} \boldsymbol{\cdot} \hspace{1mm} ; \boldsymbol{\Omega}_i), i=1,\ldots,n_g$ denote the function(s) acting on the input of the $i^{\mathrm{th}}$ layer of the encoder and decoder networks, respectively, (or equivalently the output of the corresponding $(i-1)^{\mathrm{th}}$ layer). As will be explained later in Sections \ref{ssec:enc} and \ref{ssec:dec}, some layers encompass a number of functions, depending on their objective, which will collectively form $\mathbf{h}_i(\hspace{1mm} \boldsymbol{\cdot} \hspace{1mm} ; \boldsymbol{\Theta}_i)$ or $\mathbf{g}_i(\hspace{1mm} \boldsymbol{\cdot} \hspace{1mm} ; \boldsymbol{\Omega}_i)$. In \eqref{eq:general_enc} and \eqref{eq:general_dec}, $\boldsymbol{\Theta}_i, i = 1,\ldots,n_h$ and $\boldsymbol{{\Omega}}_i, i = 1,\ldots,n_g$, denote the weights and biases of the $i^{\mathrm{th}}$ layer of the encoder and decoder networks, respectively. The set of all the weights and biases of the autoencoder, i.e., $\theta:=\{ \boldsymbol{\Theta}_1,\ldots,\boldsymbol{\Theta}_{n_h} \}$ and $\omega:=\{ \boldsymbol{{\Omega}}_1,\ldots,\boldsymbol{{\Omega}}_{n_g} \}$, are trained to minimize an appropriately defined error norm between the input to the encoder and the output of the decoder. In this manuscript, we use an equal number of layers for the encoder and decoder, i.e., $n_h=n_g=n_{\ell}$.

\begin{figure}[!htb]
    \centering
    \centerline{\includegraphics[scale=.5]{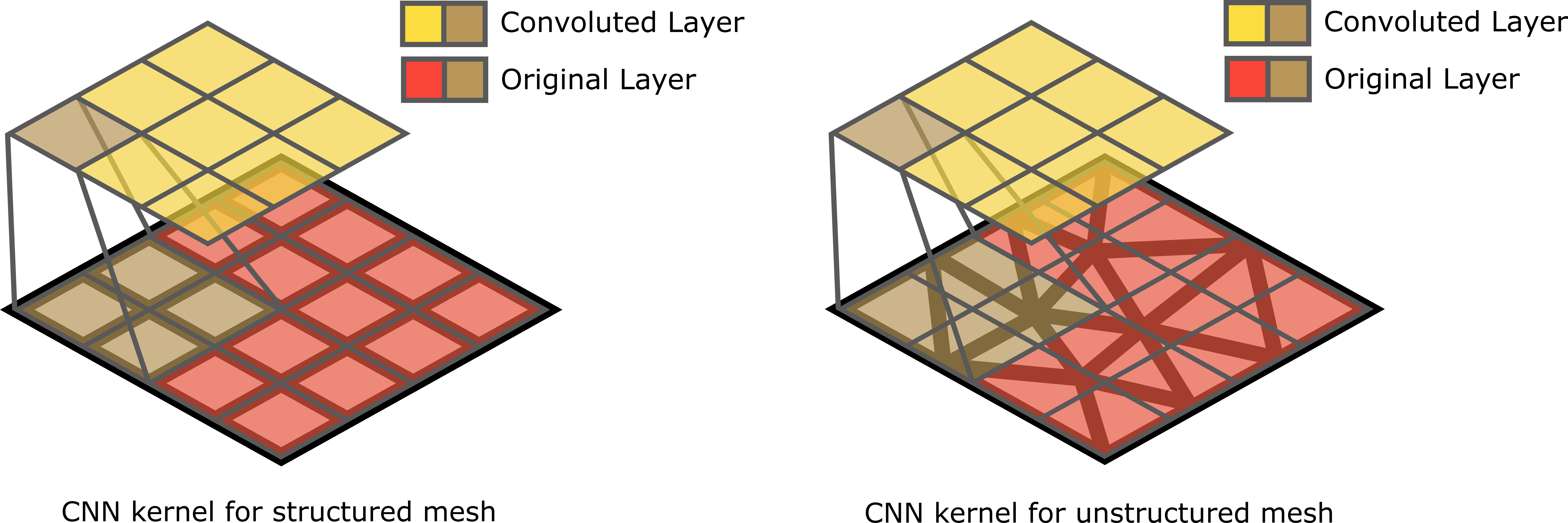}}
    \captionsetup{justification=centering}
    \caption{Visualization of CNN kernel attempting to perform dimensional compression upon a structured mesh (left) and an unstructured mesh (right). Note that the nature of a structured mesh enables direct implementation into a CNN kernel for dimensionality reduction. Specifically, a structured mesh is essentially a pixel-based image commonly found in the CNN literature. Alternatively, the unstructured mesh is not readily formulated as a pixel-based image, meaning that the CNN kernel cannot be immediately applied to the unstructured mesh.}
    \label{fig:cnn_for_structured_unstructured}
\end{figure}

Due to their remarkable ability to filter grid-based information, in an extensive amount of literature on autoencoder-based PMOR \cite{hasegawa2020mlrom,wiewel2019latent,kim2019deepfluids,lee2020deeplspg,lee2021deepconservation, fukami2020convolutional}, CNNs have been heavily relied upon as the backbone for the development of autoencoder architectures. However, since CNNs were primarily developed to analyze pixel-based images, they are dependent upon the inputs being a structured grid. As a result, PMOR methods leveraging CNNs are not readily applicable to unstructured meshes (see Figure 
\ref{fig:cnn_for_structured_unstructured}).

\begin{figure}[!htb]
    \centering
    \centerline{\includegraphics[scale=.75]{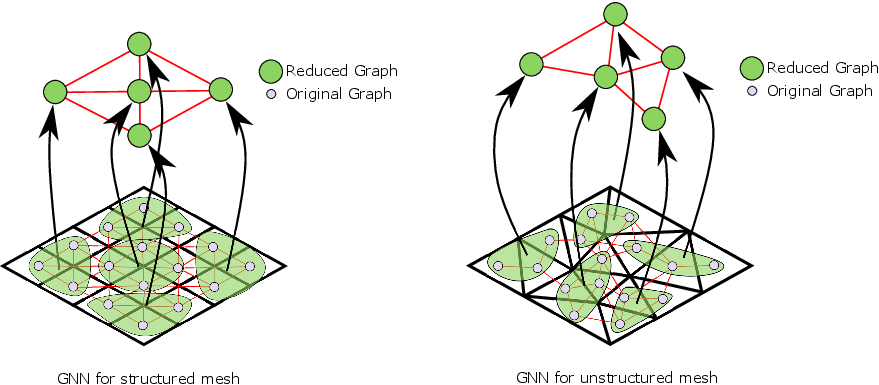}}
    \captionsetup{justification=centering}
    \caption{Visualization of graph autoencoder performing dimensional compression upon a structured mesh (left) and an unstructured mesh (right). Note that, unlike CNNs, the GNN framework readily accepts both structured and unstructured meshes by modeling collocation points as graph nodes and prescribing edges between the collocation points close to each other in the domain.}
    \label{fig:gnn_for_structured_unstructured}
\end{figure}

A common approach to enable the use of CNNs on unstructured meshes is to interpolate the unstructured mesh onto a structured grid, thereby creating a pixel-based representation suitable for deployment in CNN-based autoencoders
\cite{fresca2021comprehensive}. As will be demonstrated in Section \ref{sec:experiments}, this strategy often fails to have strong generalization performance. Our proposed method overcomes the need for structured meshes for the sake of autoencoder-based PMOR such that both structured and unstructured meshes can be inputs to the proposed autoencoder. Because GNNs have been developed for non-Euclidean data, such as unstructured meshes, we aim to leverage GNNs to perform dimensionality reduction (see Figure \ref{fig:gnn_for_structured_unstructured}). Given that unstructured meshes are commonly used in engineering applications, our approach can be widely extended to applications with arbitrary topology. Our proposed architecture follows an outline that is similar to graph U-nets \cite{gao2019unet}, multiscale graph autoencoders \cite{barwey2023multiscale}, and graph convolutional autoencoders for parameterized PDEs \cite{pichi2024graph}, wherein a hierarchy of graphs is generated, each with fewer nodes than the previous level. 

\subsection{Graph theory} \label{ssec:graphTheory}
Extensive reading on graph theory can be found in the works of \cite{hamilton2020graph} and \cite{battaglia2018graph}, but a brief overview is provided in this section to provide sufficient background for our graph autoencoder architecture. A graph is a tuple $\mathcal{G} = \{ \mathcal{V}, \mathcal{E} \}$, where $\mathcal{V}$ denotes the node set, $\vert \mathcal{V} \vert$ denotes the number of nodes in the graph, and $\mathcal{E}$ denotes the edge set, which is chosen to represent user-prescribed relationships between the nodes in the node set. Depending on the application, the graph (and the associated node and edge sets) can be used to represent a wide variety of concepts. For example, molecules can be modeled as a graph by representing atoms as nodes and bonds as edges \cite{Bongini2021molecule}, while social networks can be modeled as a graph by representing people as nodes and friendships as edges \cite{newman2002random}.

\begin{figure}[!htb] 
    \centering
    \includegraphics[scale=.25]{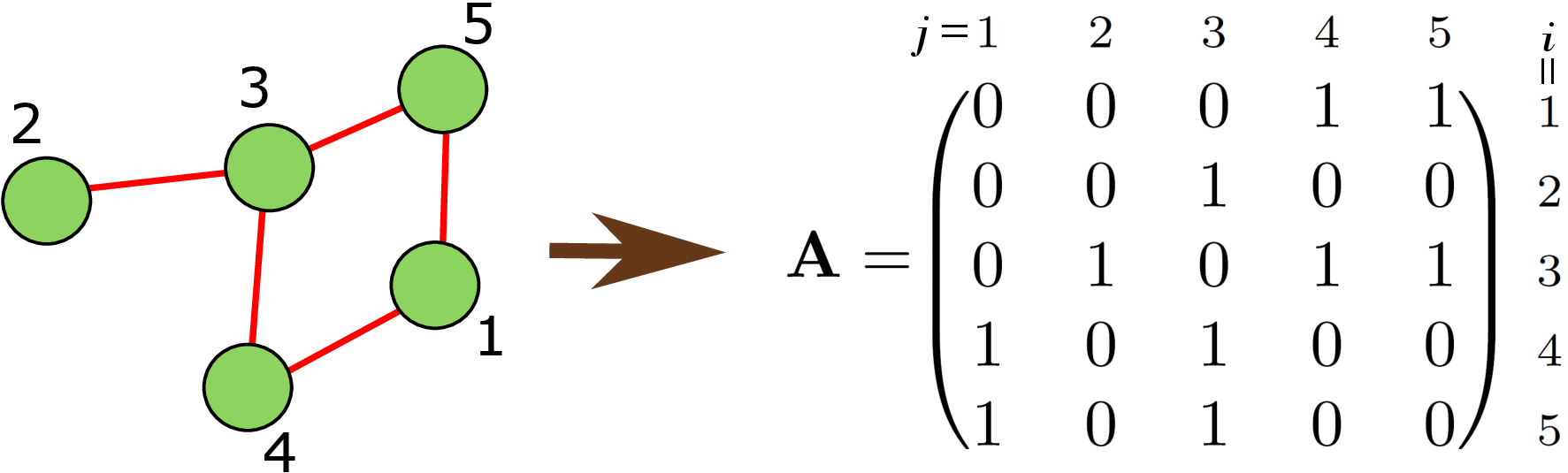}
    \caption{Formation of the adjacency matrix from a given graph.}
    \label{fig:adjacencyMatrix}
\end{figure}

The adjacency matrix, $\mathbf{A} = [a_{ij}] \in \mathbb{R}^{\vert \mathcal{V}\vert \times \vert \mathcal{V}\vert}$, is another way to represent the edge set of a graph. Consider the case where the nodes are indexed by a number, $i=1,\cdots,\vert \mathcal{V}\vert$. If nodes $i$ and $j$ in the graph are connected via an edge, i.e., if for $i,j \in \mathcal{V}$, we have $(i,j) \in \mathcal{E}$, the corresponding entry in the adjacency matrix is $a_{ij}=1$. Otherwise, we have $a_{ij}=0$. In this manuscript, we consider exclusively undirected graphs, meaning for any edge in the graph, $(i,j) \in \mathcal{E}$, we also have $(j,i) \in \mathcal{E}$. With this formulation, our adjacency matrix will be symmetric. A visualization of the construction of the adjacency matrix for a given graph is found in Figure \ref{fig:adjacencyMatrix}.

A node feature matrix, $\mathbf{X}\in \mathbb{R}^{\vert \mathcal{V} \vert \times N_F}$, can be utilized to prescribe feature information to the node set of the graph, where the $i^{\mathrm{th}}$ row of $\mathbf{X}$ denotes the node feature vector of the $i^{\mathrm{th}}$ node in the graph, and $N_F\in \mathbb{N}$ denotes the number of features prescribed to each node. 

\section{Dimension reduction via graph autoencoders} \label{sec:graphAutoencoder}

In this section, we develop the specifics of the graph autoencoder used in GD-LSPG. Upon spatial discretization of the physical domain, each collocation point (either a cell center in the FVM mesh or a mesh node in the FEM, for example) is represented by a node. We take the node set $\mathcal{V}$ to represent the collocation points in the discretized domain, i.e., $\vert \mathcal{V} \vert = N_c$. To emulate the manner in which CNNs filter information from neighboring grid points in the spatial discretization, we take the edge set, $\mathcal{E}$, to connect the node representation of collocation points within a user-defined radius of each other, i.e., 

\begin{equation} \label{eq:radius_graph}
    \mathcal{E} = \mathbf{Radius\_Graph}(\mathbf{Pos},r) = \{ \hspace{1mm}\forall\,(j,k): \hspace{2mm} j, k \in \mathcal{V}, \hspace{1mm} \vert \vert \mathbf{Pos}_j - \mathbf{Pos}_k \vert \vert \leq r \hspace{1mm} \},
\end{equation}
where $\mathbf{Pos}\in\mathbb{R}^{N_c \times n_d}$ is the matrix denoting the spatial positions of the node-representation of the collocation points in the discretization. Row $j$ of the matrix (i.e., $\mathbf{Pos}_j$) denotes the position of node $j\in\mathcal{V}$, $n_d \in \mathbb{N}$ denotes the spatial dimensionality of the modeled problem, $j$ and $k$ denote the indices of the corresponding nodes in the node set, $r\in\mathbb{R}$ denotes the user-defined radius, and $\vert \vert \boldsymbol{\cdot} \vert \vert : \mathbb{R}^{n_d} \rightarrow \mathbb{R}_+$ denotes the Euclidean norm. The adjacency matrix is used to represent the edge set in a matrix format. Alternatively, one could choose to employ an edge set defined by a fixed number of nearest neighbors or an edge set defined by the neighboring collocation points in the mesh. We choose here to define the edge set by the $\mathbf{Radius\_Graph}$ function \eqref{eq:radius_graph}, as it allows for user control over the number of edges prescribed to each node and is a notion that is not lost under the hierarchical compression used in the graph autoencoder described in Section \ref{ssec:SpectralClustering} (unlike an edge set defined by neighboring collocation points in the mesh), does not prescribe unnecessary edges or skewed graph topology near boundaries (unlike a fixed number of nearest neighbors), and ensures that the graph is undirected (a requirement for spectral clustering).

The feature matrix, $\mathbf{X}\in\mathbb{R}^{N_c\times n_q}$ is a matrix with the number of rows equal to the number of collocation points in the discretization, i.e., $N_c$, and the number of columns equal to the number of state variables in the governing PDE, $n_q$. In other words, the feature matrix $\mathbf{X}$ is the matrix version of the state vector $\mathbf{x}\in\mathbb{R}^{N}$ (with $N=n_q N_c$) introduced in Section \ref{sec:problemFormulation}. As a result, the formulation has a direct mapping between the state vector $\mathbf{x}$ and the node feature matrix $\mathbf{X}$ ($\mathrm{\mathbf{Matricize}}: \mathbf{x} \mapsto \mathbf{X}$, with $\mathrm{\mathbf{Matricize}}: \mathbb{R}^{N_c n_q} \rightarrow \mathbb{R}^{N_c \times n_q}$) and a direct mapping between the node feature matrix $\mathbf{X}$ and the state vector $\mathbf{x}$ ($\mathrm{\mathbf{Vectorize}}: \mathbf{X} \mapsto \mathbf{x}$, with $\mathrm{\mathbf{Vectorize}}: \mathbb{R}^{N_c \times n_q} \rightarrow \mathbb{R}^{N_c n_q}$).

Once a graph representation of a solution state is constructed, it can be encoded to a latent representation with a graph autoencoder following the general form of \eqref{eq:general_enc}-\eqref{eq:general_dec}. In the subsequent sections, we present the graph autoencoder used in the GD-LSPG framework and the specifics of the architecture of the encoder and the decoder. First, Section \ref{ssec:SpectralClustering} presents a hierarchical spectral clustering algorithm used by the autoencoder to generate a hierarchy of reduced graphs to emulate the compressive abilities of CNNs. Next, Section \ref{ssec:enc} details the encoder architecture and its deployment of the hierarchy of reduced graphs to create a low-dimensional embedding of the input graph. Then, Section \ref{ssec:dec} details the decoder architecture and its deployment of the hierarchy of reduced graphs in reverse order to reconstruct the original input graph from its latent representation. In our graph autoencoder, we include an additional layer with no trainable parameters for preprocessing and postprocessing in the encoder (Section \ref{ssec:preprocessing}) and the decoder (Section \ref{ssec:postprocessing}), respectively. Finally, Section \ref{ssec:trainAE} presents the training strategy deployed to optimize the training parameters of the encoder and decoder. Figure \ref{fig:ae_arch} provides a visual representation of the graph autoencoder deployed in GD-LSPG, with $n_{\ell}=3$ for demonstration purposes. 

\afterpage{
\clearpage
\begin{landscape}

\begin{figure}[!htb]
    \centering
    \centerline{\includegraphics[scale=1.0]{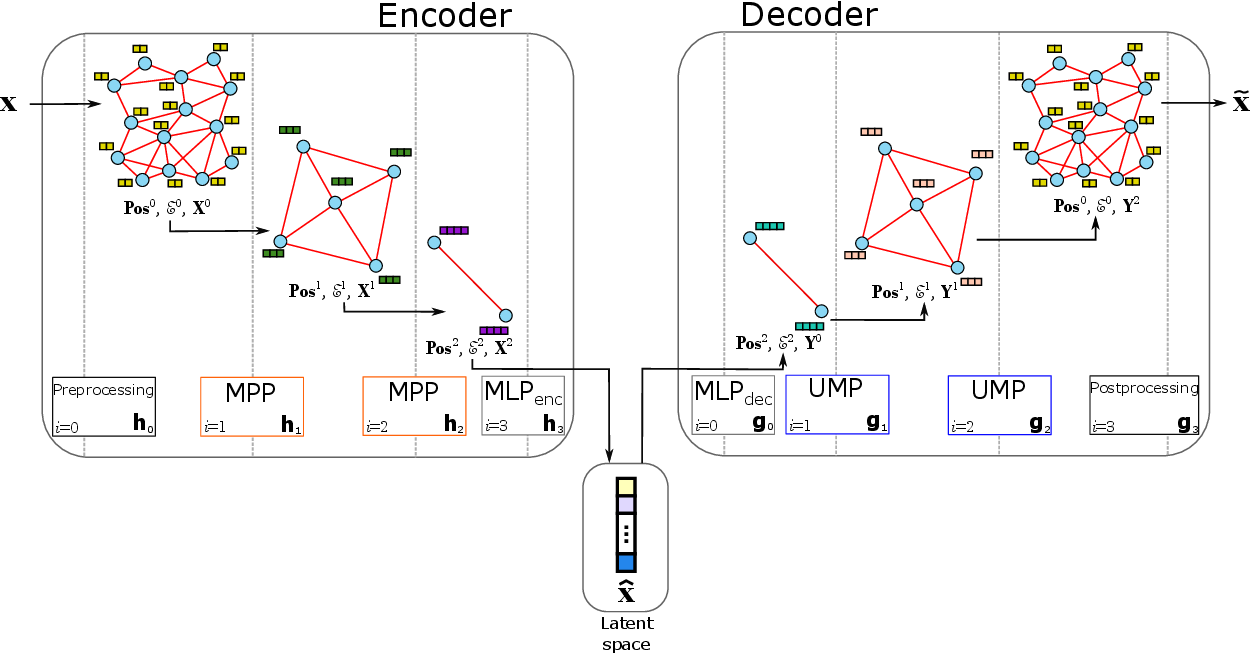}}
    \captionsetup{justification=centering}
    \caption{Graph autoencoder architecture deployed in GD-LSPG. Vertical dotted lines represent the separation between layers in the hierarchy of reduced graphs. Boxes over the vertical dotted lines at the bottom of the figure represent specific layers of the autoencoder described in this section. The encoder is the trained mapping between the high-dimensional state $\mathbf{x}$ and the low-dimensional latent space $\hat{\mathbf{x}}$. The encoder is comprised of a preprocessing layer to model the input state vector as a graph, followed by a series of trained message passing and pooling (MPP) layers to reduce the number of nodes in the graph, then a flattening and fully-connected/multilayer perceptron (MLP) layer. The resulting low-dimensional state vector is sent to the decoder, which is the trained mapping between the low-dimensional latent space $\hat{\mathbf{x}}$ and the reconstructed high-dimensional state $\tilde{\mathbf{x}}$. The decoder is comprised of a fully-connected/MLP layer, followed by a series of unpooling and message passing (UMP) layers to increase the number of nodes in the graph, and then a postprocessing layer to prepare the output for deployment in the time integration scheme. Note that the superscript $i$ in $\mathbf{Pos}^i$ and $\mathcal{E}^i$ represents the graph number in the hierarchy of graphs with $i=0$ denoting the original graph representing the discretized mesh.}
    \label{fig:ae_arch}
\end{figure}
\end{landscape}
}

\subsection{Generating a hierarchy of reduced graphs with spectral clustering} \label{ssec:SpectralClustering}
To compute a hierarchy of reduced graphs for the autoencoder used in GD-LSPG, at each level in the hierarchy, we aim to partition the graph into a pre-defined number of non-overlapping sets of strongly connected nodes. We then use the partitions to aggregate each cluster of nodes together into a single node at the next layer of the hierarchy, thereby reducing the number of nodes in the graph and the total dimension of the graph. In this study, we have chosen spectral clustering for two reasons. First, spectral clustering leverages knowledge of graph topology to compute cluster assignments, thereby considering the physical domain's inherent geometry during clustering. Second, its implementation is straightforward because it relies on the same edge sets used during message passing at each level of the hierarchy of reduced graphs. The interested reader is directed to \cite{vonluxburg2007spectral} for further discussion on the advantages of spectral clustering for graph-structured data. 

We consider the case where the encoder and the decoder each have $n_{\ell} \in \mathbb{N}$ layers. As will be presented in the subsequent sections, the encoder and decoder both have a fully-connected/multi-layer perceptron (MLP) layer along with $n_{\ell}-1$ layers with compressed graphs. Therefore, in this section, we aim to produce a hierarchy of reduced graphs composed of $n_{\ell}-1$ reduced graphs that result in a hierarchy of $n_{\ell}$ graphs, including the input graph of the discretized FOM (i.e., graph 0). The graphs in the encoder and the decoder will have the same topology but with reverse ordering. This means that the $i^{\mathrm{th}}$ graph in the hierarchy of the graphs of the encoder, $i=0,\cdots,n_{\ell}-1$, will be equivalent to the $(n_{\ell}-i-1)^{\mathrm{th}}$ graph of the decoder (refer to Figure \ref{fig:ae_arch}). In other words, the first graph of the decoder is the $(n_{\ell}-1)^{\mathrm{th}}$ (final) graph of the encoder, and the final graph of the decoder is the zeroth (original) graph of the encoder. Hence, we focus on building the hierarchy of the graphs for the encoder. This task can be achieved by minimizing the number of `broken' edges in the graph topology of the $(i-1)^{\mathrm{th}}$ layer to form the clusters for the $i^{\mathrm{th}}$ layer's graph. The graph representation of the FOM, $\mathcal{G}^0=\left(\mathcal{V}^0,\mathcal{E}^0\right)$ are given from \eqref{eq:radius_graph} using the discretized mesh of the FOM. In addition, the number of nodes in the layers $i=1,\cdots,n_{\ell}-1$, i.e., $\left\{\vert \mathcal{V}^1\vert,\vert \mathcal{V}^2\vert,\cdots,\vert \mathcal{V}^{n_{\ell}-1}\vert\right\}$, along with the radius used in \eqref{eq:radius_graph}, i.e., $\left\{ r^0, r^1, \ldots, r^{n_{\ell-1}} \right\}$, are hyperparameters prescribed \textit{a priori}, dictating the amount of reduction performed and the number of edges at each layer in the hierarchy. These hyperparameters in our graph autoencoder framework can be viewed to be equivalent to the kernel size and stride used to construct a CNN-based autoencoder.

At layer $i\in\{1,\cdots,n_{\ell}-1\}$ of the encoder, we aim to reduce the number of nodes from $\vert \mathcal{V}^{i-1} \vert$ in layer $i-1$ to $\vert \mathcal{V}^{i} \vert$ in layer $i$, with $\vert \mathcal{V}^{i} \vert < \vert \mathcal{V}^{i-1} \vert$, by clustering nodes that are strongly connected. This action is carried out by forming $\vert \mathcal{V}^{i} \vert$ clusters, i.e., $\mathcal{A}_1^{i-1},\mathcal{A}_2^{i-1},\cdots,\mathcal{A}_{\vert\mathcal{V}^i\vert}^{i-1}$ with the following conditions,
\begin{align}\label{eq:cluster_cond}
    \begin{cases}
        \vert \mathcal{A}_j^{i-1}\vert \geq 1, & j=1,\cdots,\vert\mathcal{V}^i\vert\\
        \mathcal{A}_j^{i-1}\subset  \mathcal{V}^{i-1}, &j=1,\cdots,\vert\mathcal{V}^i\vert\\
        \mathcal{A}_j^{i-1}\cap \mathcal{A}_k^{i-1}=\emptyset, &j\neq k,\; j,k=1,\cdots,\vert\mathcal{V}^i\vert\\
        \mathcal{A}_1^{i-1}\cup\mathcal{A}_2^{i-1}\cup\cdots \cup \mathcal{A}_{\vert\mathcal{V}^i\vert}^{i-1}=\mathcal{V}^{i-1},
    \end{cases}
\end{align}
which ensures that all clusters are a non-empty subsets of the node set of layer $i-1$, the intersection of any two distinct clusters is the empty set, and the union of all clusters is equal to the node set of layer $i-1$. In \eqref{eq:cluster_cond}, $\vert \boldsymbol{\cdot} \vert$ denotes the cardinality of the set.
To define clusters, $\mathcal{A}_1^{i-1},\mathcal{A}_2^{i-1},\cdots,\mathcal{A}_{\vert\mathcal{V}^i\vert}^{i-1}$, we minimize the  $\mathrm{\mathbf{RatioCut}}$ function \cite{wei1989ratiocut},

\begin{equation}\label{eq:RatioCut}
    \mathrm{\mathbf{RatioCut}}: \left( \mathcal{V}^{i-1} ,\,\mathcal{E}^{i-1} \right) \mapsto
        \frac{1}{2} \sum_{k=1}^{\vert \mathcal{V}^{i} \vert}\frac{ \vert \forall(u,v) \in \mathcal{E}^{i-1} : u \in \mathcal{A}_k^{i-1}, v \in \mathcal{\bar{A}}_k^{i-1} \vert}{\vert \mathcal{A}_k^{i-1} \vert},
\end{equation}
that tends to measure the number of broken edges for the given cluster choice, where $\mathcal{E}^{i-1}$ denotes the edge set of the graph at the $(i-1)^{\mathrm{th}}$ level in the hierarchy, $(u,v)$ represents any existing edge in $\mathcal{E}^{i-1}$ connecting nodes $u$ and $v$, $\mathcal{A}_k^{i-1} \subset \mathcal{V}^{i-1}$ denotes a subset of nodes in the graph at the $(i-1)^{\mathrm{th}}$ level in the hierarchy and $\mathcal{\bar{A}}_k^{i-1}=\mathcal{V}^{i-1}\backslash\mathcal{A}_k^{i-1}$ denotes the complement of the set $\mathcal{A}_k^{i-1}$ at the same level. Minimizing the $\mathrm{\mathbf{RatioCut}}$ from \eqref{eq:RatioCut} results in clusters of locally-connected nodes with relatively equal sizes \cite{hamilton2020graph}.

The number of distinct ways we can choose $\vert\mathcal{V}^{i}\vert$ clusters from $\vert\mathcal{V}^{i-1}\vert$ nodes while satisfying conditions of \eqref{eq:cluster_cond} is determined from the Stirling number of the second kind \cite{rennie1969stirlingNumbers}, $S(n,k)=\frac{1}{k!}\sum_{j=0}^k{(-1)^j \,\binom{k}{j}\left(k-j\right)^n}$ with $n=\vert \mathcal{V}^{i-1}\vert$ and $k=\vert \mathcal{V}^{i}\vert$ resulting in an NP-hard minimization problem \cite{hamilton2020graph,vonluxburg2007spectral}. As discussed in \cite{vonluxburg2007spectral}, spectral clustering introduces a relaxation on the minimization problem to eliminate its discrete nature. The departure from a discrete set allows the user to perform an eigenvalue analysis on the graph to generate the clusters. Specifically, spectral clustering groups nodes together based on their spectral features defined by the eigenvectors of the Laplacian using any one of a wide variety of standard clustering techniques. In this study, $K-$means clustering \cite{macqueen1967kmeans} is chosen as it is commonly used for this application in the literature \cite{hamilton2020graph,vonluxburg2007spectral}. The algorithm for building the hierarchy of reduced graphs is summarized in Algorithm \ref{alg:spectral}.

\begin{figure}[!htb]
    \centering
    \centerline{\includegraphics[scale=.55]{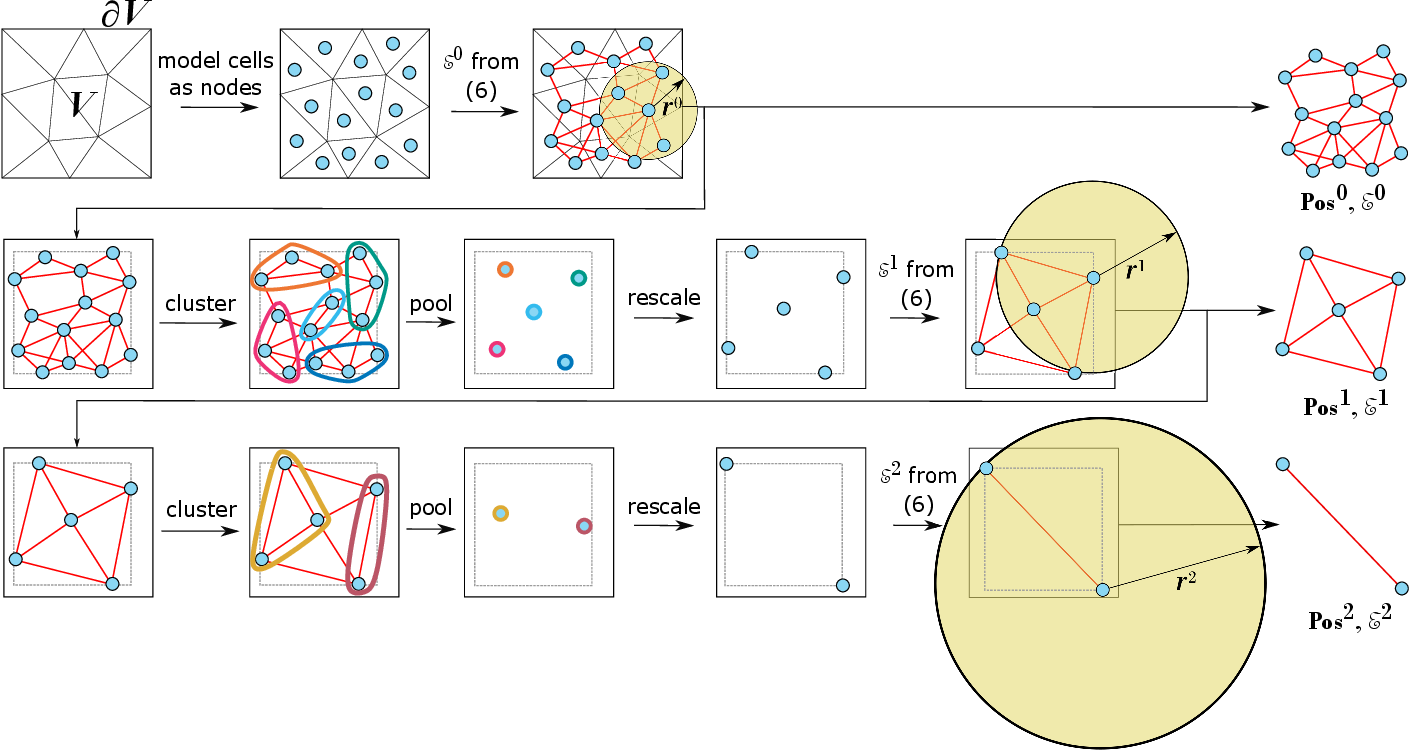}}
    \captionsetup{justification=centering}
    \caption{A visual representation of Algorithm \ref{alg:spectral} for generating a hierarchy of reduced meshes. In this figure, it is assumed that the collocation points are cells in the FVM discretization, however the same principles apply for discretization used in various other methods in computational mechanics. The input mesh (top left, generated for a domain $V$ and its boundary $\partial V$ represented by the solid black box) is modeled as the layer ``0'' graph in which the collocation points of the discretized domain form the node set $\mathcal{V}^0$, and the edge set $\mathcal{E}^0$ (and subsequently the associated adjacency matrix $\mathbf{A}^0$) are determined from \eqref{eq:radius_graph} by determining the nodes that fall within $r^0$ distance of each other. At layer $i$ ($i=0,\cdots,n_{\ell}-2$), nodes are partitioned into clusters using a spectral clustering algorithm to achieve a dimensionally-reduced graph from that of layer $i$. The nodal positions of the graph of layer $i+1$, obtained from the arithmetic mean position of node clusters from layer $i$, are rescaled to ensure that the maximum and minimum coordinates of the nodes in graph layer $i+1$ are equal to those of layer $i$, where the maximum and minimum values of the $x$ and $y$ coordinates are represented by the grey dotted box. Finally, the edge set of the reduced graph of layer $i+1$, i.e., $\mathcal{E}^{i+1}$, is determined from \eqref{eq:radius_graph}.}
    \label{fig:mesh_decomp}
\end{figure}

The method of spectral clustering from \cite{hamilton2020graph} is leveraged in this study and makes up steps $1-4$ of Algorithm \ref{alg:spectral}, where $\mathbf{Pos}^i \in \mathbb{R}^{\vert \mathcal{V}^i \vert \times n_d}$ denotes the matrix of spatial coordinates for the graph at the $i^{\mathrm{th}}$ level in the hierarchy, $\mathbf{A}^i$ is the adjacency matrix of the $i^{\mathrm{th}}$ layer in the hierarchy generated by the edge set, $\mathcal{E}^i$, of layer $i$, previously defined in \eqref{eq:radius_graph}, $r^i \in \mathbb{R}_+$ denotes a user-prescribed radius to be used in \eqref{eq:radius_graph}, $\mathbf{S}^i\in \mathbb{R}^{ \vert \mathcal{V}^{i+1} \vert \times \vert \mathcal{V}^{i} \vert}$ denotes the assignment matrix of the $i^{\mathrm{th}}$ layer of the hierarchy which is used to assign each node in layer $i$ to a cluster in layer $i+1$, and thus a portion of a single node at the layer $i+1$ in the hierarchy. The assignment matrix (defined to act as an arithmetic mean on the features of the nodes in each cluster) is used to cluster and decrease the number of nodes in the graph at each step in the hierarchy of reduced graphs. In Algorithm \ref{alg:spectral}, $\mathbf{D}^i \in \mathbb{R}^{\vert \mathcal{V}^{i} \vert \times \vert \mathcal{V}^{i} \vert}$ is the diagonal degree matrix representing the number of edges connected to each node in the $i^{\mathrm{th}}$ layer of the hierarchy, $\mathbf{L}^i = \mathbf{D}^i - \mathbf{A}^i$ is the Laplacian of the graph associated with the $i^{\mathrm{th}}$ layer in the hierarchy, $\mathbf{B}^i \in \mathbb{R}^{\vert \mathcal{V}^{i} \vert \times \left(\vert \mathcal{V}^{i+1} \vert - 1\right)}$ denotes the spectral node feature matrix formed by the eigenvectors associated with the $\vert \mathcal{V}^{i+1} \vert - 1$ smallest eigenvalues of $\mathbf{L}^i$, excluding the smallest. 
According to \cite{vonluxburg2007spectral}, the smallest eigenvalue of the unnormalized Laplacian is simply zero and can therefore be neglected. As dimensional compression is performed in the hierarchy of reduced graphs, the nodes of the graphs deeper in the hierarchy tend to become closer together due to the nature of the positions of each node being computed based on the arithmetic mean of the positions of their corresponding cluster in the previous layer. To avoid the natural accumulation of the nodes to a smaller spatial domain, a rescaling operator, i.e.,
$\mathbf{Rescale}: \mathbb{R}^{\vert \mathcal{V}^{i+1} \vert \times n_d} \rightarrow \mathbb{R}^{\vert \mathcal{V}^{i+1} \vert \times n_d}$ is used at each layer to rescale $\mathbf{Pos}^{i+1}$ such that the maximum and minimum values of the coordinates match that of the previous layer in the hierarchy. This algorithm is visually represented in Figure \ref{fig:mesh_decomp}. 

The construction of the hierarchy of reduced graphs is performed in the offline stage. While the hierarchy of graphs will be utilized in the encoder and decoder, the architecture of the encoder and decoder does not influence the spectral clustering step. The graph autoencoder must use only the exact hierarchy of graph topologies seen during training, i.e., if the original mesh is changed or if the hyperparameters of the hierarchy of reduced graphs are modified, the graph autoencoder must be retrained. 

\begin{algorithm}
\caption{Hierarchical spectral clustering for graph reduction} \label{alg:spectral}
\KwData{$\mathbf{Pos}^0,n_{\ell},\vert \mathcal{V}^{1} \vert, \ldots, \vert \mathcal{V}^{n_{\ell}-1} \vert, r^0, \ldots, r^{n_{\ell}-1}$}
\KwResult{$\mathcal{E}^0,\ldots,\mathcal{E}^{n_{\ell}-1}, \mathbf{S}^0,\ldots, \mathbf{S}^{n_{\ell}-2}, \mathbf{Pos}^1, \ldots, \mathbf{Pos}^{n_{\ell}-1}$}
Initialize $i \leftarrow 0$\;
Initialize $\mathcal{E}^0 \leftarrow \mathbf{Radius\_Graph}\left(\mathbf{Pos}^0, r^0 \right)$ \;
\While{$i < n_{\ell}-1$} {
    \begin{enumerate}

    \item Compute the adjacency matrix, $\mathbf{A}^i$, from $\mathcal{E}^i$\;
    
    \item Compute the graph Laplacian, $\mathbf{L}^i \leftarrow \mathbf{D}^i - \mathbf{A}^i$ \;

    \item Form the spectral node feature matrix, $\mathbf{B}^i \in \mathbb{R}^{\vert \mathcal{V}^{i} \vert \times \left(\vert \mathcal{V}^{i+1} \vert - 1\right)}$, with the eigenvectors associated with the $\vert \mathcal{V}^{i+1} \vert - 1$ smallest eigenvalues of $\mathbf{L}^i$ (excluding the smallest) as its columns\;

    \item Obtain node clusters $\mathcal{A}^{i}_1, \mathcal{A}^{i}_2, \ldots, \mathcal{A}^{i}_{\vert \mathcal{V}^{i+1} \vert}$ by performing $K-$means clustering on the spectral node features (i.e., the rows of $\mathbf{B}^i$)\;

    \item Generate the assignment matrix, $\mathbf{S}^i \in \mathbb{R}^{\vert \mathcal{V}^{i+1} \vert \times \vert \mathcal{V}^{i} \vert}$, such that $\mathbf{S}^i_{jk} = \frac{1}{\vert \mathcal{A}^i_j \vert}$ if $k \in \mathcal{A}^i_j$, and $\mathbf{S}^i_{jk} = 0$, otherwise.

    \item $\mathbf{Pos}^{i+1} \leftarrow \mathbf{Rescale} \left( \mathbf{S}^i \, \mathbf{Pos}^i \right)$\;

    \item Compute the edge set for layer $i+1$ based on nearest neighbors within the radius $r^{i+1}$, $\mathcal{E}^{i+1} \leftarrow \mathbf{Radius\_Graph}\left(\mathbf{Pos}^{i+1}, r^{i+1} \right)$ \;
    
    \item $i \leftarrow i + 1$\;
    
    \end{enumerate}
}
\end{algorithm}

\subsection{Encoder architecture} \label{ssec:enc}

The encoder architecture deploys the hierarchy of reduced graphs computed via the procedure from Section \ref{ssec:SpectralClustering}. The encoder consists of layers $i=0,\ldots,n_{\ell}$. The zeroth layer ($i=0$), outlined in Section \ref{ssec:preprocessing}, is a preprocessing layer that tailors the input data $\mathbf{x}$ to the form suited for the graph autoencoder. Layers $i=1,\ldots,n_{\ell}-1$, outlined in Section \ref{ssec:mpp}, leverage the hierarchy of reduced graphs from Section \ref{ssec:SpectralClustering} to perform message passing and pooling (MPP) operations that reduce the dimension of the system. The final layer of the encoder ($i=n_{\ell}$), as outlined in Section \ref{ssec:mlpc}, utilizes an MLP  to arrive at the low-dimensional embedding $\hat{\mathbf{x}}$.

\subsubsection{Preprocessing  -- Layer 0} \label{ssec:preprocessing}
The preprocessing layer of the encoder ($i=0$) encompasses two operators, $\mathrm{\mathbf{Matricize}}$ and $\mathrm{\mathbf{Scale}}$, acting on the input vector $\mathbf{x}$. For a FOM with $n_q$ state variables, the $\mathrm{\mathbf{Matricize}}$ operator is used to convert $\mathbf{x}\in\mathbb{R}^{n_q N_c}$ to the node feature matrix $\mathbf{X}\in\mathbb{R}^{N_c\times n_q}$ in which each column of the matrix represents the nodal values of one state variable. If the FOM consists of only one state variable (i.e., $n_q=1$), $\mathbf{X}=\mathbf{x}$, and the $\mathrm{\mathbf{Matricize}}$ operator will be the identity operator. As defined in \eqref{eq:scale_preproc}, the $\mathrm{\mathbf{Scale}}$ operator acts on the resulting node feature matrix to improve the numerical stability of training, as is commonly performed in the literature \cite{lee2020deeplspg,lee2021deepconservation}, 

\begin{equation}\label{eq:scale_preproc}
    \mathrm{\mathbf{Scale}}: \mathbf{X}_{ij}^0 \mapsto \frac{\mathbf{X}_{ij}^0 - \mathcal{X}^{\mathrm{min}}_{j}}{\mathcal{X}^{\mathrm{max}}_{j} - \mathcal{X}^{\mathrm{min}}_{j}}, \qquad i=1,\ldots,N_c,\;j=1,\ldots,n_q
\end{equation}
where $\mathrm{\mathbf{Scale}}: \mathbb{R} \rightarrow [0,1]$ is an element-wise scaling operator acting on the elements of $\mathbf{X}$, and $\mathcal{X}_{j}^{\mathrm{max}}$, $\mathcal{X}_{j}^{\mathrm{min}} \in \mathbb{R}$ denote the maximum and minimum values, respectively, of the $j^{\mathrm{th}}$ feature (i.e., $j^{\mathrm{th}}$ column of matrix $\mathbf{X}$) in the solution states used to train the autoencoder, which are determined and stored before training begins. The resulting form of the preprocessing layer is

\begin{equation}
    \mathbf{h}_0: \left(\mathbf{x}; \mathbf{\Theta}_0 \right) \mapsto \mathrm{\mathbf{Scale}}\left( \cdot \right) \circ \mathrm{\mathbf{\mathbf{Matricize}}} \left( \mathbf{x} \right),
\end{equation}
where $\mathbf{h}_{0}: \mathbb{R}^{N_c n_q} \rightarrow \mathbb{R}^{N_c \times n_q}$, and $\mathbf{\Theta}_0 = \emptyset$ is the empty set, as the preprocessing layer does not have trainable weights and biases.

\subsubsection{Message passing and pooling (MPP) -- Layers 1$,\ldots,n_{\ell}-$1} \label{ssec:mpp}
The MPP layer consists of two processes, where each relies upon the hierarchy of reduced graphs computed in Section \ref{ssec:SpectralClustering}. The first operation is a message passing operation, wherein nodes connected by an edge exchange information with each other to obtain information about nearby nodes. The optimal information exchange is obtained from training the autoencoder. In the encoder, the message passing operation in layer $i$ increases the number of features associated with each node from $N_F^{i-1}\in\mathbb{N}$ to $N_F^{i}\in\mathbb{N}$. We take our message passing operation to be a mean aggregation SAGEConv from \cite{hamilton2018graphsage}, which applies updates to each node based on the arithmetic mean of its neighbors' features, i.e.,

\begin{equation}\label{eq:SAGEConv}
    \mathrm{\mathbf{MP}}^{i}_{\mathrm{enc}}: (\mathbf{X}^{i-1}; \boldsymbol{\Theta}_i) \mapsto \sigma \left( \mathbf{X}^{i-1}_j \mathbf{W}^{i}_{1} +  \left(\mathrm{mean}_{n\in\mathcal{K}^{i-1}(j)} \mathbf{X}^{i-1}_n \right) \mathbf{W}^{i}_{2} \right), \qquad j=1,\ldots,\vert \mathcal{V}^{i-1} \vert,
\end{equation}
with $\mathrm{\mathbf{MP}}^{i}_{\mathrm{enc}}: \mathbb{R}^{\vert \mathcal{V}^{i-1} \vert \times N_F^{i-1}} \times \mathbb{R}^{N_F^{i-1} \times N_F^{i}}\times \mathbb{R}^{N_F^{i-1}\times N_F^{i}} \rightarrow \mathbb{R}^{\vert \mathcal{V}^{i-1} \vert \times N_F^{i}}$, where $\mathbf{X}^{i-1} \in \mathbb{R}^{\vert \mathcal{V}^{i-1} \vert \times N_F^{i-1}}$ denotes the input node feature matrix to the $i^{\mathrm{th}}$ layer, the subscripts $j$ and $n$ denote the $j^{\mathrm{th}}$ and $n^{\mathrm{th}}$ rows of $\mathbf{X}^{i-1}$, $\mathbf{W}^{i}_{1}$, $\mathbf{W}^{i}_{2} \in \mathbb{R}^{N_F^{i-1}\times N_F^{i}}$ denote the weights with $\mathbf{\Theta}_i = \{\mathbf{W}^{i}_{1}, \mathbf{W}^{i}_{2}\}$ denoting the set of weights for the $i^{\mathrm{th}}$ MPP layer, $\mathcal{K}^{i-1}(j)$ denotes the set of nodes connected to node $j$ based on the adjacency matrix $\mathbf{A}^{i-1}$, where $j\in \mathbb{N}$ denotes the $j^{\mathrm{th}}$ node in the graph at layer $i-1$, and $\sigma: \mathbb{R} \rightarrow \mathbb{R}$ denotes the element-wise activation function, chosen here to be the exponential linear unit (ELU) due to its continuously differentiable property \cite{clevert2016elu}. The SAGEConv function described in \eqref{eq:SAGEConv} includes a loop over all nodes $j \in \mathcal{V}^{i-1}$, where for each node, the $j^{\mathrm{th}}$ row of the output $\mathbf{\bar{X}}^{i-1}$ of the message passing operation is calculated. The output of \eqref{eq:SAGEConv} has the same number of rows as its input, $\mathbf{X}^{i-1}$, but can have a different number of features (i.e., $N_F^i$ is not necessarily equal to $N_F^{i-1}$).

\begin{figure}[!htb]
    \centering
    \centerline{\includegraphics[scale=.55]{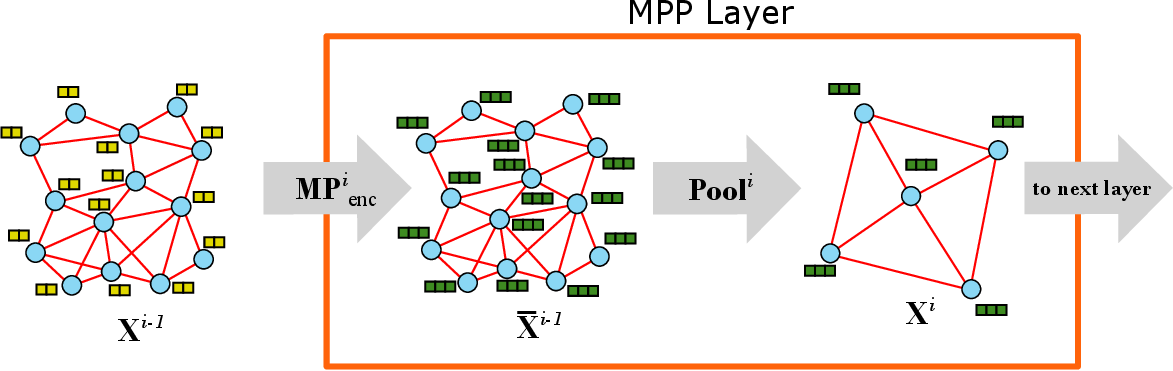}}
    \captionsetup{justification=centering}
    \caption{MPP layer used in the encoder of our graph autoencoder. The layer accepts a graph as input and performs message passing to exchange information between locally connected nodes. Next, the graph nodes are pooled together based on their clusters from the hierarchical spectral clustering algorithm. This pooling operation reduces the number of nodes in the graph to perform dimensional compression.}
    \label{fig:MPP_layer}
\end{figure}

The next step of the MPP layer is a pooling operation. In the pooling operation, the assignment matrices from Section \ref{ssec:SpectralClustering} are used to reduce the number of nodes in a graph. By construction, the assignment matrices are equivalent to an arithmetic mean operation. As a result, we use them to compute the arithmetic mean feature vector of each cluster to get $\mathbf{X}^i$, i.e.,

\begin{equation}
    \boldsymbol{\mathrm{Pool}}^i : \left(\mathbf{\bar{X}}^{i-1} \right) \mapsto \mathbf{S}^{i-1} \mathbf{\bar{X}}^{i-1},
\end{equation}
with $\mathrm{\mathbf{Pool}}^i: \mathbb{R}^{\vert \mathcal{V}^{i-1} \vert \times N_F^{i}} \rightarrow \mathbb{R}^{\vert \mathcal{V}^{i} \vert \times N_F^{i}}$, where $\mathbf{\bar{X}}^{i-1}\in \mathbb{R}^{\vert \mathcal{V}^{i-1} \vert \times N_F^{i}}$ denotes the output of the message passing operation, $\mathrm{\mathbf{MP}}^i_{\mathrm{enc}}$, and $ \mathbf{S}^{i-1} \in \mathbb{R}^{\vert \mathcal{V}^{i} \vert \times \vert \mathcal{V}^{i-1} \vert}$ is the assignment matrix precomputed by the spectral clustering algorithm in Section \ref{ssec:SpectralClustering}. The full MPP layer takes the form,

\begin{equation}
    \mathrm{\mathbf{h}}_{i}: \left( \mathbf{X}^{i-1}; \mathbf{\Theta}_i \right) \mapsto \mathrm{\mathbf{Pool}}^{i} \left( \cdot \right) \circ \mathrm{\mathbf{MP}}^{i}_{\mathrm{enc}} \left( \mathbf{X}^{i-1} ; \boldsymbol{\Theta}_i \right),
\end{equation}
with $\mathbf{h}_i: \mathbb{R}^{\vert \mathcal{V}^{i-1} \vert \times N_F^{i-1}} \times \mathbb{R}^{N_F^{i-1} \times N_F^{i}}\times \mathbb{R}^{N_F^{i-1}\times N_F^{i}} \rightarrow \mathbb{R}^{\vert \mathcal{V}^i \vert \times N_F^i}$. Hence, the MPP layer, as visually represented in Figure \ref{fig:MPP_layer}, decreases the number of nodes in a given graph and increases the number of features associated with each node.

\subsubsection{Fully-connected layer: compression -- Layer $n_{\ell}$} \label{ssec:mlpc}
In the final layer of the encoder ($i=n_{\ell}$), we first flatten the input matrix $\mathbf{X}^{n_{\ell}-1} \in \mathbb{R}^{\vert \mathcal{V}^{n_{\ell}-1} \vert \times N_F^{n_{\ell}-1}} $ to a vector-representation, i.e., $\mathrm{\mathbf{Flatten}}: \mathbf{X}^{n_{\ell}-1} \mapsto \mathbf{\mathbf{\bar{x}}}^{n_{\ell}-1}$, where $\mathbf{\bar{x}}^{n_{\ell}-1} \in \mathbb{R}^{\vert \mathcal{V}^{n_{\ell}-1} \vert N_F^{n_{\ell}-1}}$. Here, we note that the $\mathrm{\mathbf{Flatten}}$ operator is similar to the $\mathrm{\mathbf{Vectorize}}$ operator, but with dimensions different than the node feature matrix of the full-order solution. Next, a fully-connected/MLP layer is applied to the flattened state to compress it to a low-dimensional vector representation, i.e., 

\begin{equation}
    \mathrm{\mathbf{MLP}}_{\mathrm{enc}}: \left( \mathbf{\bar{x}}^{n_{\ell}-1}; \boldsymbol{\Theta}_{n_{\ell}} \right) \mapsto \mathbf{W}^{n_{\ell}} \mathbf{\bar{x}}^{n_{\ell}-1},
\end{equation}
with $\mathrm{\mathbf{MLP}}_{\mathrm{enc}}: \mathbb{R}^{\vert \mathcal{V}^{n_{\ell}-1} \vert N_F^{n_{\ell}-1}} \times \mathbb{R}^{M \times \vert \mathcal{V}^{n_{\ell}-1} \vert N_F^{n_{\ell}-1}} \rightarrow \mathbb{R}^M$, where $\mathbf{W}^{n_{\ell}} \in \mathbb{R}^{M \times \vert \mathcal{V}^{n_{\ell}-1} \vert N_F^{n_{\ell}-1}}$ denote the weights with $\mathbf{\Theta}_{n_{\ell}} = \{\mathbf{W}^{n_{\ell}}\}$. Note that no activation function is applied to the output, as the inclusion of an activation function was empirically found to be prone to vanishing gradients during time integration when performed in the manner outlined in Section \ref{sec:TimeIntegration}. Furthermore, no bias term is included, as it was empirically found to be unnecessary. The final layer of the encoder architecture takes the form
\begin{equation}
    \mathbf{h}_{n_{\ell}}: \left( \mathbf{X}^{n_{\ell}-1}; \mathbf{\Theta}_{n_{\ell}} \right) \mapsto \mathrm{\mathbf{MLP}}_{\mathrm{enc}}\left( \cdot \, ; \boldsymbol{\Theta}_{n_{\ell}}\right)  \circ \mathrm{\mathbf{Flatten}} \left(\mathbf{X}^{n_{\ell}-1}\right),
\end{equation}
with $\mathbf{h}_{n_{\ell}}: \mathbb{R}^{\vert \mathcal{V}^{n_{\ell}-1} \vert \times N_F^{n_{\ell}-1}} \times \mathbb{R}^{M \times \vert \mathcal{V}^{n_{\ell}-1} \vert N_F^{n_{\ell}-1}} \rightarrow \mathbb{R}^{M}$. The output of this layer, $\mathbf{\hat{x}} \in \mathbb{R}^M$, is the low-dimensional latent representation of the solution state. 

\subsection{Decoder architecture} \label{ssec:dec}

Much like the encoder, the decoder architecture deploys the hierarchy of reduced graphs from Section \ref{ssec:SpectralClustering}. The decoder consists of layers $i=0,\ldots,n_{\ell}$. The zeroth layer ($i=0$), outlined in Section \ref{ssec:mpe}, utilizes an MLP to reconstruct a small graph from the low-dimensional latent representation, $\hat{\mathbf{x}}$. Layers $i=1,\ldots,n_{\ell}-1$, outlined in Section \ref{ssec:ump}, leverage the hierarchy of reduced graphs from Section \ref{ssec:SpectralClustering} in reverse order to perform unpooling and message passing (UMP). The final layer of the decoder ($i=n_{\ell}$), outlined in Section \ref{ssec:postprocessing}, is a postprocessing layer that restructures the output graph into a state vector for deployment in the time integration scheme.

\subsubsection{Fully-connected layer: expansion  -- Layer 0} \label{ssec:mpe}
The zeroth layer of the decoder ($i=0$) entails two functions. It first applies a fully-connected/MLP layer to the latent representation,

\begin{equation}
    \mathrm{\mathbf{MLP}}_{\mathrm{dec}}: \left( \mathbf{\hat{x}}; \boldsymbol{\Omega}_0 \right) \mapsto \sigma \left( \mathbcal{W}^{0} \mathbf{\hat{x}} \right) ,
\end{equation}
with $\mathrm{\mathbf{MLP}}_{\mathrm{dec}}: \mathbb{R}^M \times \mathbb{R}^{\vert \mathcal{V}^{n_{\ell}-1} \vert N_F^{n_{\ell}-1} \times M } \rightarrow \mathbb{R}^{\vert \mathcal{V}^{n_{\ell}-1} \vert N_F^{n_{\ell}-1}}$, where $\mathbcal{W}^{0} \in \mathbb{R}^{\vert \mathcal{V}^{n_{\ell}-1} \vert N_F^{n_{\ell}-1} \times M }$ denote the weights of the MLP layer of the decoder, with $\mathbf{\Omega}_{0} = \{\mathbcal{W}^{0}\}$, and $\sigma: \mathbb{R} \rightarrow \mathbb{R}$ denotes the element-wise activation function. Additionally, note that, much like in the encoder, the fully-connected/MLP layer does not have a bias, as it was empirically noticed to be unnecessary. An unflattening operator is then applied to the output of the fully-connected layer, $\bar{\mathbf{y}}^0$, to generate a node feature matrix corresponding to the $n_{\ell}-1$ graph in the hierarchy of reduced graphs, $\mathrm{\mathbf{Unflatten}}: \mathbf{\bar{y}}^0 \mapsto \mathbf{Y}^{0}$, with $\mathrm{\mathbf{Unflatten}}: \mathbb{R}^{\vert \mathcal{V}^{n_{\ell}-1} \vert N_F^{n_{\ell}-1}} \rightarrow \mathbb{R}^{\vert \mathcal{V}^{n_{\ell}-1} \vert \times N_F^{n_{\ell}-1}}$, where $\mathbf{\bar{y}}^0 \in \mathbb{R}^{\vert \mathcal{V}^{n_{\ell}-1} \vert N_F^{n_{\ell}-1}}$ denotes the output of $\mathrm{\mathbf{MLP}}_{\mathrm{dec}}$ and $\mathbf{Y}^0 \in \mathbb{R}^{\vert \mathcal{V}^{n_{\ell}-1} \vert \times N_F^{n_{\ell}-1}}$. We note that the unflattening operator is similar to the $\mathrm{\mathbf{Matricize}}$ operator introduced previously but applied to a vector with a size different from the full-order state vector. Ultimately, the first layer of the decoder takes the form,

\begin{equation}
    \mathbf{g}_0: \left(\mathbf{\hat{x}}; \boldsymbol{\Omega}_0\right) \mapsto \mathrm{\mathbf{Unflatten}} \left( \cdot \right) \circ \mathrm{\mathbf{MLP}}_{\mathrm{dec}}\left( \mathbf{\hat{x}}; \boldsymbol{\Omega}_0 \right),
\end{equation}
where $\mathbf{g}_{0}: \mathbb{R}^{M} \times \mathbb{R}^{\vert \mathcal{V}^{n_{\ell}-1} \vert N_F^{n_{\ell}-1} \times M } \rightarrow \mathbb{R}^{\vert \mathcal{V}^{n_{\ell}-1} \vert \times N_F^{n_{\ell}-1}}$.

\subsubsection{Unpooling and message passing (UMP) -- Layers 1$ ,\ldots,n_{\ell}-$1} \label{ssec:ump}
\begin{figure}[!htb]
    \centering
    \centerline{\includegraphics[scale=.55]{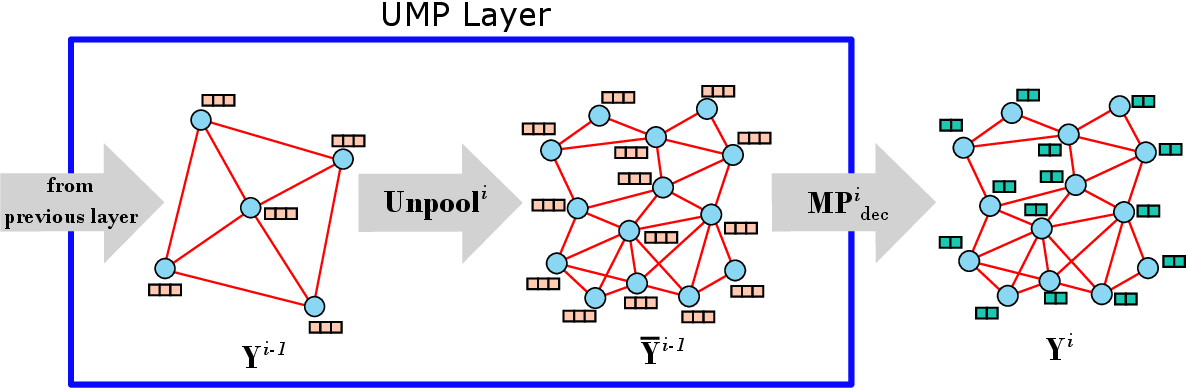}}
    \captionsetup{justification=centering}
    \caption{UMP layer used in the decoder of our graph autoencoder. The layer accepts a graph as an input and performs an unpooling operation to re-introduce nodes into the graph, thus increasing the dimension of the graph. Next, message passing is performed on the unpooled graph to exchange information between locally connected nodes.}
    \label{fig:UMP_layer}
\end{figure}

The next layers in the decoder architecture ($i=1,...,n_{\ell}-1$) consist of UMP layers. The first step in a UMP layer is to perform an unpooling operation, wherein nodes are reintroduced to the graph, and their feature vectors are interpolated. In layer $i$ of the decoder with $i=1,\cdots,n_{\ell}-1$, the unpooling operation receives the graph of layer $n_{\ell}-i$ as an input and outputs the graph of layer $n_{\ell}-i-1$ in the hierarchy of $n_{\ell}-1$ graphs of the encoder. For example, in Figure \ref{fig:ae_arch} with $n_{\ell}=3$, the input and output to the second layer of the decoder ($i=2$) are the graphs of 1st layer ($n_{\ell}-i=1$) and the zeroth layer ($n_{\ell}-i-1=0$) of the hierarchy of graphs in the encoder, respectively. For ease of notation, we introduce $\hat{i}=n_{\ell}-i$ as a counter used to denote the hierarchy of reduced graphs in the opposite order as the encoder. In the unpooling operation of layer $i$ of the decoder, a node's features of graph $\hat{i}$ are interpolated using the $k$-nearest neighbors of the node features of graph $\hat{i}-1$,

\begin{equation} \label{eq:unpooling}
    \mathrm{\mathbf{Unpool}}^i : \mathbf{Y}^{i-1} \mapsto \frac{\sum_{n\in \mathcal{N}^{\hat{i}-1}(j)} \mathbf{w}\left( \mathbf{Pos}^{\hat{i}-1}_j, \mathbf{Pos}^{{\hat{i}}}_n \right) \mathbf{Y}_n^{{i-1}}}{\sum_{n\in \mathcal{N}^{\hat{i}-1}(j)} \mathbf{w} \left(\mathbf{Pos}^{{\hat{i}-1}}_j, \mathbf{Pos}^{{\hat{i}}}_n \right)}, \qquad j=1,\cdots,\vert\mathcal{V}^{\hat{i}-1}\vert
\end{equation}
where,
\begin{equation} \label{eq:unpooling_weight}
    \mathbf{w} : \left(\mathbf{Pos}^{{\hat{i}-1}}_j,\mathbf{Pos}^{{\hat{i}}}_n \right) \mapsto \frac{1}{\vert \vert \mathbf{Pos}^{{\hat{i}-1}}_j - \mathbf{Pos}^{{\hat{i}}}_n \vert \vert},
\end{equation}
with $\mathrm{\mathbf{Unpool}}^i: \mathbb{R}^{\vert \mathcal{V}^{\hat{i}} \vert \times N_F^{\hat{i}}} \rightarrow \mathbb{R}^{\vert \mathcal{V}^{\hat{i}-1} \vert \times N_F^{\hat{i}}}$, $\mathcal{N}^{\hat{i}-1}(j)$ is the $k-$nearest neighbors in 
$\mathcal{V}^{\hat{i}}$ of the $j^{\mathrm{th}}$ node in $\mathcal{V}^{\hat{i}-1}$, with $k\in\mathbb{N}$ denoting the number of nearest neighbors used for interpolation. $\mathbf{Pos}^{\hat{i}-1}_j \in \mathbb{R}^{n_d}$ is the spatial position of the $j^{\mathrm{th}}$ node at the $(\hat{i}-1)^{\mathrm{th}}$ layer of the hierarchy of reduced graphs, $\mathbf{Pos}^{\hat{i}}_n \in \mathbb{R}^{n_d}$ is the spatial position of the $n^{\mathrm{th}}$ node in the $\hat{i}^{\mathrm{th}}$ layer in the hierarchy of reduced graphs, $\mathbf{w}: \mathbb{R}^{n_d} \times \mathbb{R}^{n_d} \rightarrow \mathbb{R}_+$ denotes the spatial interpolation function, and $\vert \vert \boldsymbol{\cdot} \vert \vert : \mathbb{R}^{n_d} \rightarrow \mathbb{R}_+$ denotes the Euclidean norm. Much like the SAGEConv function \eqref{eq:SAGEConv}, the unpooling of \eqref{eq:unpooling} is performed by looping over all nodes, $j \in \mathcal{V}^{\hat{i}-1}$, to compute the rows $j=1,\cdots,\vert \mathcal{V}^{\hat{i}-1}\vert$ of the output of the unpooling operation, $\mathbf{\bar{Y}}^{i-1} \in \mathbb{R}^{\vert \mathcal{V}^{\hat{i}-1} \vert \times N_F^{\hat{i}}}$. Next, a message passing operation is applied to the outputs of the unpooling operation,
\begin{equation}\label{eq:MP_dec}
    \mathrm{\mathbf{MP}}^{i}_{\mathrm{dec}}: \left( \mathbf{\bar{Y}}^{i-1}; \boldsymbol{\Omega}_{i} \right) \mapsto \sigma \left(\mathbf{\bar{Y}}^{i-1}_j \mathbcal{W}^{i}_1 + \left(\mathrm{mean}_{n\in\mathcal{K}^{\hat{i}-1}(j)} \mathbf{\bar{Y}}^{i-1}_n \right) \mathbcal{W}^{i}_{2} \right), \qquad j=1,\cdots,\vert \mathcal{V}^{\hat{i}-1}\vert,
\end{equation}
with $\mathrm{\mathbf{MP}}^i_{\mathrm{dec}}: \mathbb{R}^{\vert \mathcal{V}^{\hat{i}-1} \vert \times N_F^{\hat{i}}} \times \mathbb{R}^{N_F^{\hat{i}}\times N_F^{\hat{i}-1}} \times \mathbb{R}^{N_F^{\hat{i}}\times N_F^{\hat{i}-1}} \rightarrow \mathbb{R}^{\vert \mathcal{V}^{\hat{i}-1} \vert \times N_F^{{\hat{i}-1}}}$, where $\mathbcal{W}^{i}_{1} , \mathbcal{W}^{i}_{2} \in \mathbb{R}^{N_F^{\hat{i}}\times N_F^{\hat{i}-1}}$ denote the weights with $\mathbf{\Omega}_i = \{\mathbcal{W}^{i}_{1}, \mathbcal{W}^{i}_{2}\}$ denoting the set of weights for the $i^{\mathrm{th}}$ UMP layer, $\mathcal{K}^{\hat{i}-1}(j)$ denotes the set of nodes connected to node $j$ in the graph of $(\hat{i}-1)^{\mathrm{th}}$ layer based on the adjacency matrix $\mathbf{A}^{\hat{i}-1}$, where the subscripts $j$ and $n$ denote the $j^{\mathrm{th}}$ and $n^{\mathrm{th}}$ nodes, respectively, and $\sigma: \mathbb{R} \rightarrow \mathbb{R}$ denotes the element-wise activation function. According to \eqref{eq:MP_dec}, the output of $\mathrm{\mathbf{MP}}^i_{\mathrm{dec}}$ is determined in a row-wise manner. Ultimately, the UMP layer takes the form,

\begin{equation}
    \mathrm{\mathbf{g}}_{i}: \left( \mathbf{Y}^{i-1}; \mathbf{\Omega}_i \right) \mapsto \mathrm{\mathbf{MP}}^{i}_{\mathrm{dec}}\left( \cdot ; \boldsymbol{\Omega}_i \right)  \circ \mathrm{\mathbf{Unpool}}^{i} \left( \mathbf{Y}^{i-1} \right),
\end{equation}
with $\mathbf{g}_i: \mathbb{R}^{\vert \mathcal{V}^{\hat{i}} \vert \times N_F^{\hat{i}}} \times \mathbb{R}^{N_F^{\hat{i}}\times N_F^{\hat{i}-1}} \times \mathbb{R}^{N_F^{\hat{i}}\times N_F^{\hat{i}-1}} \rightarrow \mathbb{R}^{\vert \mathcal{V}^{\hat{i}-1} \vert \times N_F^{\hat{i}-1}}$. Hence, the UMP layer increases the number of nodes in a given graph and decreases the number of features associated with each node, and it gives the node feature matrix $\mathbf{Y}^i$ as the output. The UMP layer is visually represented in Figure \ref{fig:UMP_layer}.  

\subsubsection{Postprocessing -- Layer $n_{\ell}$} \label{ssec:postprocessing}
To represent the output of the decoder as a state vector for appropriate deployment in the time integration scheme, a postprocessing step is applied as the final layer ($i=n_{\ell}$). First, the $\mathrm{\mathbf{InvScale}}$ operator is applied to invert the original $\mathrm{\mathbf{Scale}}$ operation,

\begin{equation}
    \mathrm{\mathbf{InvScale}}: \mathbf{Y}_{ij}^{n_{\ell}-1} \mapsto \mathbf{Y}_{ij}^{n_{\ell}-1} \left( {\mathcal{X}^{\mathrm{max}}_{j} - \mathcal{X}^{\mathrm{min}}_{j}} \right) + \mathcal{X}^{\mathrm{min}}_{j},
\end{equation}
where $\mathrm{\mathbf{InvScale}}: \mathbb{R} \rightarrow \mathbb{R}$ is an element-wise scaling operator. Next, a $\mathrm{\mathbf{Vectorize}}$ operator is applied to reshape the output of the decoder to a state vector. Ultimately, the postprocessing step takes the form,

\begin{equation}
    \mathbf{g}_{n_{\ell}}: \left(\mathbf{Y}^{n_{\ell}-1}; \mathbf{\Omega}_{n_{\ell}} \right) \mapsto \mathrm{\mathbf{Vectorize}}\left( \cdot \right) \circ \mathrm{\mathbf{\mathbf{InvScale}}} \left( \mathbf{Y}^{n_{\ell}-1} \right),
\end{equation}
where $\mathbf{g}_{n_{\ell}}: \mathbb{R}^{N_c \times n_q} \rightarrow \mathbb{R}^{N_c  n_q}$, and $\mathbf{\Omega}_{n_{\ell}} = \emptyset$ is the empty set, as there are no trainable parameters in the postprocessing layer. The output of the decoder $\tilde{\mathbf{x}}\in\mathbb{R}^N$ is a reconstruction of the original state vector $\mathbf{x}$.

\subsection{Training the autoencoder} \label{ssec:trainAE}
The components of the autoencoder that require training are the message passing operations and fully-connected/MLP layers with $\theta=\left\{\mathbf{\Theta}_1,\,\mathbf{\Theta}_2,\,\cdots,\,\mathbf{\Theta}_{n_{\ell}}\right\}$, $\omega=\left\{\mathbf{\Omega}_0,\,\mathbf{\Omega}_1,\,\cdots,\,\mathbf{\Omega}_{n_{\ell-1}}\right\}$ as trainable parameters. To train these, we adopt the same loss function as \cite{lee2020deeplspg,lee2021deepconservation}, which is the $L^2$-norm of the reconstructed solution state,

\begin{equation} \label{eq:loss}
    \mathcal{L} =  \sum_{i=1}^{N_{\mathrm{train}}} \Big\vert \Big\vert \mathbf{x}^i - \mathbf{Dec}\left( \cdot \right) \circ \mathbf{Enc}\left( \mathbf{x}^i \right) \Big\vert \Big\vert ^2 _2,
\end{equation}
where $\mathbf{x}^i \in \mathbb{R}^N$ is the $i^{\mathrm{th}}$ solution state in the training set and $N_{\mathrm{train}} \in \mathbb{N}$ denotes the total number of training solution states generated by the FOM. 

%\section{Time integration} \label{sec:TimeIntegration}
\section{Projection scheme and interpretability} \label{sec:TimeIntegration}

Time-stepping for ROMs using autoencoders has been achieved by a variety of methods, including training neural networks to compute time updates \cite{fresca2022poddlrom,kim2019deepfluids,maulik2021advectionrom,fresca2021comprehensive,Regazzoni2019MLforFastDiffEqs}, identifying low-dimensional systems of ODEs \cite{fries2022lasdi,he2023glasdi}, and projecting the semi-discretized equations of the FOM onto a nonlinear manifold \cite{lee2020deeplspg,lee2021deepconservation,kim2022masked,chen2022crom}. To examine the autoencoder's ability to embed the physics of the FOM, we choose to perform time integration using the time-discrete residual-minimizing LSPG projection \cite{carlberg2011lspg,lee2020deeplspg,lee2021deepconservation,carlberg2017galerkinvslspg}. Specifically, GD-LSPG leverages the graph autoencoder to project the governing equations onto a low-dimensional latent space, thus performing time integration on the latent state variables. 

\subsection{Least-squares Petrov-Galerkin projection} \label{ssec:lspg}
In this section, we summarize the main points from the literature to provide sufficient background on the LSPG projection and how it is leveraged in this scope of work and direct the reader to \cite{carlberg2011lspg,carlberg2013gnat,lee2020deeplspg,lee2021deepconservation,carlberg2017galerkinvslspg} for further background and properties of the LSPG projection.

To illustrate, we set the initial conditions of the low-dimensional state vector to be the encoding of the initial conditions of the high-dimensional system, i.e., $\mathbf{\hat{x}}\left(0;\boldsymbol{\mu}) = \mathbf{Enc}(\mathbf{x}(0;\boldsymbol{\mu})\right)$, and approximate the full-order state vector of the solution of the system, \eqref{eq:ode1}, to be,

\begin{equation} \label{eq:approx}
    \mathbf{\tilde{x}}\left(t; \boldsymbol{\mu} \right) = \mathbf{Dec}\left(\mathbf{\hat{x}}\left(t ; \boldsymbol{\mu} \right)\right),
\end{equation}
where $\mathbf{\tilde{x}}:\mathbb{R}_+ \times \mathcal{D} \rightarrow \mathbb{R}^N$ denotes the predicted solution state. As is performed across the literature \cite{carlberg2011lspg,lee2020deeplspg,lee2021deepconservation}, we next substitute \eqref{eq:approx} into \eqref{eq:ode2argmin} and project the residual onto a test basis, $\mathbf{\Psi} \in \mathbb{R}^{N \times M}$, to prevent the system from becoming overdetermined. Ultimately, we arrive at the following minimization problem:

\begin{equation} \label{eq:res_min_full}
    \mathbf{\hat{x}}(t; \boldsymbol{\mu}) = \underset{\boldsymbol{\hat{\xi}} \in \mathbb{R}^M}{\argmin} \Big\vert \Big\vert (\mathbf{\Psi}(\boldsymbol{\hat{\xi}}; \boldsymbol{\mu}))^T \mathbf{r} \left( \mathbf{Dec}\left(\boldsymbol{\hat{\xi}}\left(t ; \boldsymbol{\mu} \right)\right) \right) \Big\vert \Big\vert ^2_2,
\end{equation}
where $\boldsymbol{\hat{\xi}}\in\mathbb{R}^M$ is the sought-after low-dimensional solution state at time $t$. Under this approximation, the boundary conditions are not explicitly preserved. However, boundary conditions are promoted implicitly by the chosen time-discrete residual minimizing projection scheme. Taking the test basis, $\boldsymbol{\Psi}:\mathbb{R}^M \times \mathcal{D} \rightarrow \mathbb{R}^{N \times M}$, to be,  
\begin{equation} \label{eq:dlspg_psi}
    \mathbf{\Psi}: (\boldsymbol{\hat{\xi}}; \boldsymbol{\mu}) \mapsto \left(\frac{\partial \mathbf{r}}{\partial \mathbf{x}} \Big \vert_{\mathbf{Dec}\left(\boldsymbol{\hat{\xi}}\left(t ; \boldsymbol{\mu} \right)\right)} \right) \left( \frac{\mathrm{d}\mathbf{Dec}}{\mathrm{d}\boldsymbol{\hat{\xi}}} \Big \vert_{ \boldsymbol{\hat{\xi}}\left(t ; \boldsymbol{\mu}\right)} \right),
\end{equation}
a nonlinear manifold LSPG projection is obtained (with $\frac{\mathrm{d}\mathbf{Dec}}{\mathrm{d}\boldsymbol{\hat{\xi}}} \Big \vert_{ \boldsymbol{\hat{\xi}}\left(t ; \boldsymbol{\mu}\right)}$ denoting the Jacobian of the decoder), and an iterative Newton solver can be used to minimize \eqref{eq:res_min_full}. Such a solver takes the form,

\begin{equation} \label{eq:gn_update}
    \left(\mathbf{\Psi}( \mathbf{\hat{x}}^{n(j)}; \boldsymbol{\mu})\right)^T \mathbf{\Psi}( \mathbf{\hat{x}}^{n(j)}; \boldsymbol{\mu}) \left(\mathbf{\hat{x}}^{n(j+1)} - \mathbf{\hat{x}}^{n(j)}\right) = - \beta^{(j)} \left(\mathbf{\Psi} \left( \mathbf{\hat{x}}^{n(j)}; \boldsymbol{\mu}\right)\right)^T \mathbf{r} \left(\mathbf{Dec}(\mathbf{\hat{x}}^{n(j)}); \boldsymbol{\mu} \right),
\end{equation}
where the superscript $n(j)$ denotes the $j^{\mathrm{th}}$ iteration of $n^{\mathrm{th}}$ time step, $\beta^{(j)}\in\mathbb{R}_+$ is the step size chosen to satisfy Wolfe conditions \cite{nocedal1999numerical}. An initial guess at each time step is chosen to be $\mathbf{\hat{x}}^{n(0)} = \mathbf{\hat{x}}^{n-1}$, where $\mathbf{\hat{x}}^{n-1}$ denotes the converged solution from the previous time step, $n-1$. The solution is updated iteratively until the $L^2$-norm of the reduced-state residual for the current iteration falls below a user-prescribed fraction of that of the initial guess at the time step $n=1$ (which is the encoded initial condition, $\mathbf{\hat{x}}^0$), 

\begin{equation} \label{eq:convergence}
    \text{Convergence criterion} : \frac{\Big \vert \Big \vert \mathbf{\hat{r}}(\mathbf{\hat{x}}^{n(j)}; \boldsymbol{\mu}) \Big \vert \Big \vert_2}{\Big \vert \Big \vert \mathbf{\hat{r}}(\mathbf{\hat{x}}^{1(0)}; \boldsymbol{\mu}) \Big \vert \Big \vert_2} \leq \kappa,
\end{equation}
where $\kappa \in [0, 1]$ is the user-defined tolerance, and the reduced state residual $\hat{\mathbf{r}}$ is obtained from the projection of the residual,
\begin{equation} \label{eq:lowDimRes}
    \mathbf{\hat{r}}: (\boldsymbol{\hat{\xi}}; \boldsymbol{\mu}) \mapsto \left(\mathbf{\Psi} \left( \boldsymbol{\hat{\xi}}; \boldsymbol{\mu}\right)\right)^T \mathbf{r} \left(\mathbf{Dec}(\boldsymbol{\hat{\xi}}); \boldsymbol{\mu} \right).
\end{equation}
It is important to note that LSPG projections of this form require knowledge of the FOM solver and direct access to the residual and its Jacobian, meaning GD-LSPG can only be performed for methods in computational mechanics where these information are available.

\subsection{Interpretability of the latent state vector} \label{ssec:interpretability}
In this section, we build from the LSPG projection to provide further insight into the interpretability of the proposed graph autoencoder. Although there is no unified definition for interpretability in scientific machine learning, in our study, we take it to mean, ``the ability to explain or present in understandable terms to a human,'' (the definition from \cite{doshi2017interpretable}). It is known that the latent space for autoencoders is commonly difficult to interpret due to the entanglement of the latent state variables \cite{eivazi2022betavae,kang2022betavariational}. More broadly, in recent years, interpretability has emerged as a major focus area in the field of scientific machine learning \cite{baker2019osti}.

Here, we demonstrate that the Jacobian of the decoder used in the nonlinear manifold LSPG projection scheme (Section \ref{ssec:lspg}) bears a direct analogy to the POD modes. To illustrate this relationship, consider the case where the decoder is an affine POD projection (see \ref{appendix:POD}),

\begin{equation}
    \mathbf{\tilde{x}}\left(t; \boldsymbol{\mu} \right) = \mathbf{Dec}\left(\mathbf{\hat{x}}\left(t ; \boldsymbol{\mu} \right)\right) := \boldsymbol{\Phi} \mathbf{\hat{x}}\left(t ; \boldsymbol{\mu} \right),
\end{equation}
where $\boldsymbol{\Phi} \in \mathbb{R}^{N \times M}$ denotes the POD modes, it can be shown that the Jacobian of the decoder is simply the POD modes themselves, i.e.,
\begin{equation} \label{eq:PODjacobian}
    \frac{\mathrm{d}\mathbf{Dec}}{\mathrm{d}\mathbf{\hat{x}}} \Big \vert_{ \mathbf{\hat{x}}\left(t ; \boldsymbol{\mu}\right)} = \boldsymbol{\Phi}.
\end{equation}
In this case, substituting \eqref{eq:PODjacobian} into \eqref{eq:dlspg_psi} yields a classical POD-LSPG projection. Building on this intuition, we interpret the Jacobian of the decoder in the graph autoencoder in the same manner. While the POD modes are time-invariant and independent of the latent state vector, $\mathbf{\hat{x}}\left(t ; \boldsymbol{\mu}\right)$, the Jacobian of the decoder of the graph autoencoder depends explicitly on $\mathbf{\hat{x}}\left(t ; \boldsymbol{\mu}\right)$, indicating that the corresponding modes evolve over time. Further discussion on the interpretability of the graph autoencoder is provided in Section \ref{sec:experiments} through numerical examples.

We can further relate our perspective for investigating interpretability to common strategies found in the literature, typically, classified as either \textit{global} or \textit{local} interpretability \cite{doshi2017interpretable}. Local interpretability seeks to identify the reason for a specific prediction, whereas global interpretability seeks to identify the trends in a model's behavior for all predictions. Through this lens, analyzing the Jacobian of the decoder for the graph autoencoder can be viewed as a form of local interpretability. In particular, this approach closely parallels saliency maps \cite{simonyan2013deep}, originally developed for image classification problems. Saliency maps aim to find the features in the input that are most predictive of the output by employing a first-order Taylor expansion.

\section{Numerical experiments} \label{sec:experiments}
We evaluate the efficiency and accuracy of the GD-LSPG through a series of test problems. First, to provide a baseline for comparison to the rest of the literature, we use a commonly studied 1D Burgers' model using a structured mesh \cite{Barnett2023NNAugmentedPROM,lee2020deeplspg,lee2021deepconservation,rewienski2003phd}. This allows us to benchmark the accuracy of GD-LSPG with PMOR methods that deploy CNN-based autoencoders. Second, we deploy GD-LSPG to a model solving the 2D Euler equations for two different settings. The first setting uses an unstructured mesh to solve a setup for the Riemann problem \cite{kurganov2002riemann,liska2003comparison}. Despite using an unstructured mesh, the domain is square and regular, meaning we can interpolate the unstructured mesh solution onto a structured mesh solution that can be deployed to a CNN-based autoencoder. In this setting, we demonstrate that the graph autoencoder generalizes better than interpolating to a structured mesh and using a CNN-based autoencoder. The second setting demonstrates the versatility of GD-LSPG for a problem with a more complicated geometry. In the second setting, we further evaluate the ability of the graph autoencoder when presented with noisy training data. All autoencoders are trained with PyTorch \cite{paszke2019pytorch} and PyTorch-Geometric \cite{fey2019pyg}. A detailed description of the employed autoencoder architectures and choice of hyperparameters for all examples are provided in \ref{section:architecture}. To train the models and therefore minimize \eqref{eq:loss}, the Adam optimizer \cite{kingma2014adam} is deployed to perform stochastic gradient descent with an adaptive learning rate. In this study, we use reconstruction and state prediction errors as the primary performance metrics to assess accuracy. The reconstruction error is used to assess the ROM's ability to reconstruct a precomputed full-order solution, and it is defined as

\begin{equation}\label{eq:ae_error}
    \text{autoencoder reconstruction error} = \frac{\sqrt{ \sum_{n=1}^{N_t} \Big\vert\Big\vert \mathbf{x}^n \left( \boldsymbol{\mu} \right) -  \mathbf{Dec} \circ \mathbf{Enc} \left(\mathbf{x}^n \left( \boldsymbol{\mu} \right) \right) \Big\vert\Big\vert_2^2}} {\sqrt{\sum_{n=1}^{N_t} \Big\vert\Big\vert \mathbf{x}^n \left( \boldsymbol{\mu} \right)   \Big\vert\Big\vert_2^2}},
\end{equation}
where $\mathbf{x}^n\left(\boldsymbol{\mu}\right)$ is the full-order solution at the $n^{\mathrm{th}}$ time step. On the other hand, the POD reconstruction error is evaluated from

\begin{equation}\label{eq:pod_error}
    \text{POD reconstruction error} = \frac{\sqrt{ \sum_{n=1}^{N_t} \Big\vert\Big\vert \left( \mathbf{I} - \boldsymbol{\Phi} \boldsymbol{\Phi}^T \right) \mathbf{x}^n \left( \boldsymbol{\mu} \right) \Big\vert\Big\vert_2^2}}{\sqrt{\sum_{n=1}^{N_t} \Big\vert\Big\vert \mathbf{x}^n \left( \boldsymbol{\mu} \right)  \Big\vert\Big\vert_2^2}},
\end{equation}
where $\mathbf{\Phi}\in \mathbb{R}^{N\times M}$ is the matrix of reduced basis vectors from an affine POD approximation constructed based on the method of snapshots (see \ref{appendix:POD}). 

The state prediction error is used to assess the accuracy of the ROM obtained from different methods in predicting the full-order solution,

\begin{equation}\label{eq:state_err}
    \text{state prediction error} = \frac{\sqrt{\sum_{n=1}^{N_t} \Big\vert\Big\vert \mathbf{x}^n \left( \boldsymbol{\mu} \right) - \mathbf{\tilde{x}}^n \left( \boldsymbol{\mu}\right) \Big\vert\Big\vert_2^2}}{\sqrt{\sum_{n=1}^{N_t} \Big\vert\Big\vert \mathbf{x}^n \left( \boldsymbol{\mu} \right)  \Big\vert\Big\vert_2^2}}.
\end{equation}

In all numerical examples, both dLSPG and GD-LSPG employ functorch's automatic differentiation \cite{he2021functorch} to obtain the Jacobian of the decoder.

\subsection{One-dimensional Burgers' equation} \label{ssec:burgers}

To benchmark GD-LSPG with dLSPG and POD-LSPG, we use a 1D Burgers' model on a structured mesh. Due to its close relationship to the Navier-Stokes equations \cite{chan2010class} and advection-driven behavior, the 1D Burgers' model is commonly chosen as a test case in the literature \cite{geelen2022localized,Barnett2023NNAugmentedPROM,lee2020deeplspg,lee2021deepconservation,chen2022crom}. In this study, we choose the numerical experiment originally found in \cite{rewienski2003phd} with the added parameterization from \cite{lee2020deeplspg}. The governing equation is,

\begin{equation} \label{eq:burgers}
    \begin{split}
        \frac{\partial w(x,t;\boldsymbol{\mu})}{\partial t} + \frac{\partial f (w(x,t;\boldsymbol{\mu}))}{\partial x} &= 0.02e^{\mu_2 x}, \quad \forall x\in (0,L), \forall t \in (0,T], \\
        w(0,t;\boldsymbol{\mu}) &= \mu_1, \quad \forall t \in (0,T], \\
        w(x,0; \boldsymbol{\mu}) &= 1, \quad \forall x \in (0,L),
    \end{split}
\end{equation}
where $f(w) = 0.5w^2$, $x\in \mathbb{R}$ denotes spatial position, $t\in\mathbb{R}_+$ denotes time, $L\in\mathbb{R}$ denotes the length of the 1D physical domain, and $T \in \mathbb{R}_+$ denotes the final time. We utilize the FVM by dividing the spatial domain into 256 equally-sized cells over a domain of length $L=100$, lending to a structured finite volume mesh. A backward-Euler time-integration scheme is employed, which corresponds to the cell-wise equations at the $i^{\mathrm{th}}$ cell,
\begin{equation}\label{eq:burgers_p}
    \mathbf{f}_i: (\boldsymbol{\xi}, t; \boldsymbol{\mu}) \mapsto \frac{1}{2\Delta x} \left( \left( w_{i-1}^{n+1} \right)^2 - \left( w_{i}^{n+1} \right)^2 \right) - 0.02e^{\mu_2 x_i},
\end{equation}
with $\alpha_0=1$, $\alpha_1=-1$, $\beta_0 = -\Delta t$, $\beta_1=0$, and $\tau=1$ when written in the form of \eqref{eq:ode2}. Note that since Burgers' equation involves only one state variable, i.e., $\mathbf{f}_i$ is a scalar variable. In \eqref{eq:burgers_p}, $\Delta x \in \mathbb{R}_+$ is the length of each cell in the uniform 1D mesh, $x_i \in \mathbb{R}$ is the coordinate of the center of the $i^{\mathrm{th}}$ cell in the mesh, and $\boldsymbol{\xi} = (w_1^{n+1}, w_2^{n+1}, \ldots, w_{N_c}^{n+1})^T$ is the sought-after state solution at the $(n+1)^{\mathrm{th}}$ time step. The time integration scheme uses a constant time step size $\Delta t = .07$ and a final time $T=35$ for a total of 501 time steps per solution including the initial value. To train both the CNN-based autoencoder and the graph autoencoder, as well as obtain an affine POD basis, the solution to the FOM is computed for a total of 80 parameter scenarios with the parameters $\boldsymbol{\mu} = \left(\mu_1 = 4.25 + \left( \frac{1.25}{9}\right) \mathrm{i}, \mu_2 = .015 + \left( \frac{.015}{7}\right)\mathrm{j} \right)$, for $\mathrm{i}=0,\ldots,9$ and $\mathrm{j}=0,\ldots,7$. Once trained, the autoencoders are deployed in an online setting to perform time integration for their respective ROMs. 

To generate the POD-LSPG solution, we set the tolerance $\kappa$ in \eqref{eq:convergence} to be $10^{-4}$, whereas, the tolerance is set to $10^{-3}$ for the CNN-based dLSPG and GD-LSPG. The step size, $\beta^{(j)}$, for POD-LSPG is set to $1.0$. Alternatively, in dLSPG and GD-LSPG, an adaptive step size strategy is adopted. For dLSPG, we begin with a step size of $1.0$ and reduce by $5\%$ every $5$ iterations that convergence is not achieved. Likewise, in GD-LSPG, we begin with a step size of $0.5$ and reduce by $10\%$ every $10$ iterations if convergence is not achieved.

\begin{figure}[t!]
    \centering
    \begin{tabular}{cccc}
        & {\footnotesize $M=3$} & {\footnotesize $M=10$} & {\footnotesize $\;$Reconstruction and state prediction errors}\\
        \raisebox{2.0em}{\rotatebox[origin=lb]{90}{\footnotesize\smash{$\boldsymbol \mu = (4.30, 0.021)$}}}
        & \includegraphics[height=0.22\textwidth]{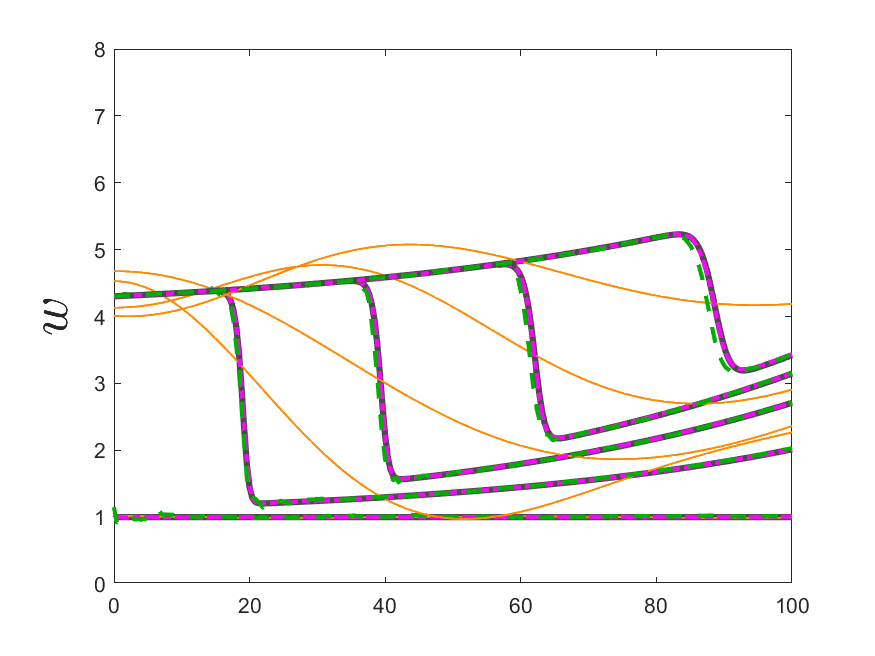}
        & \includegraphics[height=0.22\textwidth]{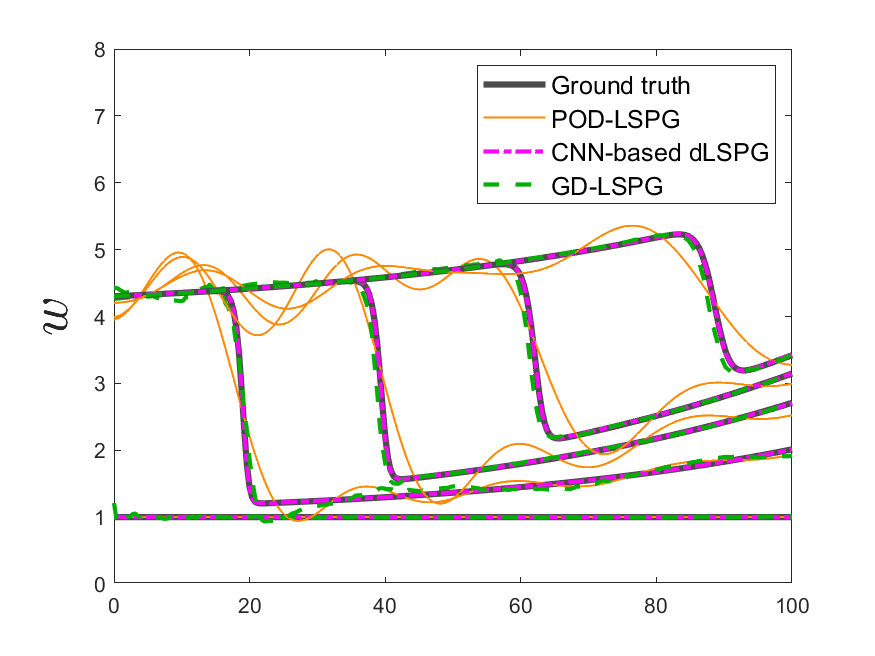}
        & \includegraphics[height=0.22\textwidth]{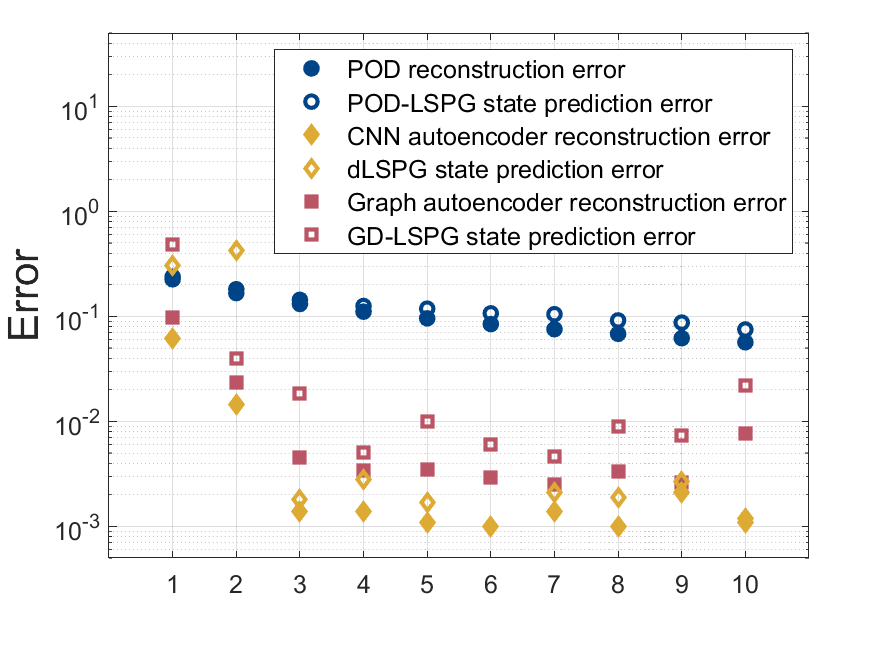}
        \\
        \raisebox{2.0em}{\rotatebox[origin=lb]{90}{\footnotesize\smash{$\boldsymbol \mu = (5.15, 0.0285)$}}}
        & \includegraphics[height=0.22\textwidth]{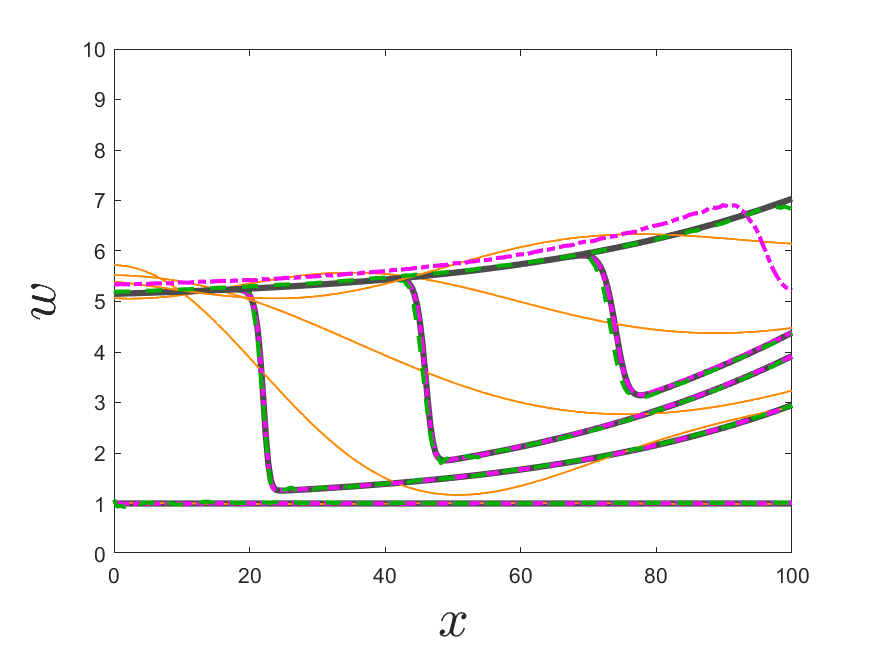}
        & \includegraphics[height=0.22\textwidth]{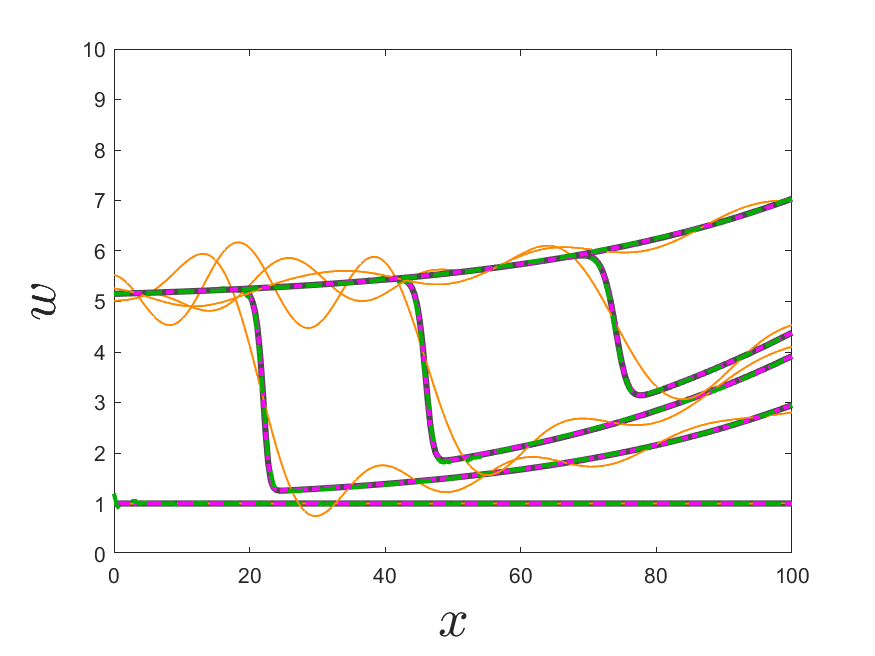}
        & \includegraphics[height=0.22\textwidth]{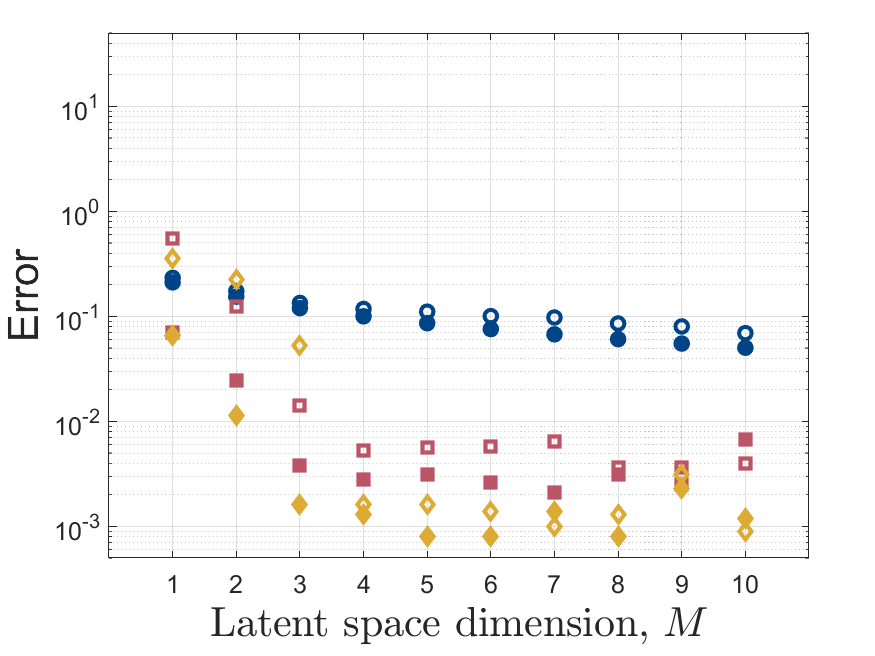}
    \end{tabular}
    \captionsetup{justification=centering}
    \caption{The left two columns represent the state solution for Burgers' model \eqref{eq:burgers} for time steps $t=0,\,7,\,14,\,21,$ and $28$ (ordered from left to right) for two latent space dimensions ($M=3,\, 10$, respectively), while the right column depicts the error metrics from \eqref{eq:ae_error}-\eqref{eq:state_err} for various PMOR methods. Figures in the first row correspond to test parameters $\boldsymbol \mu = (\mu_1 = 4.30, \mu_2 = 0.021)$, while figures in the second row correspond to test parameters $\boldsymbol \mu = (\mu_1 = 5.15, \mu_2 = 0.0285)$. GD-LSPG and dLSPG both outperform POD-LSPG in predicting the highly nonlinear behavior of the Burgers' equation. (Online version in color.)}
    \label{fig:burgers_results}
\end{figure}

Figure \ref{fig:burgers_results} depicts the solution state at various time steps for two test parameter set realizations not seen in the training set and two latent space dimensions of $M=3,\,10$. Additionally, the POD reconstruction errors from \eqref{eq:pod_error}, and autoencoder reconstruction errors from \eqref{eq:ae_error} for the CNN-based autoencoder, inspired by \cite{lee2020deeplspg,lee2021deepconservation}, and the reconstruction errors from the graph autoencoder can be found in Figure \ref{fig:burgers_results}. Using the state prediction error of \eqref{eq:state_err}, we also compare the performance of GD-LSPG to that of the traditional affine POD-LSPG \cite{carlberg2011lspg,carlberg2017galerkinvslspg}, as well as dLSPG, which leverages a CNN-based autoencoder \cite{lee2020deeplspg,lee2021deepconservation}. We emphasize that the reconstruction errors for the graph autoencoder are more than an order of magnitude smaller than that of the affine POD approximation for latent space dimensions $3$ to $10$. Likewise, the state prediction errors of GD-LSPG are roughly an order of magnitude lower than that of POD-LSPG for latent space dimensions $3$ to $10$. This outcome is due to the fact that an affine subspace is not well-suited for such nonlinear problems. Benchmarking the graph autoencoder with the traditional CNN-based autoencoder, we find that the graph autoencoder's reconstruction errors and state prediction errors to be less than an order of magnitude greater than those of the CNN-based autoencoder for the vast majority of latent space dimensions. This comparison implies that, while GD-LSPG gains adaptability and is applicable to unstructured meshes, it does not perform as well as CNN-based dLSPG for the Burgers' model with a structured mesh. However, as noticed from the solution states provided in Figure \ref{fig:burgers_results}, GD-LSPG is able to model the advection-dominated shock behavior in a manner similar to traditional CNN-based dLSPG, where traditional affine POD-LSPG tends to fail. We note that for $\boldsymbol \mu = (\mu_1 = 5.15, \mu_2 = 0.0285)$ the dLSPG solution using a latent space dimension $M=3$ exhibits an erroneous solution when the shock approaches the right side of the domain, which persistently occurred for different adaptive step size strategies. It is worth noting that high errors were reported in \cite{lee2020deeplspg} for the same latent space dimension without explicit elaboration on the main cause of such high error.
However, for latent space dimensions 4 to 10, dLSPG outperforms GD-LSPG in terms of accuracy. Additionally, the errant solution was not noticed for the parameter set $\boldsymbol \mu = (\mu_1 = 4.30, \mu_2 = 0.021)$. 

\begin{figure}[t!]
    \centering
    \begin{tabular}{ccc}
        \includegraphics[height=0.24\textwidth]{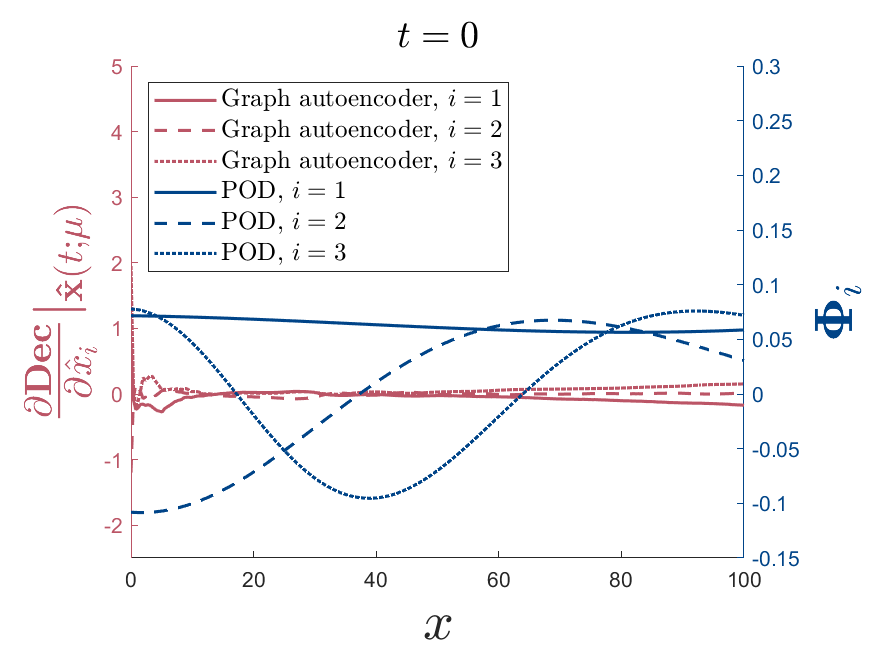}
        \hspace{-6mm}
        & \includegraphics[height=0.24\textwidth]{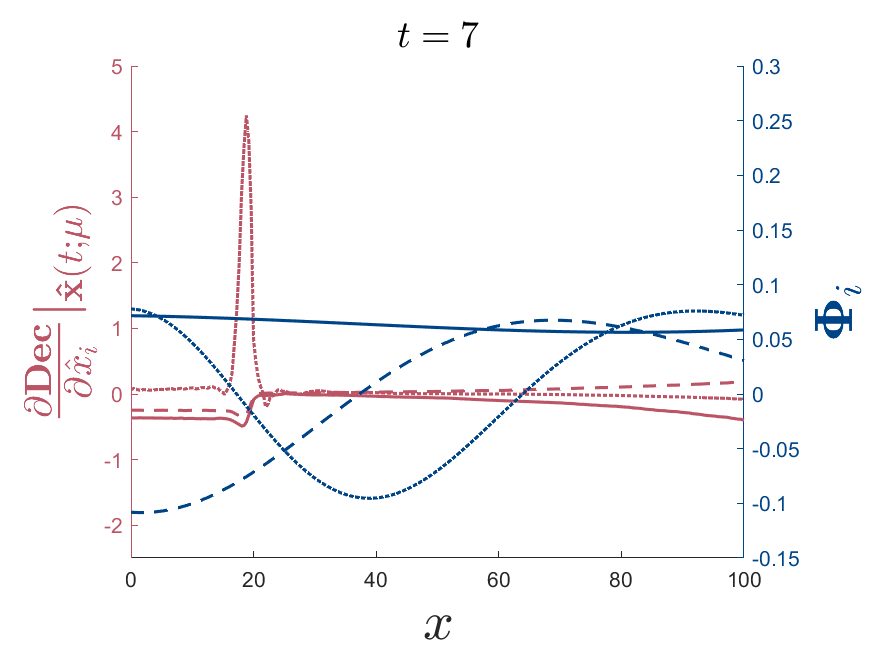}
        \hspace{-6mm}
        & \includegraphics[height=0.24\textwidth]{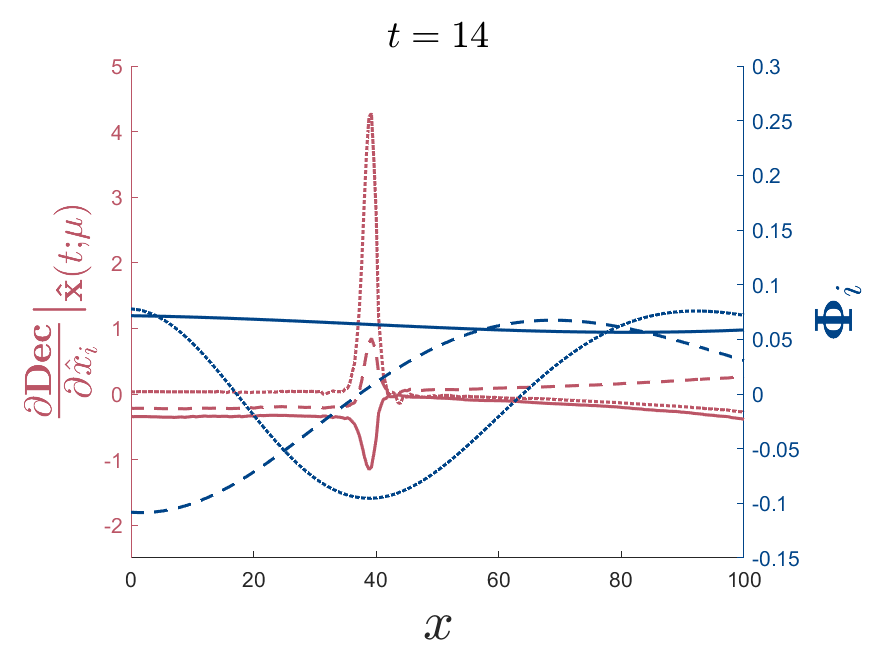} \\[6pt] \includegraphics[height=0.24\textwidth]{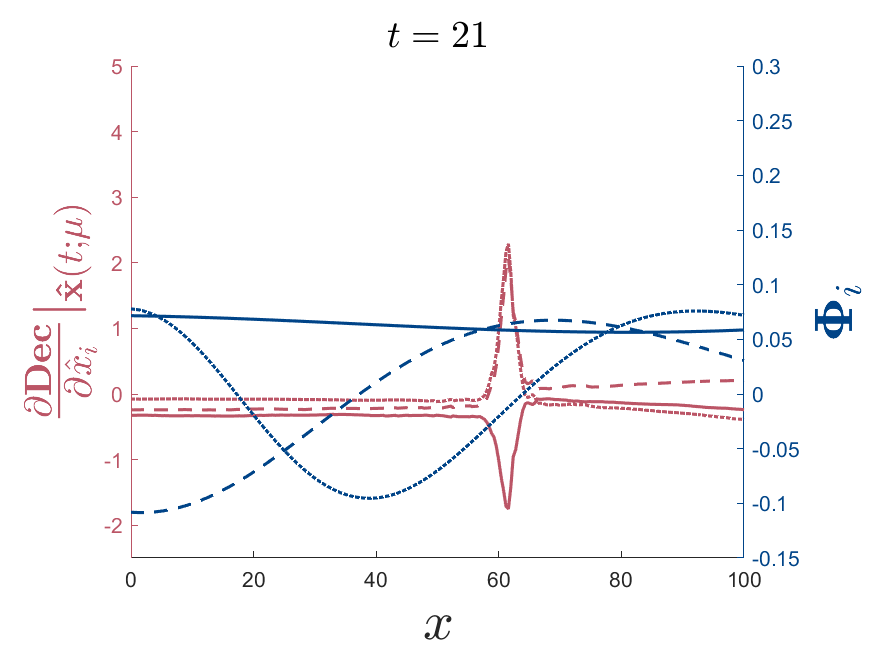}
        \hspace{-6mm}
        & \includegraphics[height=0.24\textwidth]{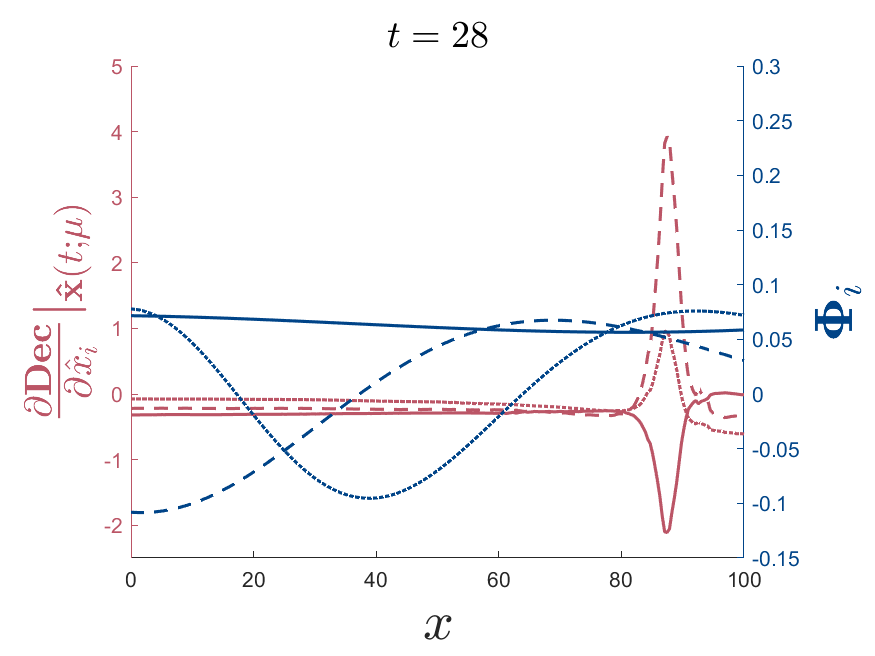}
        \hspace{-6mm}
        & \includegraphics[height=0.24\textwidth]{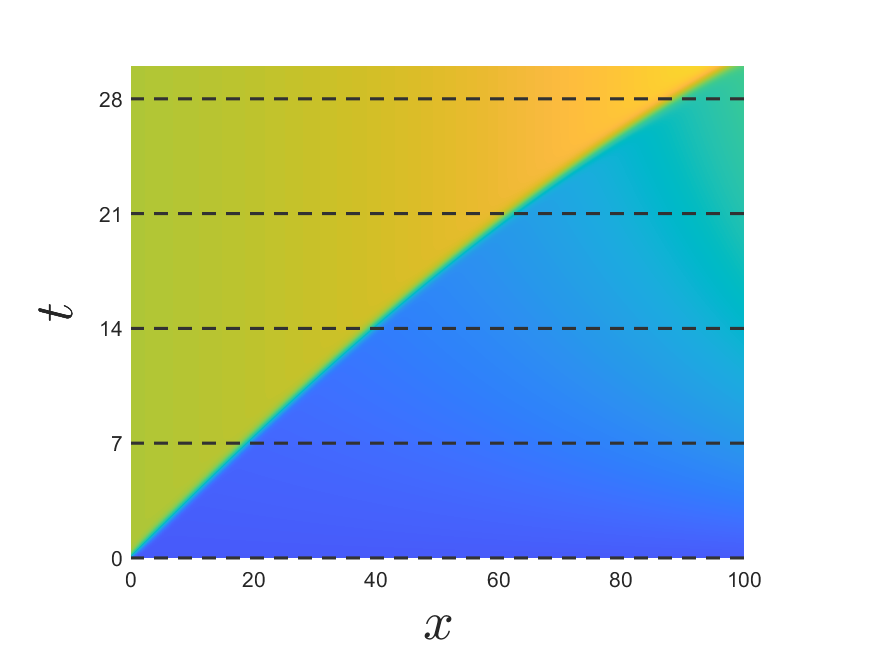}
    \end{tabular}
    \captionsetup{justification=centering}
    \caption{Jacobian of the decoder of GD-LSPG with respect to the $i^{\mathrm{th}}$ latent state variable, $\frac{\partial \mathrm{\mathbf{Dec}}}{\partial \hat{x}_i} \vert_{\mathbf{\hat{x}}(t;\boldsymbol{\mu})}$, is shown alongside the POD modes, $\boldsymbol{\Phi}_i$, for a latent space dimension $M=3$. Results are presented at time instances $t=0,\,7,\,14,\,21,$ and $28$, for the test parameter set $\boldsymbol \mu = (\mu_1 = 4.30, \mu_2 = 0.021)$. Unlike the time-invariant and highly diffusive POD modes, the Jacobian of the decoder for the graph autoencoder captures critical information about the location of the moving shock boundary. In all subplots (except for the bottom right), the left axis corresponds to the Jacobian of the decoder for latent space variable $i$, while the right axis corresponds to the the $i^{\mathrm{th}}$ POD mode. All subplots (except for the bottom right) share the same legend as the one shown for $t=0$. The bottom right figure displays the full-order solution with time and space, in which the horizontal lines highlight the selected time instances to highlight the match between the location of the moving shock boundary and the identified features by the Jacobian of the decoder. (Online version in color.)}
    \label{fig:burgers_jacobian}
\end{figure}

Next, we analyze the interpretability of the latent state vector through the Jacobian of the decoder (as presented in Section \ref{ssec:interpretability}). We reiterate that the Jacobian of the decoder for the graph autoencoder can be interpreted in the same manner as the POD modes from a classical POD-LSPG scheme. Figure \ref{fig:burgers_jacobian} presents the Jacobian of the decoder for the GD-LSPG solution to the test parameters $\boldsymbol \mu = (\mu_1 = 4.30, \mu_2 = 0.021)$ for $M=3$ as well as the POD modes used to obtain the POD-LSPG solution. Whereas the POD modes remain fixed for all time steps, the Jacobian of the decoder for the graph autoencoder demonstrates highly interpretable time-varying mode shapes. It is evident that the latent state variables from the graph autoencoder capture information about the moving shock boundary.

\begin{figure}[t!]
    \centering
    \begin{tabular}{cccc}
        & {\footnotesize POD-LSPG} & {\footnotesize dLSPG} & {\footnotesize GD-LSPG}\\
        \raisebox{3.8em}{\rotatebox[origin=lb]{90}{\footnotesize\smash{$M=3$}}}\captionsetup{justification=centering,width=\textwidth} 
        & \includegraphics[height=0.22\textwidth]{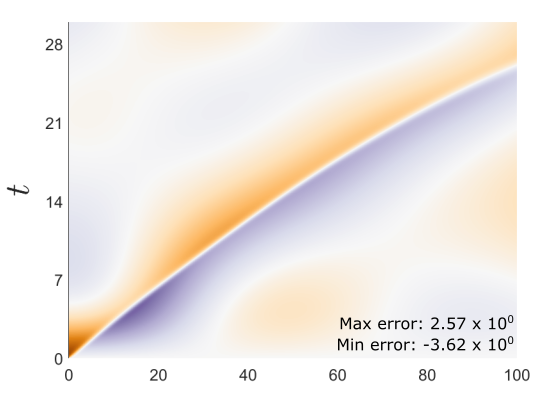}
        & \includegraphics[height=0.22\textwidth]{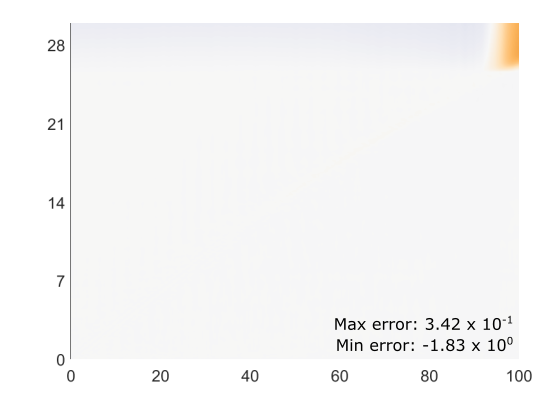}
        & \includegraphics[height=0.22\textwidth]{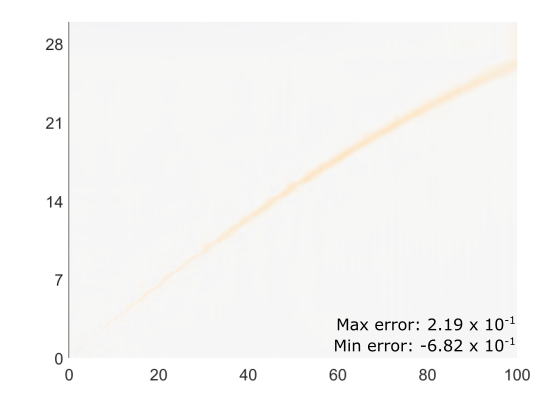}
        \\
        \raisebox{3.7em}{\rotatebox[origin=lb]{90}{\footnotesize\smash{$M=10$}}}
        \captionsetup{justification=centering,width=\textwidth} 
        & \includegraphics[height=0.22\textwidth]{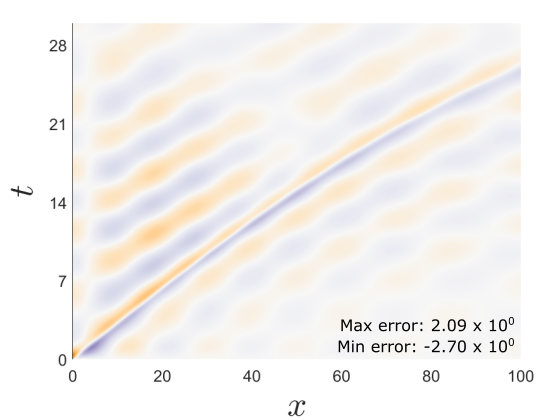}
        & \includegraphics[height=0.22\textwidth]{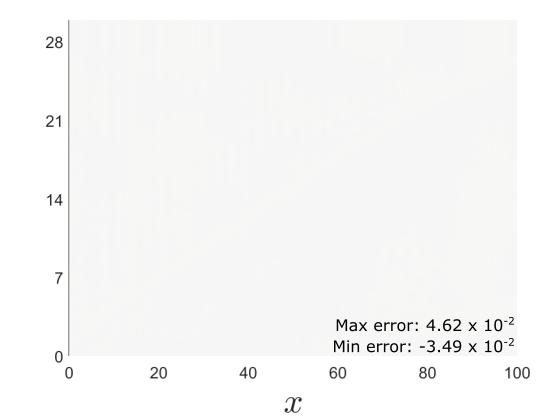}
        & \includegraphics[height=0.22\textwidth]{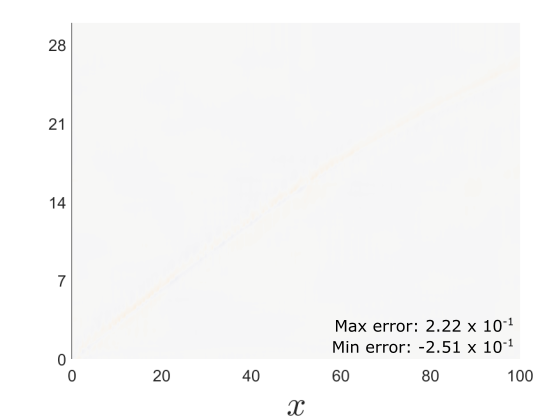}
        \\
        & \multicolumn{3}{c}{\includegraphics[scale=.31]{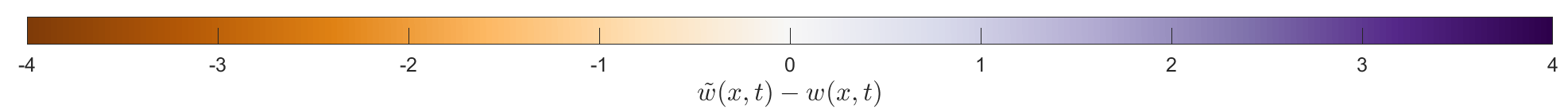}}
    \end{tabular}
    \captionsetup{justification=centering}
    \caption{Local error for dLSPG, GD-LSPG, and POD-LSPG for the parameter set $\boldsymbol{\mu} = (\mu_1 = 5.15, \mu_2 = 0.0285)$. The top and bottom rows correspond to solutions generated with latent space dimensions $M=3$ and $M=10$, respectively. The local error is simply taken to be $\tilde{w}(x,t) - w(x,t)$, or the difference between the predicted solution state and the ground truth solution state. The POD-LSPG solution introduces considerable error throughout the domain. Alternatively, dLSPG and GD-LSPG introduce lower-order localized errors, primarily around the shock. The slight phase difference between shocks in the predicted solution and the ground truth solution is the main error contributor. (Online version in color.)}
    \label{fig:430_021_tx}
    
\end{figure}

Figure \ref{fig:430_021_tx} depicts the difference between the ROM prediction of the full-order state vector and the FOM results with space and time. It can be seen that both dLSPG and GD-LSPG provide an improved ability to model the shock behavior of \eqref{eq:burgers} over POD-LSPG. Additionally, we once again see that the dLSPG solution for $\boldsymbol \mu = (\mu_1 = 5.15, \mu_2 = 0.0285)$ at latent space dimension $M=3$ struggles to converge when the shock approaches the right side of the domain. Finally, it is apparent that the main source of error for GD-LSPG is a slight phase lag between the ground truth location of the shock and GD-LSPG's prediction of the shock location. Hence, on a structured mesh, GD-LSPG provides an improvement over traditional affine POD-LSPG \cite{carlberg2011lspg,carlberg2013gnat} in a manner comparable to that of dLSPG \cite{lee2020deeplspg,lee2021deepconservation,kim2022masked}.

\begin{figure}[t!]
    \centering
    \includegraphics[width=.9\textwidth]{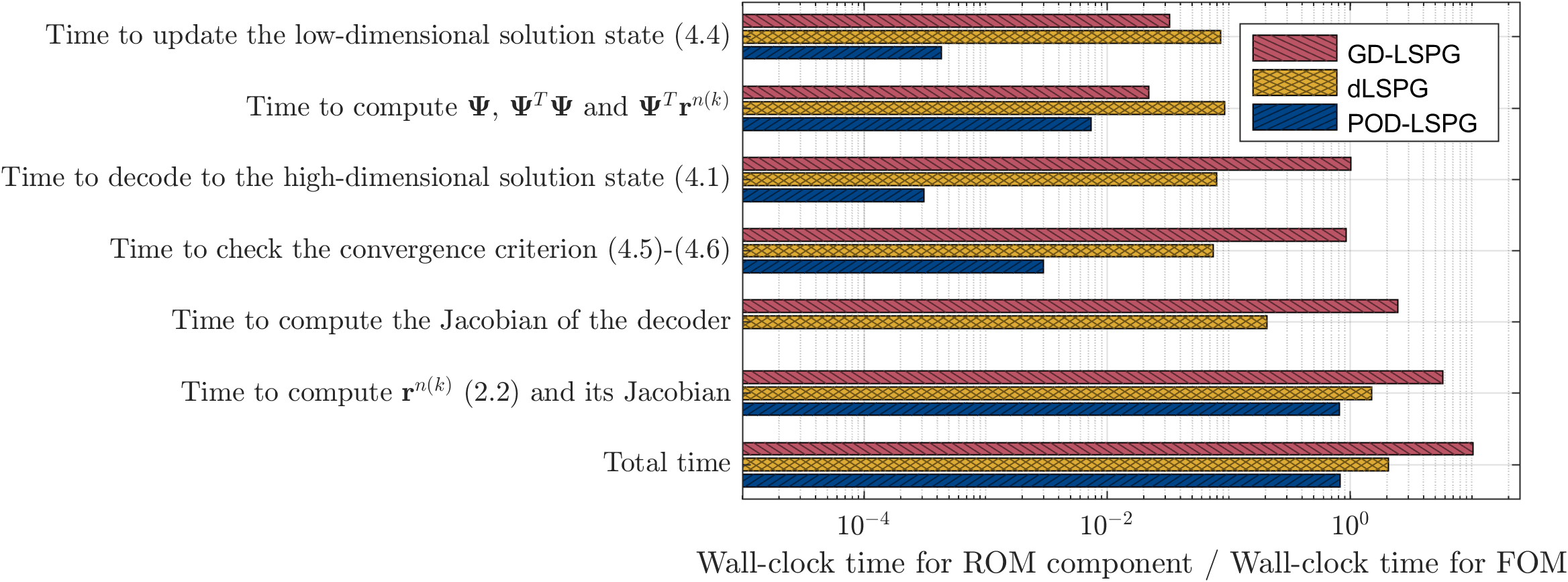}
    \captionsetup{justification=centering}
    \caption{Wall-clock time breakdown of individual components of GD-LSPG, dLSPG, and POD-LSPG with a latent space dimension $M=3$ normalized to the wall-clock time associated with the FOM solution. Solutions were generated for the parameter set $\boldsymbol{\mu} = (\mu_1 = 4.30, \mu_2 = 0.021)$. The POD-LSPG approach does not have to compute the Jacobian of the decoder at each iteration. The breakdown reveals the most expensive components of the dLSPG and GD-LSPG methods include the time to get the high-dimensional residual, Jacobian of the high-dimensional residual, and Jacobian of the decoder. The wall-clock time for the FOM was 38.63 seconds. (Online version in color.)}
    \label{fig:costBreakdown}
    
\end{figure}

To assess the computational cost of the GD-LSPG method and compare it to POD-LSPG and dLSPG, we provide an analysis of the 1D Burgers' model. All operations in this section are performed in PyTorch using a single Intel\textsuperscript{\textregistered} Xeon\textsuperscript{\textregistered} Platinum 8358 CPU @ 2.60GHz ICE LAKE core. We generate the solution five times using each ROM and report the average time for each component of the ROM procedure (normalized by the average time of the FOM) in Figure \ref{fig:costBreakdown}. We present the time to get $\mathbf{r}^{n(k)}$ from \eqref{eq:ode2} and evaluate its Jacobian, the time to get the Jacobian of the decoder, the time to check the convergence criterion \eqref{eq:convergence}-\eqref{eq:lowDimRes}, the time to decode to the high-dimensional space \eqref{eq:approx}, the time to compute $\boldsymbol{\Psi}$, $\boldsymbol{\Psi}^T \boldsymbol{\Psi}$, and $\boldsymbol{\Psi}^T \mathbf{r}^{n(k)}$, and the time to update the low-dimensional solution state \eqref{eq:gn_update}. The most time-consuming components of dLSPG and GD-LSPG are the time associated with computing the Jacobian of the decoder (which is not needed for the POD-LSPG approach) and the time associated with computing the high-dimensional residual and its Jacobian. 

As noted by \cite{farhat2015ecsw}, cost savings in ROMs employing LSPG projection can be achieved under three primary scenarios. In the first scenario, although setting up the linear system in \eqref{eq:gn_update} has an operation count that scales with the dimension of the FOM, cost savings can still be achieved if the computational cost of the FOM is dominated by the cost of inverting the Jacobian of the time-discrete residual, and if the cost to set up and minimize \eqref{eq:res_min_full} is considerably lower than that of setting up and minimizing \eqref{eq:ode2argmin}. In the second scenario, the ROM is sufficienctly stable to be solved with a much larger time step than the FOM, thereby requiring fewer evaluations of \eqref{eq:gn_update} than \eqref{eq:ode2argmin}. In the third (and most common) scenario, a hyper-reduction scheme is employed to sparsely sample terms in the residual to approximate the minimization of \eqref{eq:res_min_full} with substantially fewer computations. POD-LSPG achieves approximately 20\% cost savings for the latent space dimension $M=3$, primarily due to satisfying the first scenario. However, this comes at the expense of reduced solution accuracy. Since dLSPG and GD-LSPG do not fall under any of the identified scenarios for cost savings, they do not achieve cost savings with respect to the FOM in this setting. However, the primary goal of this study is not computational efficiency, but rather to assess the extent of dimensionality reduction capabilities of GD-LSPG while accurately capturing the advection-dominated nonlinear behavior inherent in the numerical examples.

\subsection{Two-dimensional Euler equations} \label{ssec:2DEuler}

In our second and third numerical experiments, we consider the FVM deployed to solve the two-dimensional Euler equations in two distinct settings. In our first setting, a Riemann problem setup for a square domain is solved using an unstructured mesh, which allows us to establish the benefits of our graph autoencoder architecture. In the second setting, we demonstrate the flexibility of our graph autoencoder architecture by applying it to a bow shock generated by flow past a cylinder problem on an irregular domain modeled with an unstructured mesh. To introduce the FVM solver used for these two experiments, we provide a brief overview of the important concepts in this section and encourage the reader to consult \cite{nishikawa2008rhrs,paardekooper2017multidimensional,Ren2003Riemann,Roe1981Riemann} for further reading on Riemann solvers for the Euler equations. We begin with the two-dimensional Euler equations in the form of hyperbolic PDEs:

\begin{equation} \label{eq:FVMpde}
    \frac{\partial \mathbf{U}}{\partial t} + \frac{\partial \mathbf{F}}{\partial x} + \frac{\partial \mathbf{G}}{\partial y} = \mathbf{0},
\end{equation}

\begin{equation} \label{eq:FVMeuler}
    \mathbf{U} = 
    \begin{bmatrix}
        \rho \\
        \rho u \\
        \rho v \\
        \rho E
    \end{bmatrix}, 
    \quad    
    \mathbf{F} =
    \begin{bmatrix}
        \rho u\\
        \rho u^2 + P \\
        \rho u v \\
        \rho u H
    \end{bmatrix},
    \quad
    \mathbf{G} =
    \begin{bmatrix}
        \rho v \\
        \rho u v \\
        \rho v^2 + P  \\
        \rho v H 
    \end{bmatrix},
\end{equation}

where $\rho \in \mathbb{R}_+$ denotes density, $u \in \mathbb{R}$ and $v \in \mathbb{R}$ denote velocities in the $x$ and $y$ directions, respectively, $P \in \mathbb{R}_+$ denotes pressure, $E=\frac{1}{\gamma-1}\frac{p}{\rho}+\frac{1}{2}\left(u^2+v^2\right) \in \mathbb{R}_+$ and $H = \frac{\gamma}{\gamma-1}\frac{p}{\rho}+\frac{1}{2}\left(u^2+v^2\right) \in \mathbb{R}_+$ denote specific total energy and enthalpy, respectively, and $\gamma \in \mathbb{R}_+$ is the specific heat ratio. We integrate \eqref{eq:FVMpde} over a control volume, $\Gamma$, and apply the divergence theorem to get the form,

\begin{equation} \label{eq:FVMpde_int}
    \int_{\Gamma} \frac{\mathrm{d}} {\mathrm{dt}} \mathbf{U} \mathrm{dV} + \int_{\partial \Gamma} \mathbf{H} \cdot \mathbf{\hat{n}} \mathrm{dA} = \mathbf{0},
\end{equation}
where $\mathrm{dA}\in \mathbb{R}_+$, $\mathrm{dV}\in\mathbb{R}_+$ and $\partial \Gamma$ denote the differential surface area, the differential volume, and the surface of the control volume, respectively, $\mathbf{H} = \mathbf{F}\hat{i} + \mathbf{G} \hat{j}$, $\mathbf{\hat{n}} = n_x \hat{i} + n_y \hat{j}$ denotes the outward facing unit normal vector from the control volume, where $\hat{i}$ and $\hat{j}$ denote the Cartesian unit vectors in $x$ and $y$ directions, respectively, and $n_x\in [-1,1]$ and $n_y \in [-1,1]$ denote the components of $\mathbf{\hat{n}}$ decomposed in the $x$ and $y$ directions. 

An approximate solution for \eqref{eq:FVMpde_int} is achieved by first spatially discretizing the domain, where the surface integral term is approximated by obtaining the numerical flux passing over the cell faces in the unstructured mesh. The numerical flux is computed using a Riemann solver designed to resolve the computationally difficult nature of the hyperbolic Euler equations. In this numerical experiment, we choose a Rotated Roe, Harten, Lax, and van Leer (R-RHLL) flux from \cite{nishikawa2008rhrs} coupled with a forward Euler time integration scheme to generate a time series solution. The resulting scheme for a single finite volume cell takes the form

\begin{equation} \label{eq:FVM_fluxes}
    \mathbf{U}^{n+1}_i = \mathbf{U}^n_i - \Delta t \sum_{j\in \mathcal{M}(i)} \boldsymbol{\Pi}_{ij} \left(\mathbf{U}^n_i, \hspace{2mm} \mathbf{U}^n_j \right),
\end{equation}
where $\mathbf{U}_i^{n}, \, \mathbf{U}_j^{n} \in \mathbb{R}^4$ denote the state vector of the $i^{\mathrm{th}}$ and $j^{\mathrm{th}}$ cells at the $n^{\mathrm{th}}$ time step, respectively,
$\mathcal{M}(i)$ denotes the set of neighboring cells of the $i^{\mathrm{th}}$ cell (i.e., sharing an interface), $\boldsymbol{\Pi}_{ij}: \mathbb{R}^4 \times \mathbb{R}^4 \rightarrow \mathbb{R}^4$ denotes the function that computes the R-RHLL flux at the interface between the $i^{\mathrm{th}}$-- $j^{\mathrm{th}}$ cells \cite{nishikawa2008rhrs}. Therefore, \eqref{eq:FVM_fluxes} can be written in the residual-minimization cell-wise form of \eqref{eq:ode2} at the $i^{\mathrm{th}}$ cell with

\begin{equation} \label{eq:FVM_q}
    \mathbf{f}_i: (\mathbf{U}^{n}, t^n; \boldsymbol{\mu}) \mapsto \sum_{j\in \mathcal{M}(i)} \boldsymbol{\Pi}_{ij} \left(\mathbf{U}^n_i, \hspace{2mm} \mathbf{U}^n_j \right),
\end{equation}
and $\alpha_0=1$, $\alpha_1=-1$, $\beta_0=0$, $\beta_1=\Delta t$, $\tau=1$, and $\boldsymbol{\xi} = (\mathbf{U}_1^{n+1},\mathbf{U}_2^{n+1},\ldots,\mathbf{U}_{N_c}^{n+1})^T$ when written in the form of \eqref{eq:ode2}. We note that the minimization problem associated with the residual of the FOM \eqref{eq:FVM_fluxes} is solved explicitly using the forward Euler scheme. Consequently, the POD-LSPG solution is equivalent to the POD-Galerkin solution (see Theorem 4.2 in \cite{lee2020deeplspg}). Additionally, when the forward Euler scheme is deployed in GD-LSPG, minimizing the projection of the residual onto the low-dimensional latent space according to \eqref{eq:lowDimRes} is performed implicitly using the iterative solver described in Section \ref{sec:TimeIntegration}. 
Our implementation of the R-RHLL flux uses a triangular mesh generated by Gmsh \cite{geuzaine2009gmsh} and Numba's just-in-time compiler \cite{lam2015numba} to compile the code efficiently.

For the problems studied in this section, the ROMs are built to reconstruct the conserved state variables ($\rho, \rho u, \rho v, \rho E$), as it allows for natural integration of \eqref{eq:FVMpde}-\eqref{eq:FVM_q} into the projection scheme outlined in Section \ref{sec:TimeIntegration}. Consequently, the reported reconstruction and state prediction errors are evaluated with respect to the conserved state variables. To compare the ROM solutions with those of FOM, we focus on pressure and density fields where the latter is shown by contours.

\subsubsection{Riemann problem}

\begin{figure}[!htb]
    \centering
    \begin{subfigure}[b]{0.375\textwidth}
    \centering
    \hspace{-.5cm}
    \includegraphics[height=4cm,valign=t]{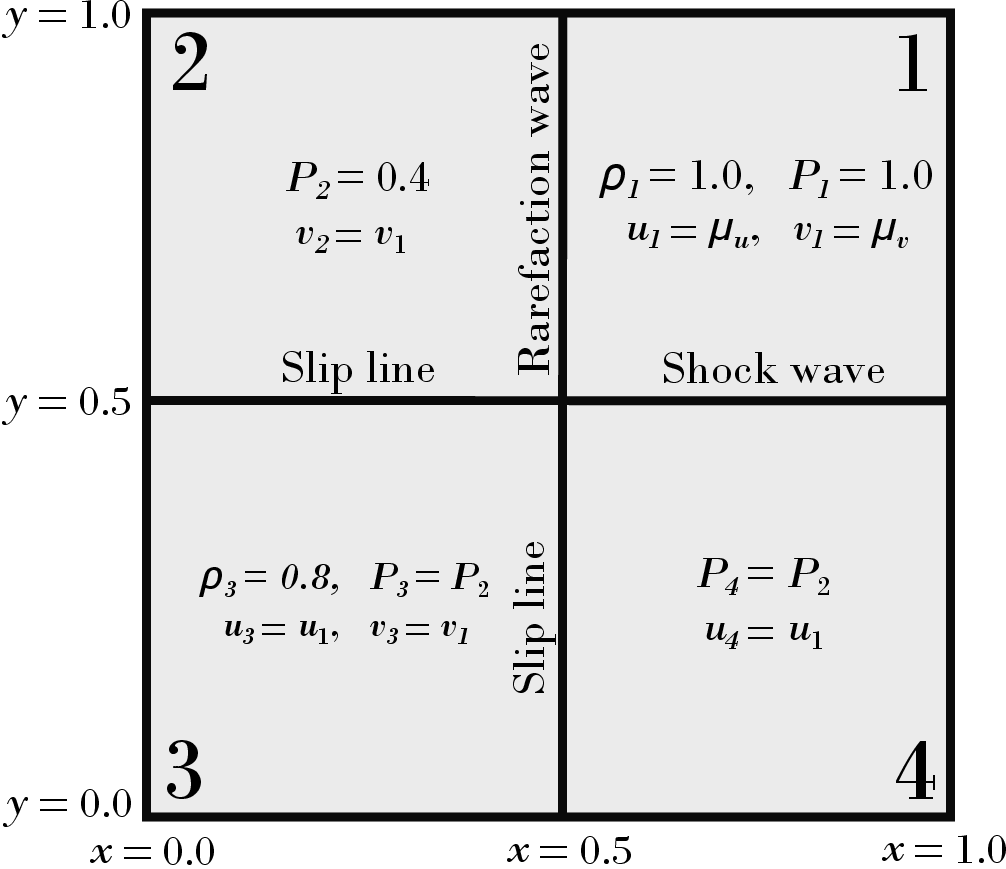}
    \caption{}
    \end{subfigure}
    \begin{subfigure}[b]{0.375\textwidth}
    \centering
    \hspace{-.5cm}
    \includegraphics[height=4cm,valign=t]{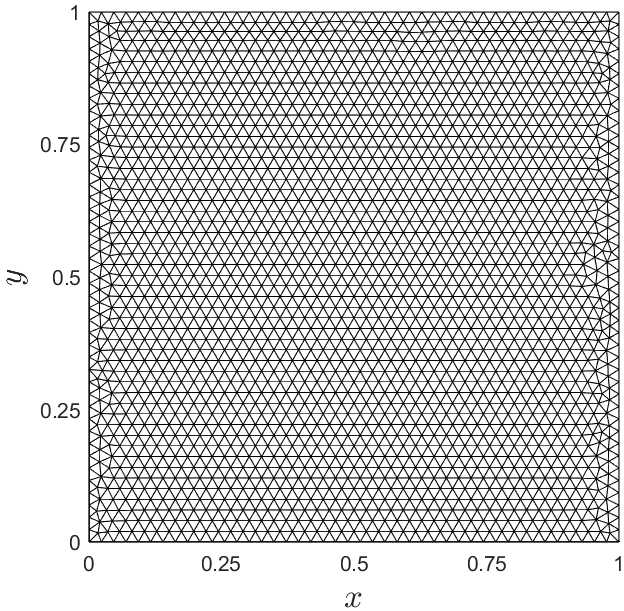}
    \caption{}
    \end{subfigure}
    \captionsetup{justification=centering}
    \caption{(a) Setup for the parametric Euler equations to be solved by a Riemann solver. The quadrants have been numbered in the figure. The problem's parameters are taken to be the initial velocities in the $x$ and $y$ directions in the top right quadrant, i.e., $\boldsymbol{\mu}=\left(\mu_u,\mu_v\right)$. Varying the initial velocity results in the shock wave and rarefaction wave propagating at different speeds and in different directions, resulting in an advection-driven flow. (b) The unstructured mesh used to solve 2D Euler equations for Riemann problem setup. Note that this unstructured finite volume mesh is not directly compatible with a CNN-based autoencoder, and will therefore require interpolation to perform dLSPG.}
    \label{fig:RiemannProb_mesh}
\end{figure}

Our first setting in which we deploy GD-LSPG for the 2D Euler equations is a Riemann problem setup. We solve \eqref{eq:FVMpde}-\eqref{eq:FVMeuler} on the domain $x \in [0, 1]$, $y \in [0, 1]$ with outflow boundary conditions that are computed via the fluxes of the cells along each boundary. The initial conditions are defined by dividing the domain into quadrants, where a different state is defined in each quadrant (see Figure \ref{fig:RiemannProb_mesh}a). In this experiment, we define the quadrants as a parameterized version of configuration G from \cite{schulzrinne1993riemann} (or configuration 15 from \cite{kurganov2002riemann,liska2003comparison,lax1998riemann}), i.e.,

\begin{equation}
    \begin{gathered}
        \rho_1 = 1.0, \quad \rho_3 = 0.8, \\
        u_1 = u_3 = u_4 = \mu_{u}, \\
        v_1 = v_2 = v_3 = \mu_{v}, \\ 
        P_1 = 1.0, \quad P_2 = P_3 = P_4 = 0.4,
    \end{gathered}
\end{equation}
where $\mu_{u},\, \mu_{v} \in \mathbb{R}$ are the model's parameters, i.e., $\boldsymbol{\mu} = ( \mu_u, \mu_v )$, and the remaining variables ($\rho_2, \rho_4, u_2, v_4$) are defined by the Rankine-Hugoniot relations and the relations for a polytropic gas. Specifically, the rarefaction wave yields the conditions,

\begin{equation}
    \begin{gathered}
    \rho_2 = \rho_1 \left( \frac{P_2}{P_1} \right) ^ {\frac{1}{\gamma}}, \\
u_2 = u_1 + \frac{2 \sqrt{\gamma}}{\gamma-1}\left( \sqrt{\frac{P_2}{\rho_2}} -  \sqrt{\frac{P_1}{\rho_1}} \right),
    \end{gathered}
\end{equation}
and the shock wave yields the conditions,
\begin{equation}
    \begin{gathered}
        \rho_4 = \rho_1 \left(\frac{\frac{P_4}{P_1} + \frac{\gamma-1}{\gamma+1} }{1+\frac{\gamma-1}{\gamma+1} \frac{P_4}{P_1}} \right), \\
        v_4 = v_1 + \sqrt{\frac{(P_4-P_1)(\rho_4-\rho_1)}{\rho_4 \rho_1}}.
    \end{gathered}
\end{equation}

We generate an unstructured mesh with $4328$ finite volume cells, where the mesh is presented in Figure \ref{fig:RiemannProb_mesh}b. Next, we perform a parametric study by varying the initial velocities in the top right quadrant via $\boldsymbol{\mu} = (\mu_u=-1.2-0.2i, \mu_v=-0.3-0.1j)$, with $i=0,\ldots,4$ and $j=0,\ldots,4$, resulting in solutions to 25 different parameter sets. We take $\Delta t = 0.001$ and $T_f=0.3$ (which ensures that the shock and rarefaction waves remain in the domain for all parameter sets), therefore collecting $301$ snapshots for each parameter set including the initial conditions. The solutions from the parametric study were used as training data to train the autoencoder. To generate the POD-LSPG solution, we set the tolerance $\kappa$ in \eqref{eq:convergence} to be $10^{-4}$, whereas, the tolerance is set to $10^{-3}$ for the CNN-based dLSPG and GD-LSPG. The step size, $\beta^{(j)}$, for all three models is set to $1.0$ at all time steps.

While this solution is modeled by an unstructured mesh, the domain is square and the discretization is mostly regular. As a result, we can interpolate the solution states on unstructured cell centers to a regular, structured counterpart mesh for direct application to a CNN-based autoencoder. In this study, we simply use a $k-$nearest neighbors interpolation with  $k=3$. Here, we establish the benefit of using GD-LSPG for the unstructured mesh as opposed to deploying the interpolated CNN-based autoencoder to dLSPG. To illustrate, we repeat the training for each autoencoder at a given latent space dimension five times, where the only variation between training processes is the random initialization of weights and biases and the random generation of mini-batches at each epoch.

\begin{figure}[ht!]
    \centering
    \begin{tabular}{ccccc|c}
        & {\footnotesize $\quad$Ground truth} & {\footnotesize $\qquad$POD-LSPG} & {\footnotesize $\quad$Interpolated dLSPG} & {\footnotesize $\qquad$GD-LSPG} & \\
        \hline & & & & & \\[-1em]
        \raisebox{3em}{\multirow{2}{*}{{\rotatebox[origin=lb]{90}{\footnotesize\smash{$\boldsymbol{\mu} = (\mu_u = -1.3, \mu_v = -0.65)$}}}}}
        & \raisebox{3em}{\multirow{2}{*}{\includegraphics[height=0.22\textwidth]{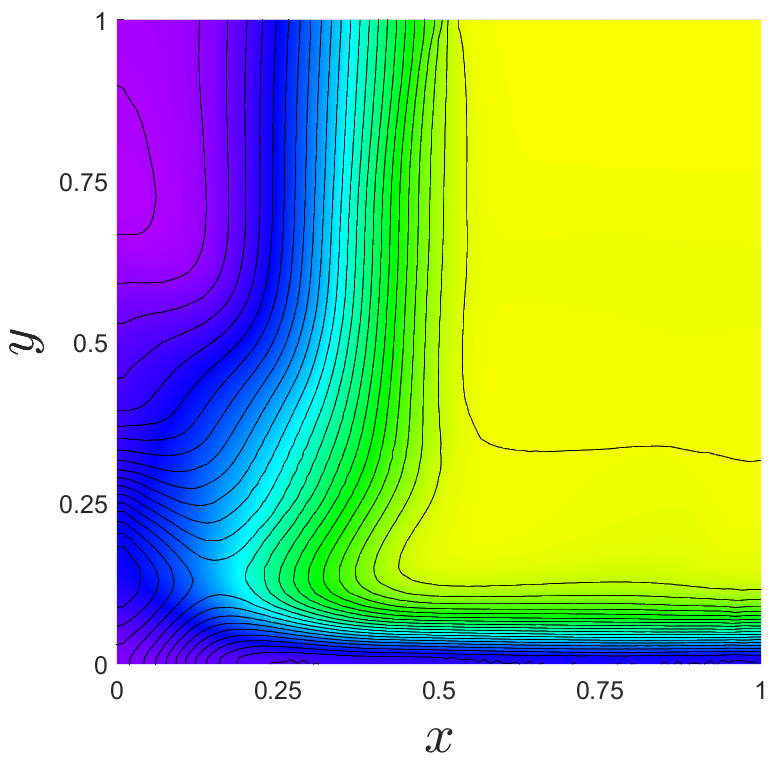}}}
        & \raisebox{3em}{\multirow{2}{*}{\includegraphics[height=0.22\textwidth]{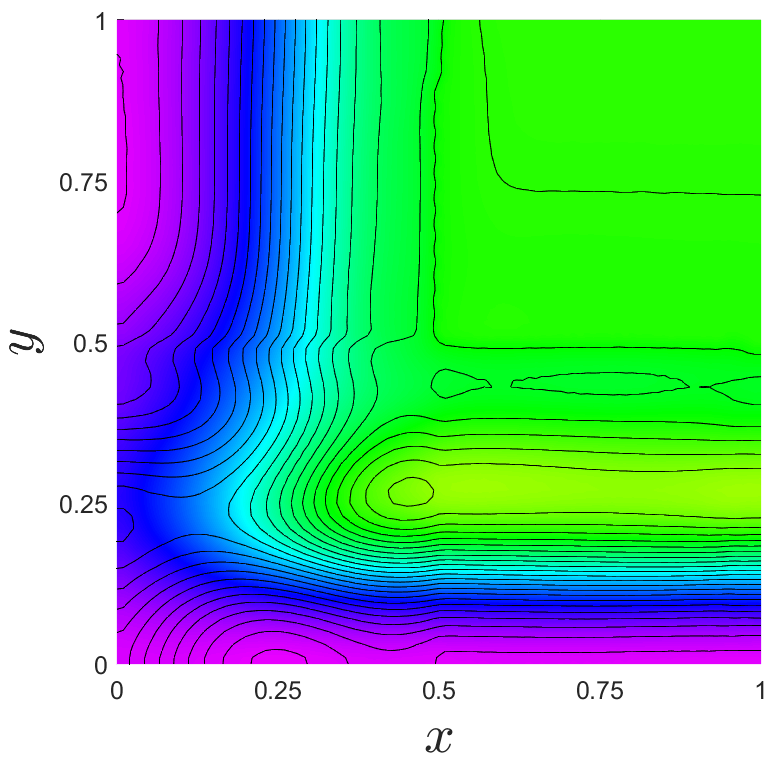}}}
        & \includegraphics[height=0.22\textwidth]{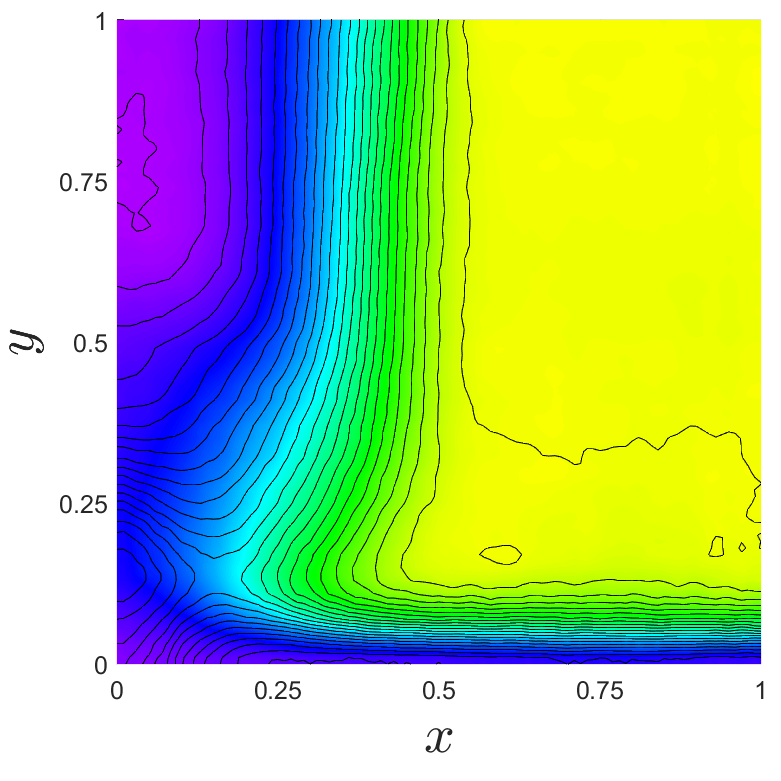}
        & \includegraphics[height=0.22\textwidth]{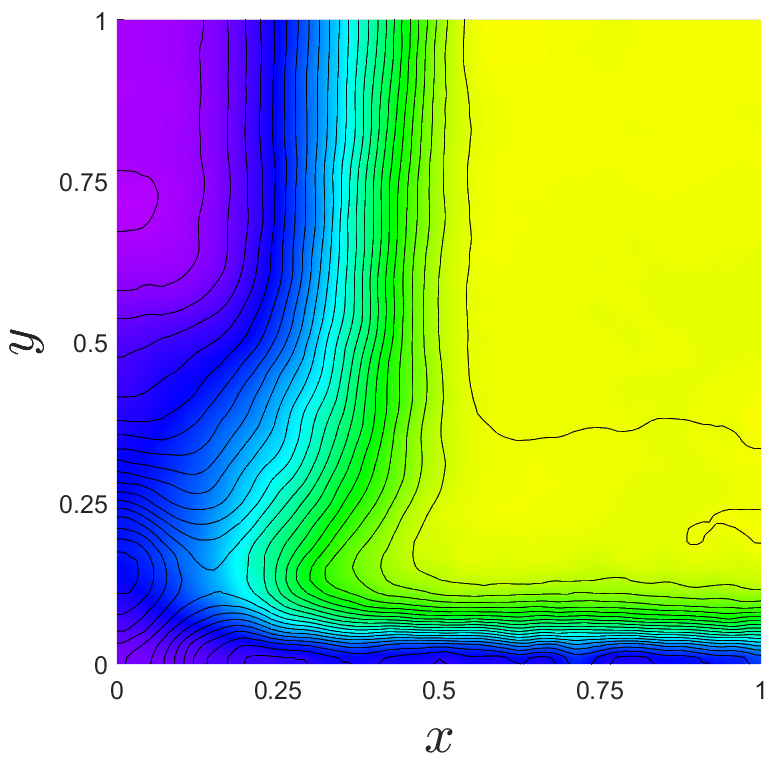}
        & \raisebox{6.5em}{\rotatebox[origin=lb]{270}{\footnotesize\smash{best-case}}}
        \\
        & 
        & 
        & \includegraphics[height=0.22\textwidth]{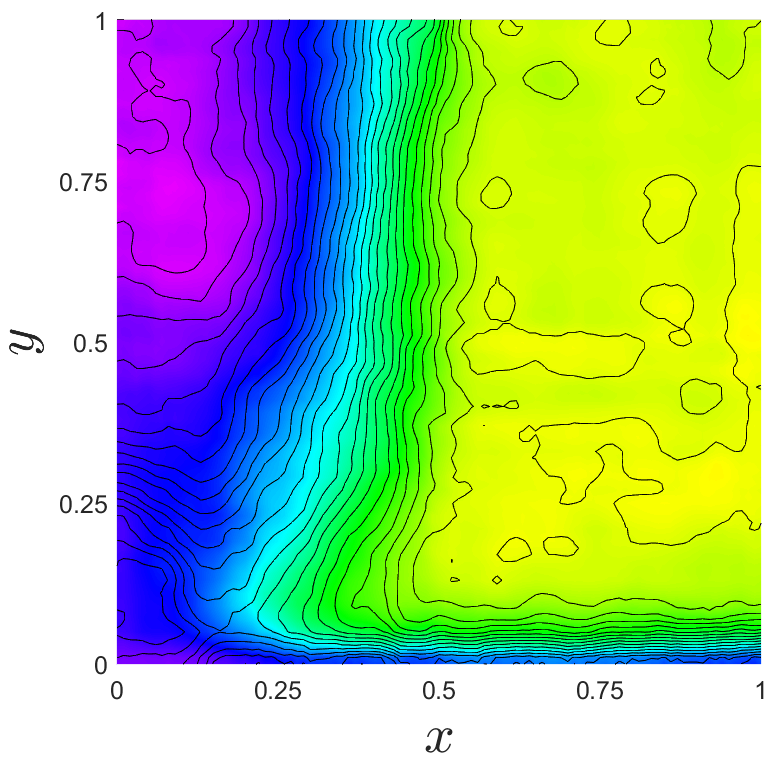}
        & \includegraphics[height=0.22\textwidth]{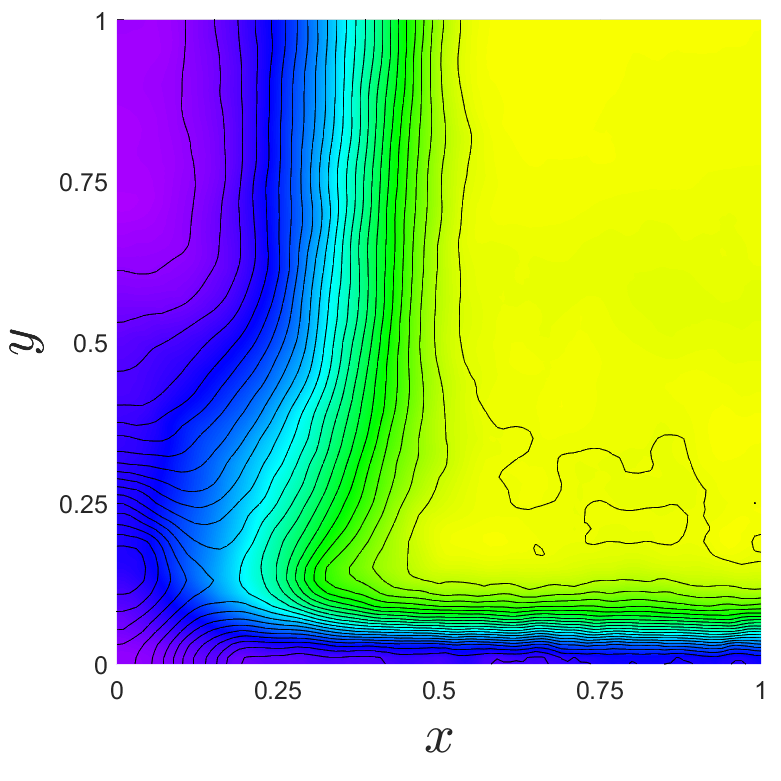}
        & \raisebox{6.5em}{\rotatebox[origin=lb]{270}{\footnotesize\smash{worst-case}}}
        \\[2pt]
        \hline & & & & & \\[-1em]
        \raisebox{3em}{\multirow{2}{*}{{\rotatebox[origin=lb]{90}{\footnotesize\smash{$\boldsymbol{\mu} = (\mu_u = -1.9, \mu_v = -0.35)$}}}}}
        & \raisebox{3em}{\multirow{2}{*}{\includegraphics[height=0.22\textwidth]{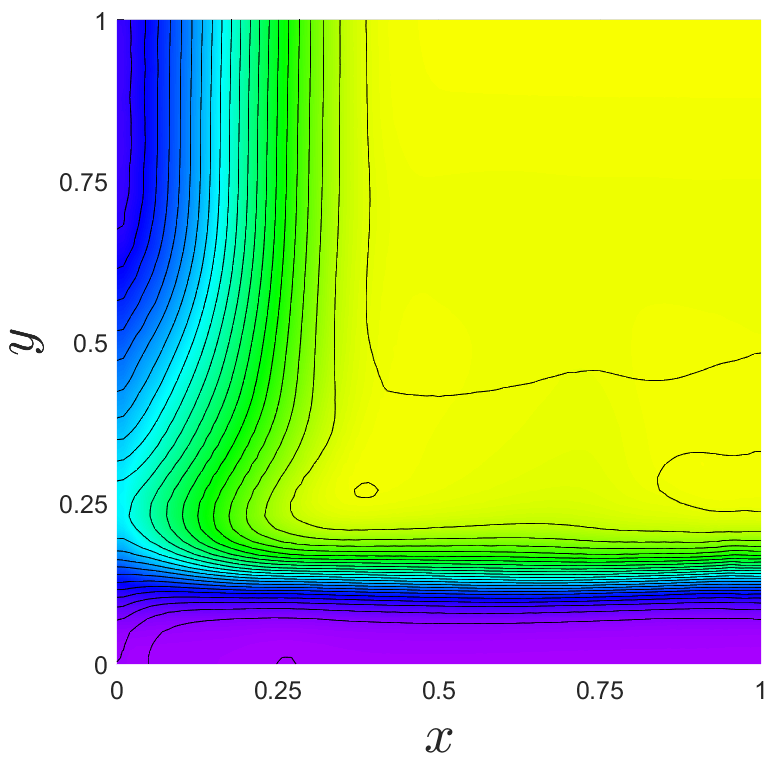}}}
        & \raisebox{3em}{\multirow{2}{*}{\includegraphics[height=0.22\textwidth]{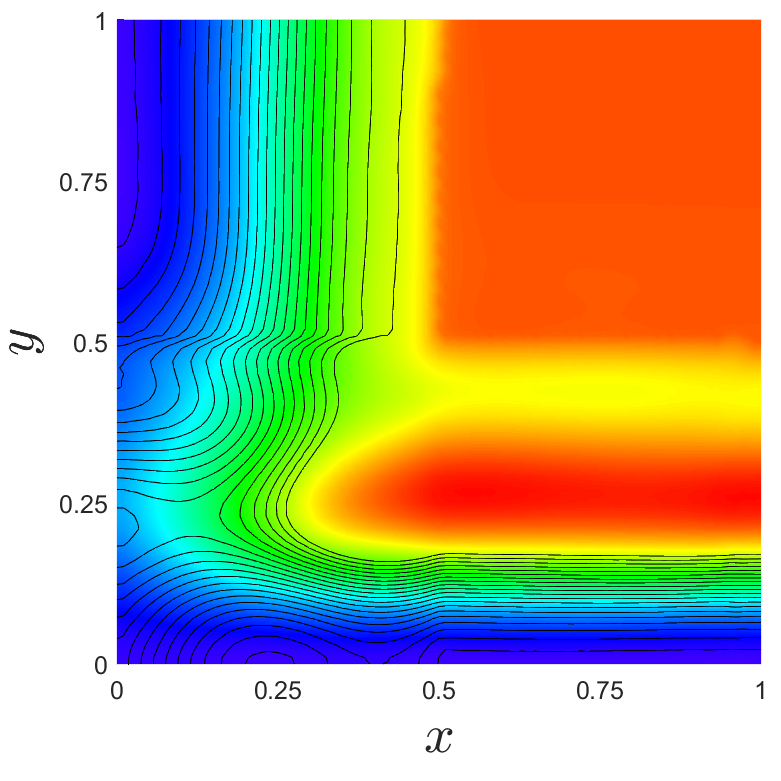}}}
        & \includegraphics[height=0.22\textwidth]{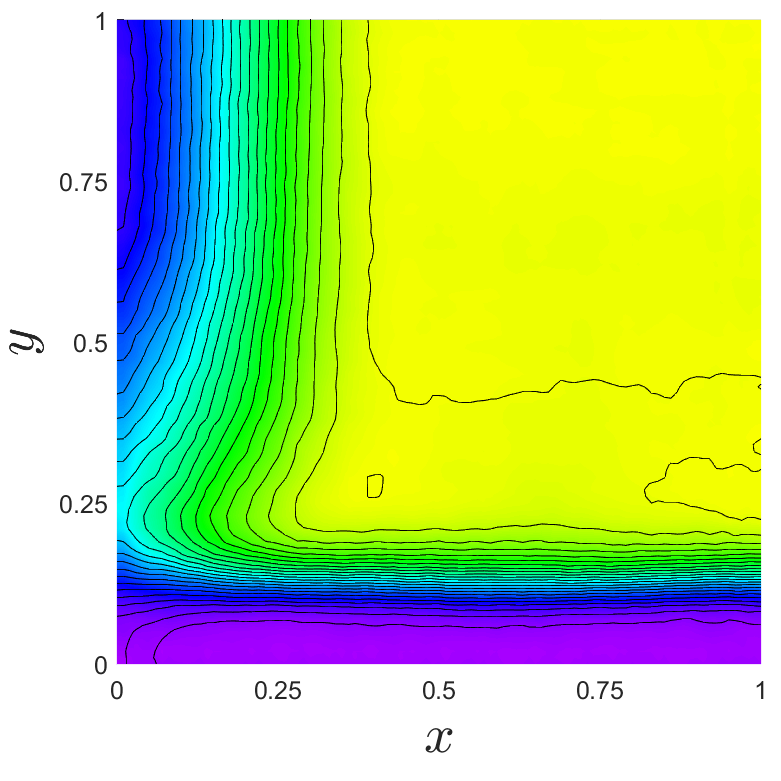}
        & \includegraphics[height=0.22\textwidth]{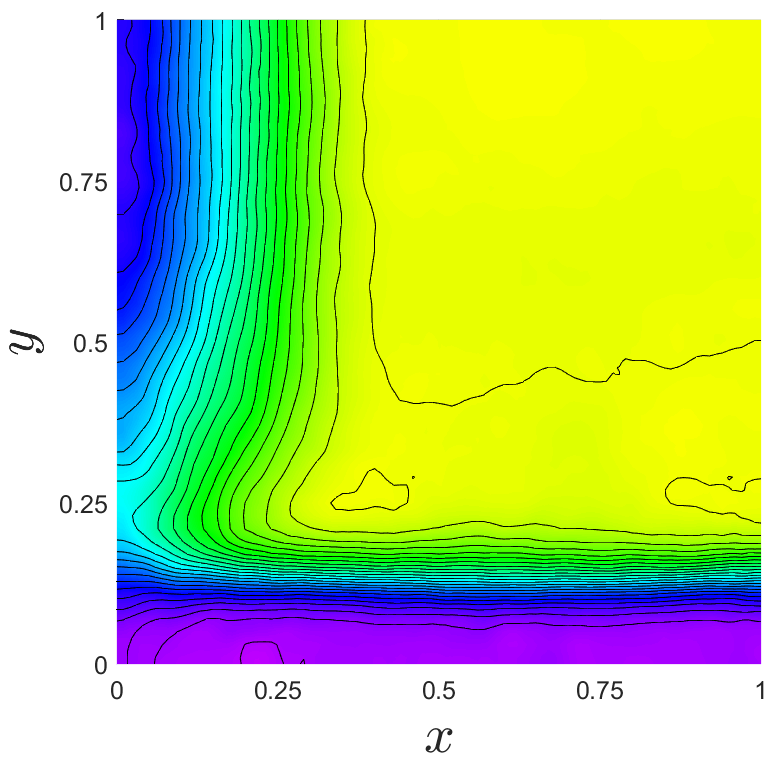}
        & \raisebox{6.5em}{\rotatebox[origin=lb]{270}{\footnotesize\smash{best-case}}}
        \\
        & 
        & 
        & \includegraphics[height=0.22\textwidth]{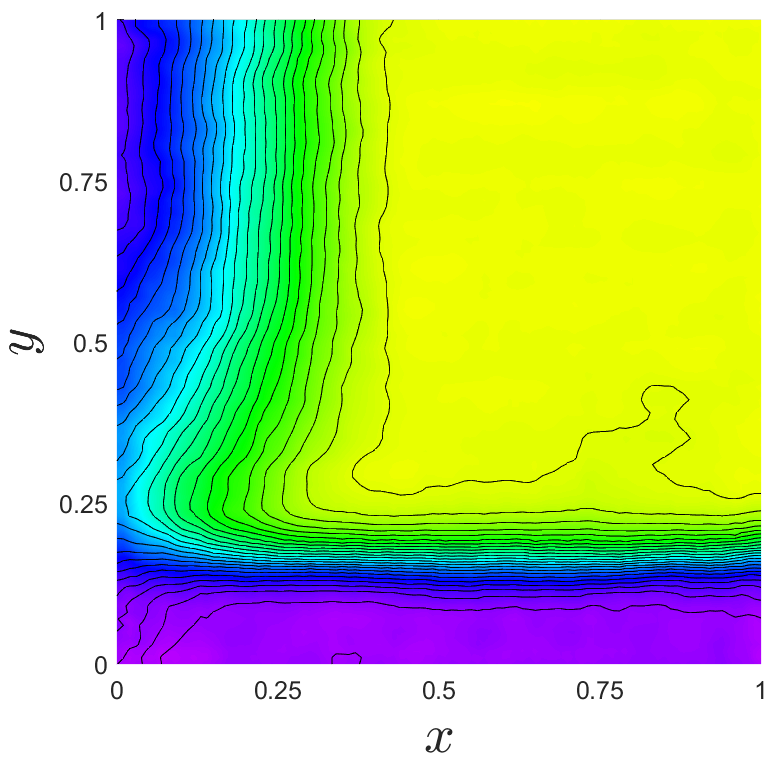}
        & \includegraphics[height=0.22\textwidth]{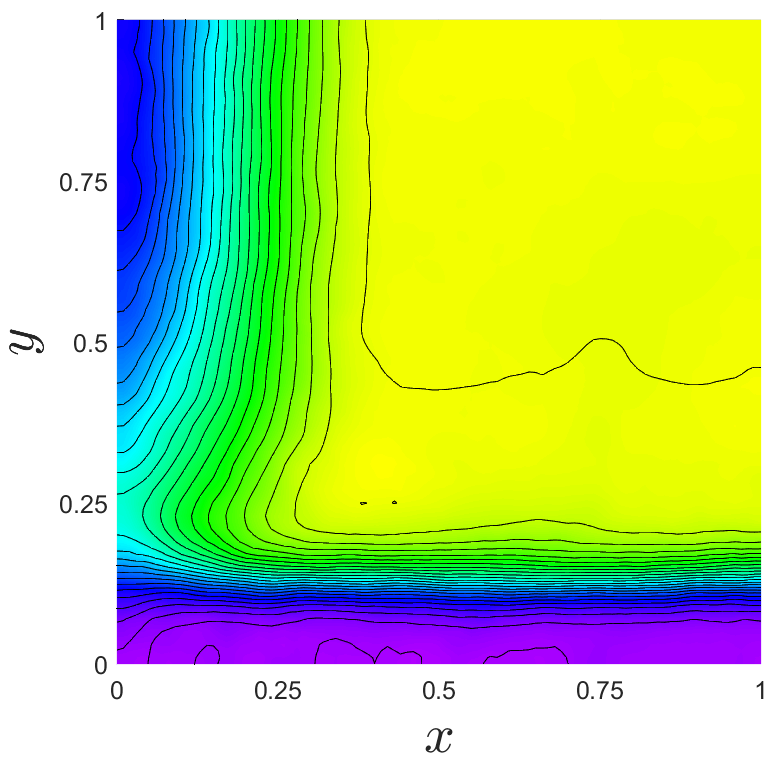}
        & \raisebox{6.5em}{\rotatebox[origin=lb]{270}{\footnotesize\smash{worst-case}}}
        \\
        & \multicolumn{5}{c}{\includegraphics[scale=.31]{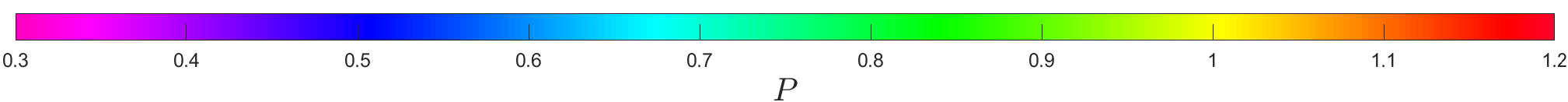}}
        
    \end{tabular}
    \captionsetup{justification=centering}
    \caption{Ground truth, POD-LSPG, and both best- and worst-case interpolated dLSPG and GD-LSPG pressure fields, denoted by color, and density fields, denoted by contours, for the 2D Euler equations. The results are shown at $t=0.3$ for test parameter sets $\boldsymbol{\mu} = (\mu_u = -1.3, \mu_v = -0.65)$ and $\boldsymbol{\mu} = (\mu_u = -1.9, \mu_v = -0.35)$ and for a latent space dimension of $M=3$. Note POD-LSPG's inability to model the shock behavior, leading to a diffusive and inaccurate solution that does not accurately model the advection-driven behavior of the problem. The best-case interpolated dLSPG and GD-LSPG solutions model the shock behavior with much higher accuracy, while the worst-case GD-LSPG solution is more accurate than the worst-case interpolated dLSPG solution. (Online version in color.)}
    \label{fig:13_65_19_35_M3}
\end{figure}

Figure \ref{fig:13_65_19_35_M3} presents the pressure field at $t=0.3$ for two test parameter sets not seen during training, $\boldsymbol{\mu} = (\mu_u=-1.3, \mu_v=-0.65)$ and $\boldsymbol{\mu} = (\mu_u=-1.9, \mu_v=-0.35)$. Results are presented for the ground truth, POD-LSPG, and both the best- and worst-case predictions for interpolated dLSPG and GD-LSPG, where the best and worst cases correspond to the lowest and highest state predictions errors among the five independently trained autoencoders, respectively. As expected, GD-LSPG captures the moving shock behavior more accurately than the affine POD-LSPG solution and also provides more accurate pressure fields in the regions separated by the shock and rarefaction waves. For the best-case error, the interpolated dLSPG solution is slightly more accurate than the GD-LSPG solution. However, for the worst-case error, the GD-LSPG solution is considerably more accurate than the interpolated dLSPG solution. This implies that, within the setting studied, the GD-LSPG framework yields more consistent results than those obtained via the interpolated dLSPG framework.

\begin{figure}[ht!]
    \centering
     \begin{tabular}{cccc}
        & {\footnotesize $\boldsymbol{\mu} = (-1.3, -0.65)$} & {\footnotesize $\boldsymbol{\mu} = (-1.9, -0.35)$} \\
        & \includegraphics[height=0.3\textwidth]{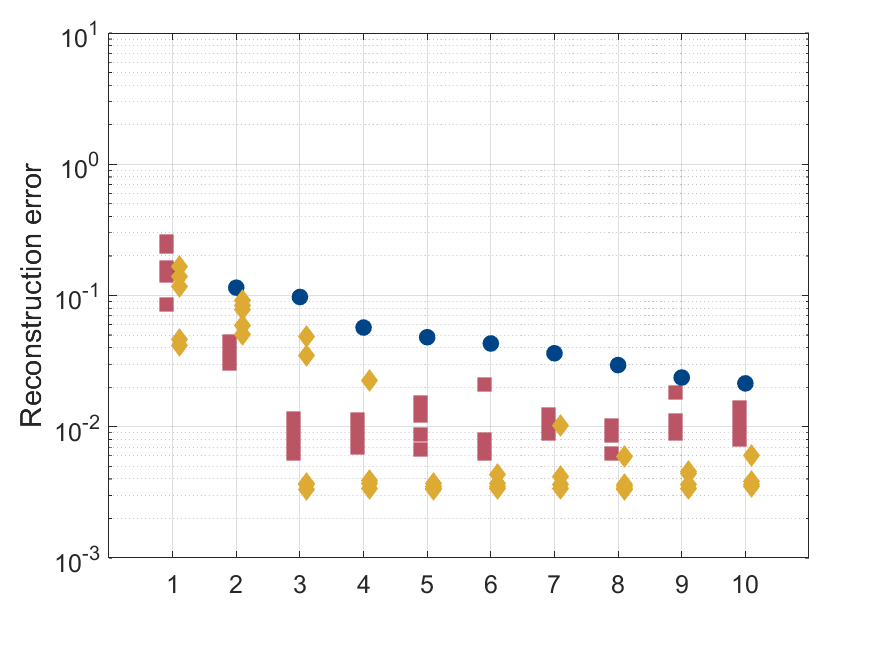}
        & \includegraphics[height=0.3\textwidth]{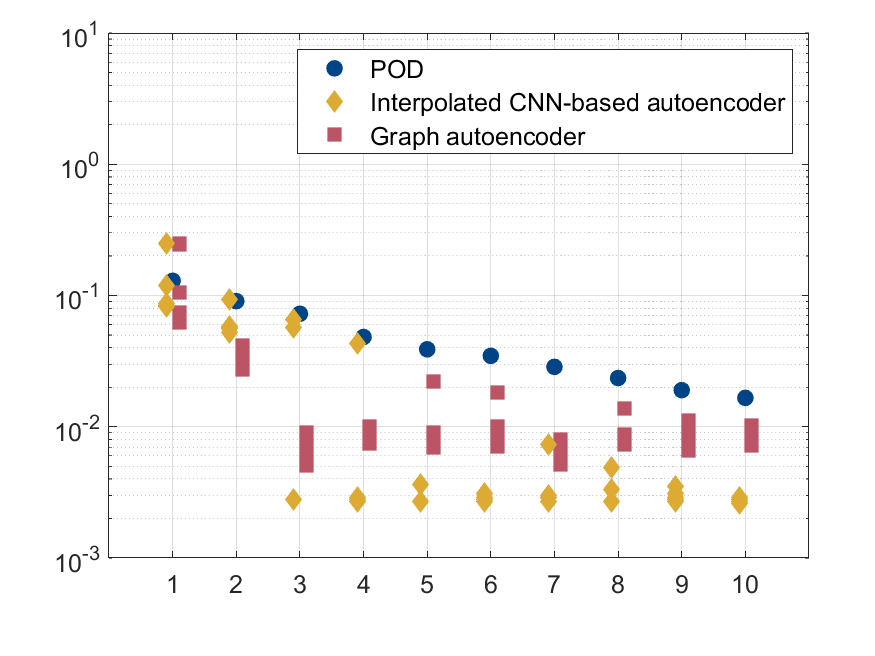}\\ &\includegraphics[height=0.3\textwidth]{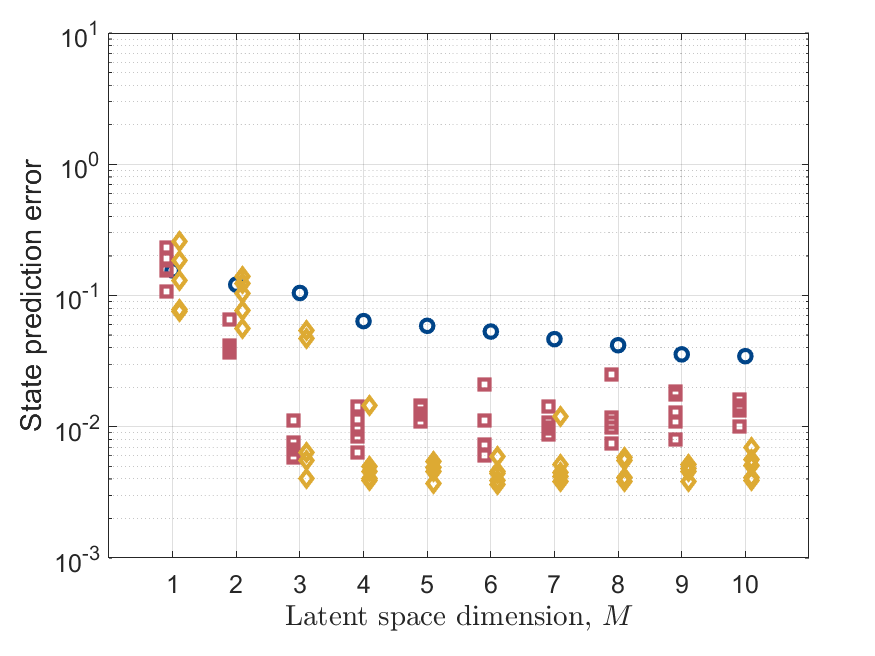}
        & \includegraphics[height=0.3\textwidth]{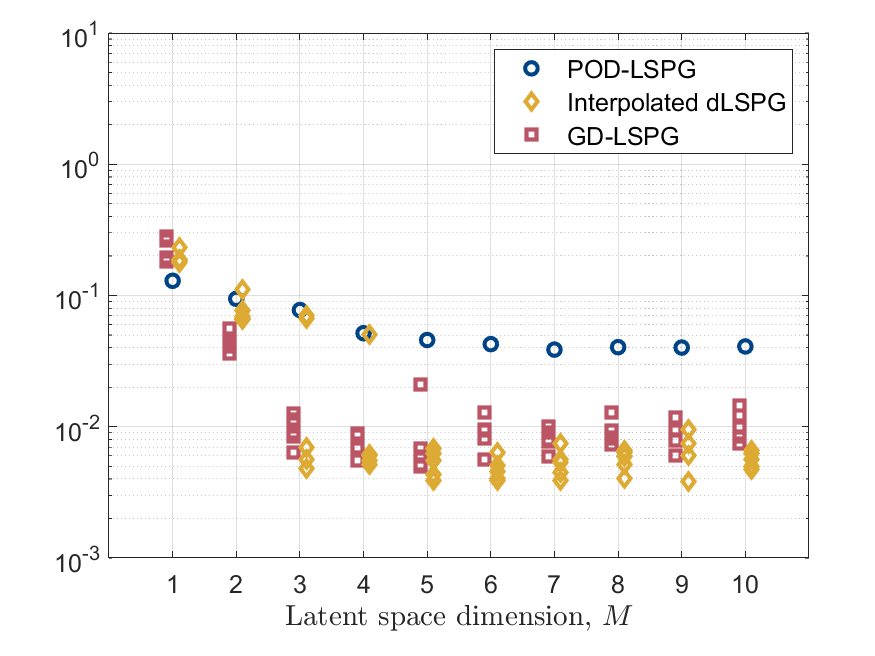}
    \end{tabular}
    \captionsetup{justification=centering}
    \caption{Reconstruction and state prediction errors for two choices of test parameter sets (each represented by a single column) for the Riemann problem setup. We repeat the training for the interpolated CNN-based autoencoder and the graph autoencoder five times, which correspond to the five points at each latent space dimension. We only obtain the POD modes once, which corresponds to only one point at each latent space dimension for POD reconstruction and POD-LSPG. The top row includes plots of POD, interpolated CNN-based autoencoder, and graph autoencoder reconstruction errors evaluated from \eqref{eq:ae_error} and \eqref{eq:pod_error} with respect to the latent space dimension, $M$. Note that the graph autoencoder performs similar level of reconstruction accuracy as the interpolated CNN-based autoencoder. The bottom row demonstrates POD-LSPG, interpolated dLSPG, and GD-LSPG state prediction errors evaluated from \eqref{eq:state_err} with respect to the latent space dimension, $M$. For small latent space dimensions, GD-LSPG (and often interpolated dLSPG) outperforms POD-LSPG in terms of accuracy. Additionally, the interpolated CNN-based autoencoder used in the dLSPG solution fails to generalize well in this application for several latent space dimensions. Note: one of the interpolated dLSPG solutions for $M=1$ failed to converge. (Online version in color.)}
    \label{fig:RP_errors}
\end{figure}

Figure \ref{fig:RP_errors} reports reconstruction and state prediction errors for five independently trained autoencoders at each latent space dimension. We consider latent space dimensions of 1 to 10 compared to the FOM dimension of $4328 \times 4 = 17312$. Specifically, for the latent space dimensions 3 and 4, some of the interpolated CNN-based autoencoders generate solutions that perform considerably worse than the worst-case GD-LSPG solutions at the same latent space dimension. This indicates that the interpolated CNN-based autoencoder is prone to failure in generalizing for parameter sets not seen during training.

\begin{figure}[t!]
    \centering
    \begin{tabular}{cccc|c}
        & {\footnotesize $\qquad$ POD-LSPG} & {\footnotesize $\qquad$ Interpolated dLSPG} & {\footnotesize $\qquad$ GD-LSPG} & \\[1pt]
        \hline & & & & \\[-1em]
        \raisebox{3em}{\multirow{2}{*}{{\rotatebox[origin=lb]{90}{\footnotesize\smash{$\boldsymbol{\mu} = (\mu_u = -1.3, \mu_v = -0.65)$}}}}}
        & \raisebox{3em}{\multirow{2}{*}{\includegraphics[height=0.23\textwidth]{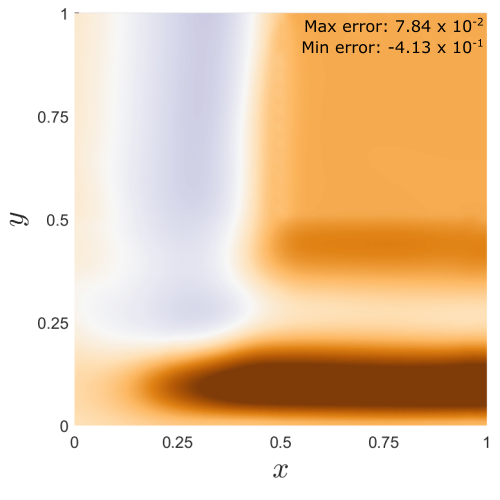}}}
        & \includegraphics[height=0.23\textwidth]{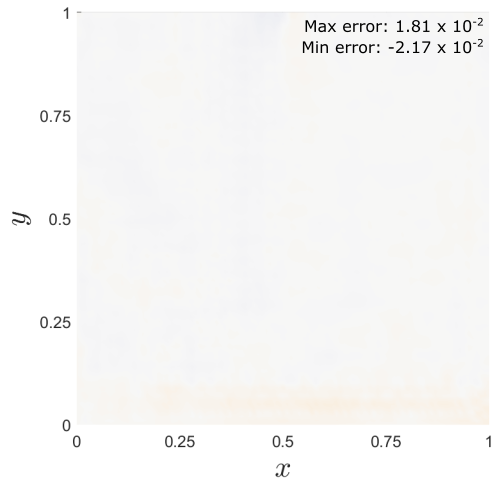}
        & \includegraphics[height=0.23\textwidth]{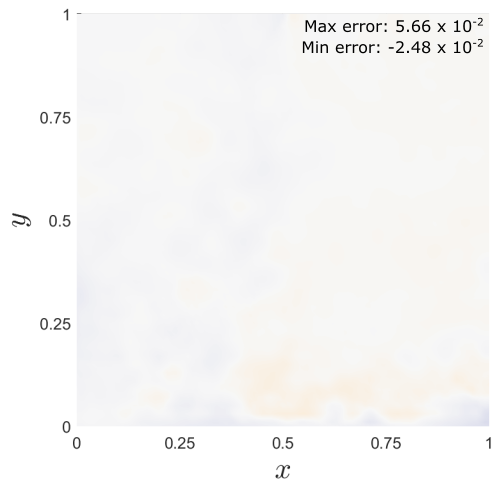}
        & \raisebox{6.5em}{\rotatebox[origin=lb]{270}{\footnotesize\smash{best-case}}}
        \\
        & & \includegraphics[height=0.23\textwidth]{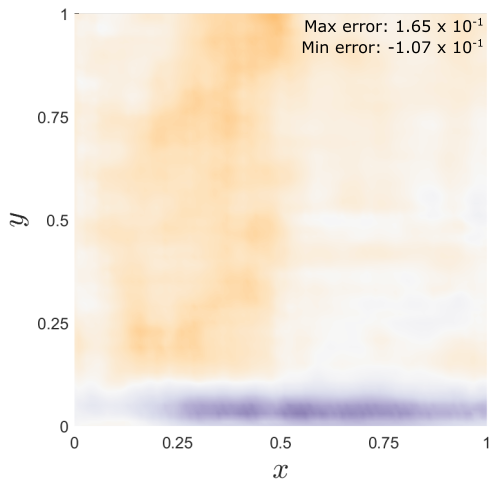}
        & \includegraphics[height=0.23\textwidth]{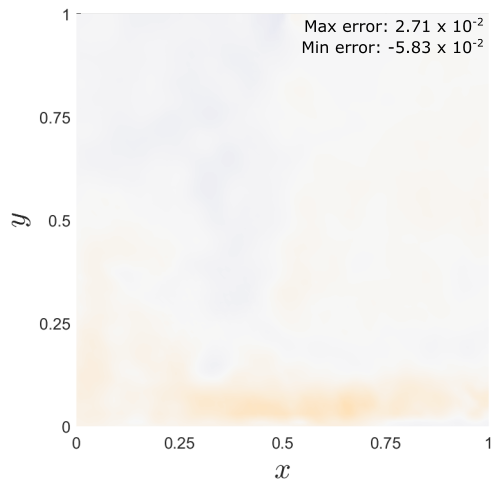}
        & \raisebox{6.5em}{\rotatebox[origin=lb]{270}{\footnotesize\smash{worst-case}}}
        \\[2pt]
        \hline & & & & \\[-1em]
        \raisebox{3em}{\multirow{2}{*}{{\rotatebox[origin=lb]{90}{\footnotesize\smash{$\boldsymbol{\mu} = (\mu_u = -1.9, \mu_v = -0.35)$}}}}}
        & \raisebox{3em}{\multirow{2}{*}{\includegraphics[height=0.23\textwidth]{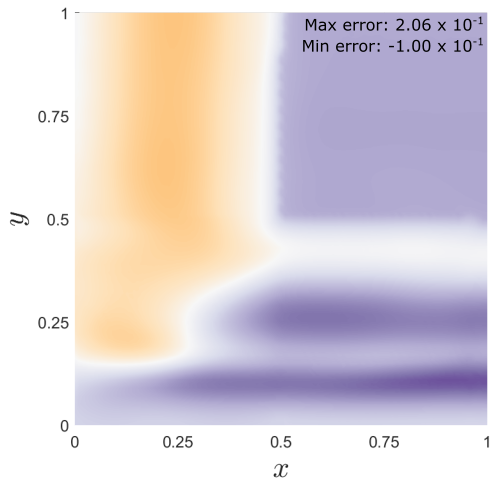}}}
        & \includegraphics[height=0.23\textwidth]{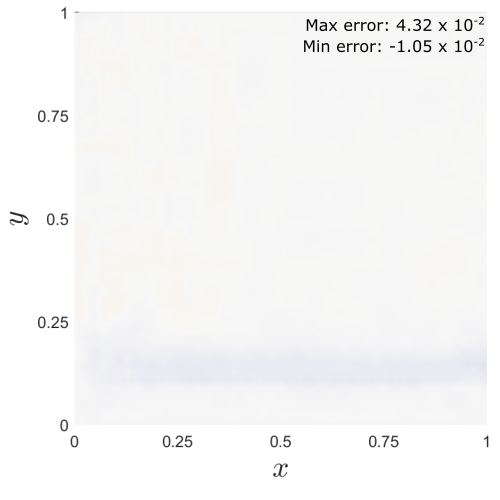}
        & \includegraphics[height=0.23\textwidth]{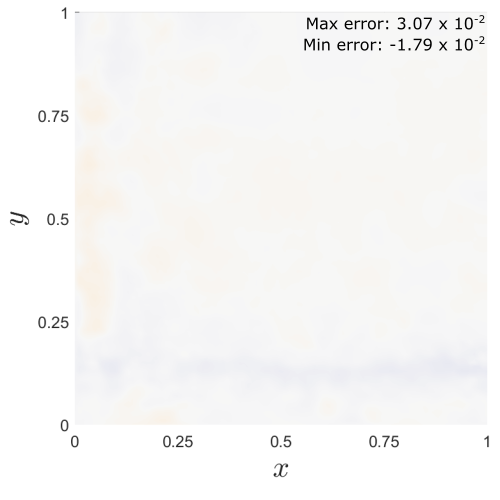}
        & \raisebox{6.5em}{\rotatebox[origin=lb]{270}{\footnotesize\smash{best-case}}}
        \\
        & & \includegraphics[height=0.23\textwidth]{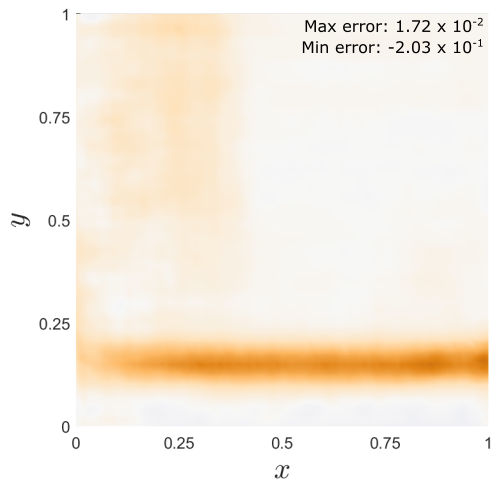}
        & \includegraphics[height=0.23\textwidth]{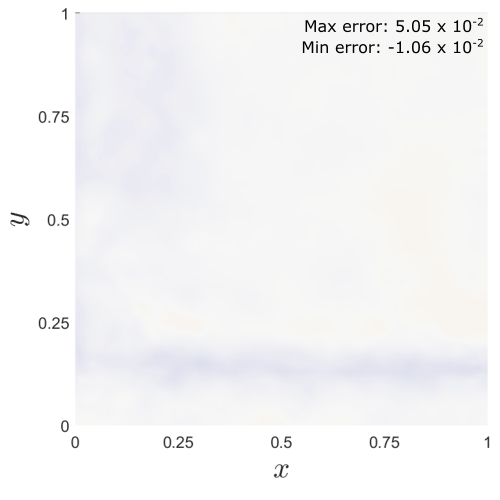}
        & \raisebox{6.5em}{\rotatebox[origin=lb]{270}{\footnotesize\smash{worst-case}}}
        \\
        \multicolumn{5}{c}{\includegraphics[scale=.28]{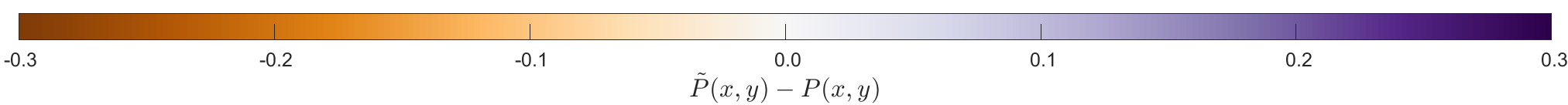}}
        
    \end{tabular}
    \captionsetup{justification=centering}
    \caption{ROM prediction errors associated with the best- and worst-case solutions for the Riemann problem setup. The left, middle and right columns represent the errors for POD-LSPG, interpolated dLSPG and GD-LSPG, respectively, at latent space dimension $M=3$. The results are depicted for the parameter sets $\boldsymbol{\mu} = (\mu_u=-1.3, \mu_v=-0.65)$ and $\boldsymbol{\mu} = (\mu_u=-1.9, \mu_v=-0.35)$ at time $t=0.3$. We define local error to be $\tilde{P}(x,y) - P(x,y)$ at a given time step. The POD-LSPG solution fails to accurately model the advection-driven nature of the solution and shows errors as high as 25\% in some parts of the domain. The best-case interpolated dLSPG and GD-LSPG solutions model the shock wave much more accurately, with only small errors localized near the shock, highlighting the slight phase misalignment between FOM and ROM shock locations as the primary source of error. The worst-case interpolated dLSPG solution has considerably higher errors than the worst-case GD-LSPG solution. (Online version in color.)}
    \label{fig:localError_13_65_19_35}
    
\end{figure}

Finally, Figure \ref{fig:localError_13_65_19_35} presents the local errors in the pressure field at parameter sets $\boldsymbol{\mu} = (\mu_u=-1.3, \mu_v=-0.65)$ and $\boldsymbol{\mu} = (\mu_u=-1.9, \mu_v=-0.35)$ at time $t=0.3$ for POD-LSPG, and the best- and worst-case interpolated dLSPG, and GD-LSPG. Here, it is evident that the POD-LSPG solution introduces considerable error to the pressure field throughout the domain. Alternatively, for best-case solutions, both interpolated dLSPG and GD-LSPG introduce relatively small errors isolated around the shock and rarefaction waves, indicating that the primary source of error for these methods is a phase lag. However, the worst-case GD-LSPG solution is considerably more accurate than the worst-case interpolated dLSPG solution, establishing a better generalization capabilities of GD-LSPG in this setting.

\subsubsection{Bow shock generated by flow past a cylinder problem} \label{ssec:bowShock}

\begin{figure}[!htb]
    \centering
    \begin{subfigure}[b]{0.375\textwidth}
    \centering
    \hspace{.5cm}
    \includegraphics[height=5.5cm]{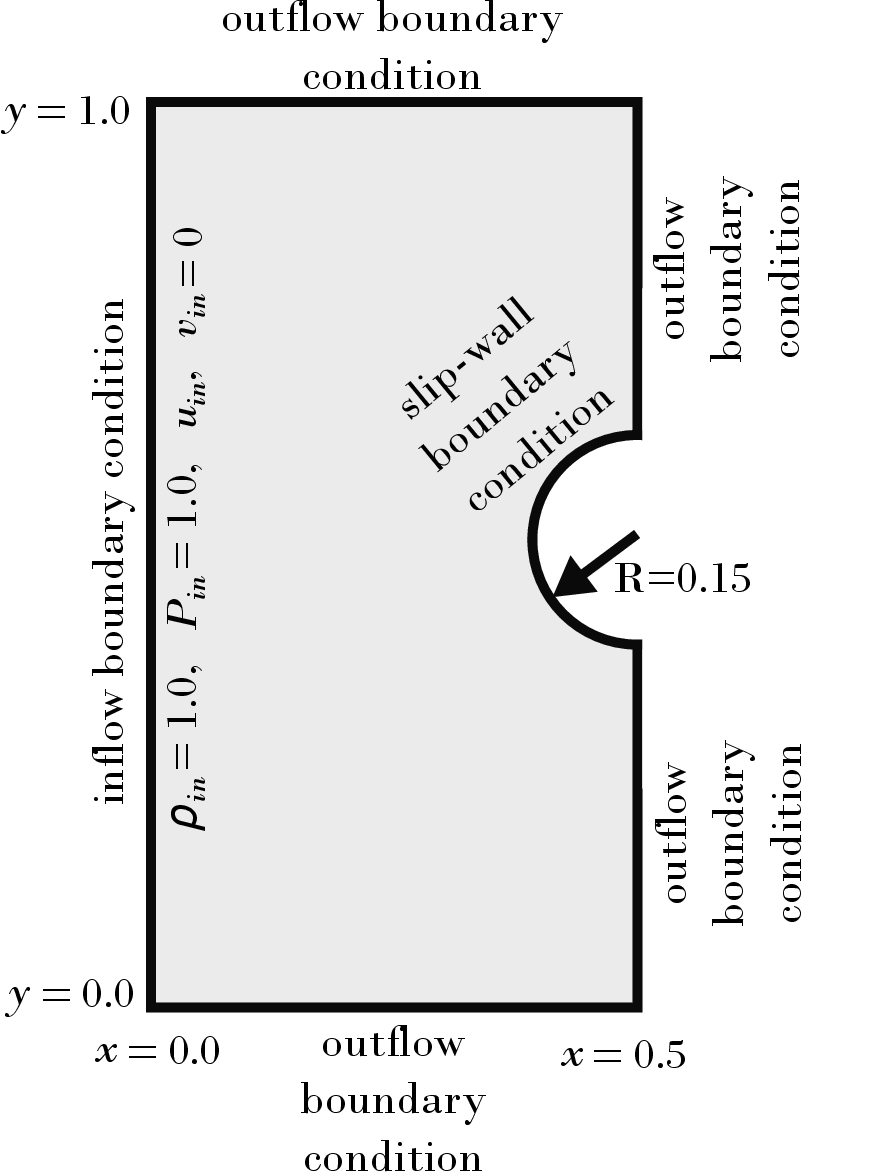}
    \caption{}
    \end{subfigure}
    \hspace{-1cm}
    \begin{subfigure}[b]{0.375\textwidth}
    \centering
    \hspace{-.75cm}
    \raisebox{.5em}{\includegraphics[height=5.2cm]{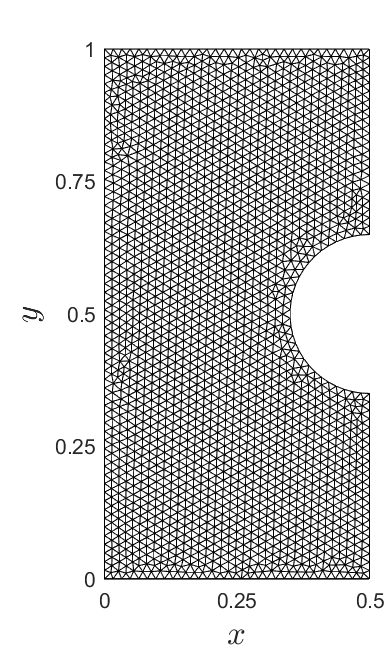}}
    
    \caption{}
    \end{subfigure}
    \captionsetup{justification=centering}
    \caption{(a) Setup for the parametric Euler equations to be solved by a Riemann solver with an unstructured finite volume mesh. A rectangular domain with the leading edge of a cylinder is modeled with the noted boundary conditions. Depending on the freestream Mach number, $\mu_{\mathrm{in}}$, which will be varied as this problem's parameter, a shock can form along the leading edge of the cylinder and propagate through the domain at different speeds. (b) Mesh used to solve 2D Euler equations for the bow shock generated by flow past a cylinder problem. In this case, the physical domain's geometry is more complex than in the Riemann problem setup, therefore benefiting from the unstructured mesh.}
    \label{fig:Cyl_setup}
\end{figure}

In our second setting, we model a bow shock generated by flow past a cylinder using an unstructured finite volume mesh for a parameterized version of a test case found in \cite{nishikawa2008rhrs} to demonstrate the flexibility of GD-LSPG. As seen in Figure \ref{fig:Cyl_setup}a, the model consists of a rectangular domain and the leading edge of a cylinder, where, for the selected range of parameter values, it is expected that a shock will form on the leading edge and propagate through the domain. Because of the domain's geometric irregularity, it is natural to employ an unstructured mesh, which is presented in Figure \ref{fig:Cyl_setup}b. Once again, we solve \eqref{eq:FVMpde}-\eqref{eq:FVMeuler} on the domain that has been spatially discretized into $4148$ finite volume cells using Gmsh \cite{geuzaine2009gmsh}. However, this time we model an inflow boundary condition on the left boundary of the domain, outflow boundary condition on the right side of the domain and the top and bottom boundaries, and a slip-wall boundary condition on the surface of the cylinder. We parameterize the model by the freestream Mach number, $\mu_{\mathrm{in}}$, i.e.,

\begin{equation}
    \begin{gathered}
        \rho_{\mathrm{in}} = 1.0, \quad P_{\mathrm{in}} = 1.0, \quad u_{\mathrm{in}} = \mu_{\mathrm{in}} \sqrt{\frac{\gamma P_{\mathrm{in}}}{\mathrm{\rho_{\mathrm{in}}}}}, \quad v_{\mathrm{in}} = 0,
    \end{gathered}
\end{equation}
where $\rho_{\mathrm{in}}$, $P_{\mathrm{in}}$, $u_{\mathrm{in}}$, and $v_{\mathrm{in}}$ denote the freestream density, freestream pressure, and freestream velocities in the $x$ and $y$ directions, respectively, and $\mu_{in} = 1.0, 1.050, \dots, 1.250$, which results in solutions to 6 different parameters. The solutions exhibit a shock that develops along the leading edge of the cylinder at varying rates. Additionally, the shock propagates at different speeds through the domain and incurs varying pressure, density, and velocity differences across the shock depending on the freestream Mach number. We take $\Delta t = 0.001$ and $T_f=1.0$, therefore collecting $1001$ snapshots for each parameter including the initial conditions.

\begin{figure}[t!]
    \centering
    \begingroup
    \footnotesize
    \begin{tabular}{c|c|cc|cc}
        & $\qquad$Ground truth & $\qquad$POD-LSPG & $\quad$GD-LSPG  & $\qquad$POD-LSPG & $\quad$GD-LSPG\\
        & $\qquad$solution & $\qquad$solution & $\quad$solution & $\qquad$local error & $\quad$local error\\
        \raisebox{7.5em}{\rotatebox[origin=lb]{90}{\smash{$t=0.25$}}}
        & \includegraphics[width=0.15\textwidth]{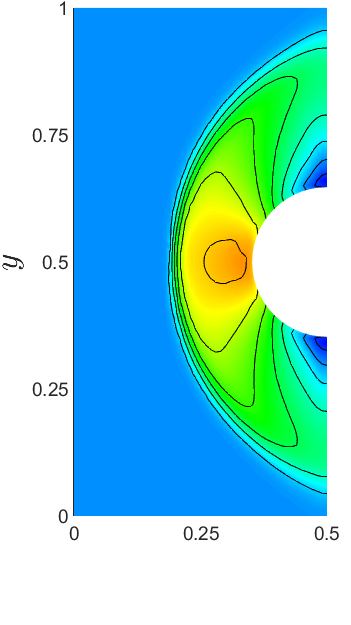} 
        & \includegraphics[width=0.15\textwidth]{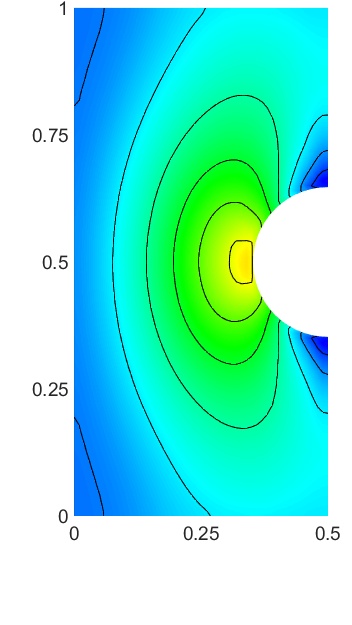}
        & \includegraphics[width=0.15\textwidth]{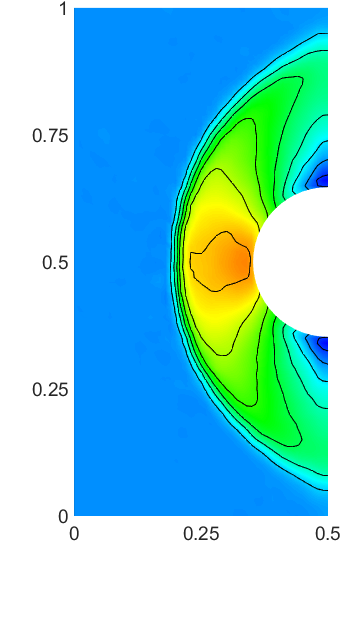} 
        & \includegraphics[width=0.15\textwidth]{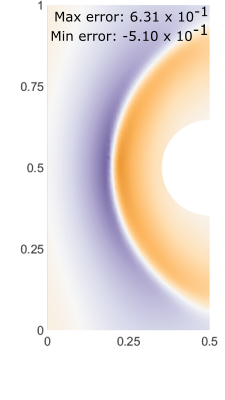} 
        & \includegraphics[width=0.15\textwidth]{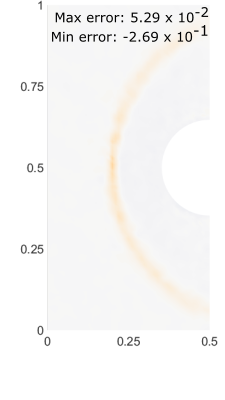}
        \\ 
        \raisebox{7.5em}{\rotatebox[origin=lb]{90}{\smash{$t=0.75$}}}
        &  \includegraphics[width=0.15\textwidth]{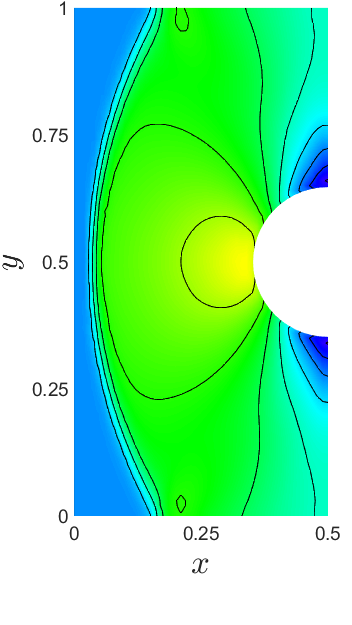} 
        &  \includegraphics[width=0.15\textwidth]{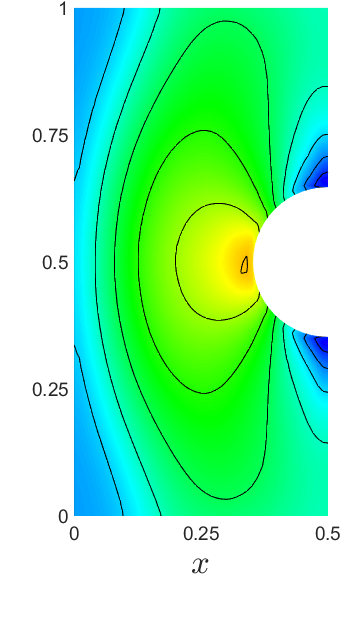} 
        & \includegraphics[width=0.15\textwidth]{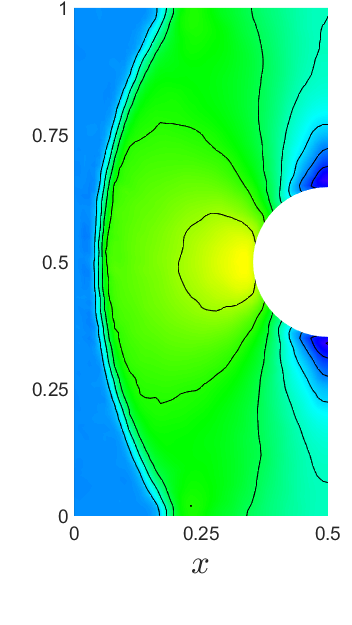} 
        & \includegraphics[width=0.15\textwidth]{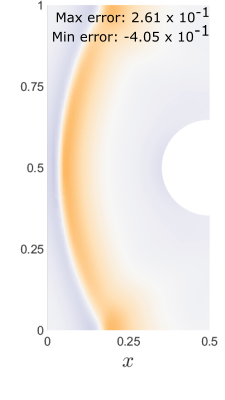}
        & \includegraphics[width=0.15\textwidth]{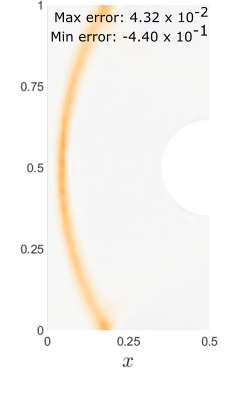}
    \end{tabular}
    \begin{tabular}{c}
        \includegraphics[scale=.35]{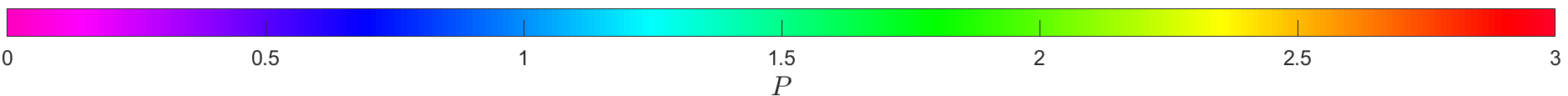}\\
       \includegraphics[scale=.35]{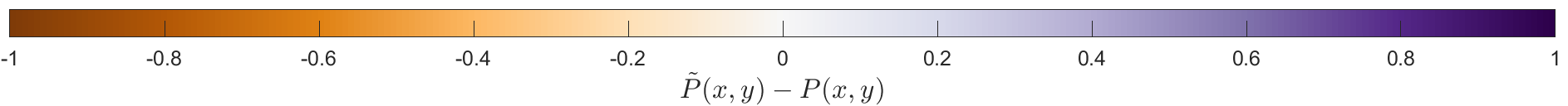}
    \end{tabular}
    \endgroup
    \captionsetup{justification=centering}
    \caption{Pressure field (denoted by color) and density field (denoted by contours) are demonstrated for the ground truth solutions (first column) as well as the POD-LSPG (second column) and GD-LSPG solutions (third column). Additionally, the local errors, $\tilde{P}(x,y) - P(x,y)$, between the corresponding ROMs and the ground truth solution for the bow shock generated by flow past a cylinder are presented in columns 4 and 5. The results are provided at two time steps, namely $t=0.25$ (top row) and $t=0.75$ (bottom row), for test parameter $\mu_{\mathrm{in}} = 1.125$. The ROMs are constructed using a latent space dimension of $M=2$. Like before, POD-LSPG struggles to accurately model the advection-driven shock wave occurring in this model. Alternatively, GD-LSPG models the shock behavior with higher accuracy where errors are mainly isolated around the shock. The top colorbar refers to the pressure field solutions while the bottom colorbar refers to the local error plots. (Online version in color.)}
    \label{fig:cyl_M2_results}
\end{figure}

In this numerical experiment, we present POD-LSPG and GD-LSPG solutions. To generate the POD-LSPG solution, we set the tolerance $\kappa$ in \eqref{eq:convergence} to be $10^{-4}$, whereas, the tolerance is set to $10^{-3}$ for GD-LSPG. The step size $\beta^{(j)}$ for both models is set to $1.0$ at all time steps. Figure \ref{fig:cyl_M2_results} demonstrates the pressure and density fields at two different time steps and at the test parameter $\mu_{\mathrm{in}}=1.125$ for the ground truth solution, POD-LSPG, and GD-LSPG, both using a latent space dimension $M=2$. Additionally, Figure \ref{fig:cyl_M2_results} provides the local errors between both ROM solutions versus the ground truth solution for the pressure field. As before, note that POD-LSPG smooths out the shock and fails to accurately model the moving discontinuity. Alternatively, GD-LSPG models the moving shock behavior much more faithfully without generating spurious high-pressure regions. It is evident that the majority of the errors for GD-LSPG are isolated around the shock. 

\begin{figure}[ht!]
    \centering
    \begin{tabular}{c|c|cc}
        & {\footnotesize POD} & \multicolumn{2}{c} {\footnotesize Graph autoencoder} \\[1pt]
        & {\footnotesize time invariant} & {\footnotesize $t=0.25$} & {\footnotesize $t=0.75$} \\ 
        \hline & & & \\[-1em]
        \raisebox{8.2em}{\multirow{2}{*}{\rotatebox[origin=lb]{90}{\footnotesize\smash{mode 1 ($i=1$)}}}}
        & \includegraphics[height=0.28\textwidth]{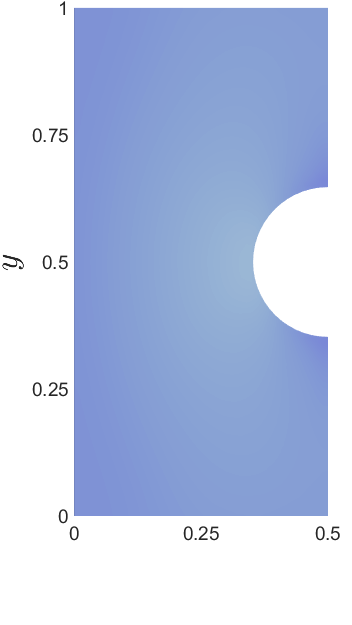}
        & \includegraphics[height=0.28\textwidth]{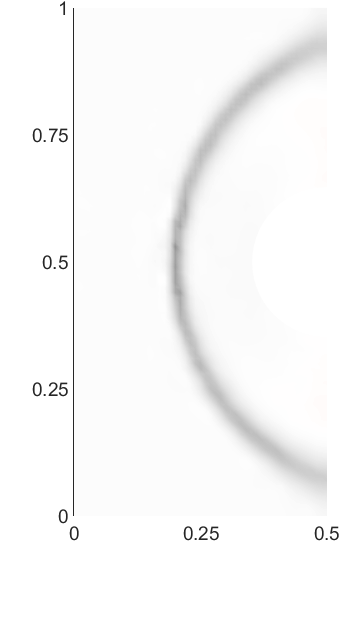}
        & \includegraphics[height=0.28\textwidth]{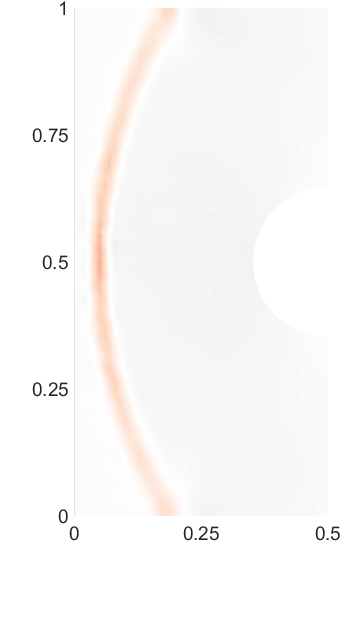} 
        \\
        \raisebox{8.2em}{\multirow{2}{*}{\rotatebox[origin=lb]{90}{\footnotesize\smash{mode 2 ($i=2$)}}}}
        & \includegraphics[height=0.28\textwidth]{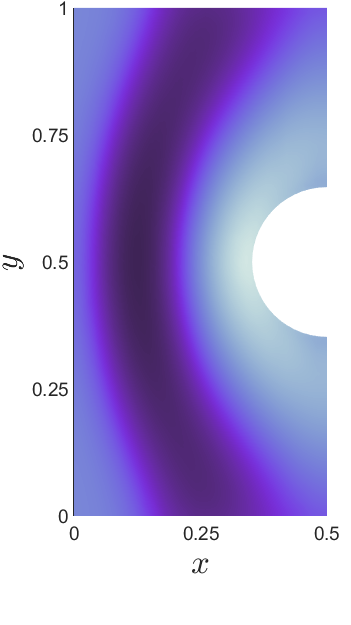}
        & \includegraphics[height=0.28\textwidth]{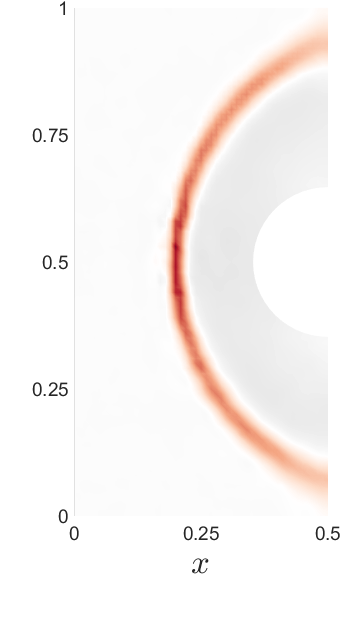}
        & \includegraphics[height=00.28\textwidth]{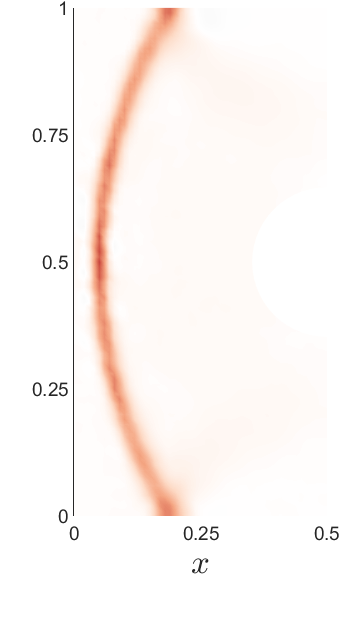}
         
    \end{tabular}
    
    \begin{tabular}{c}
        {\includegraphics[scale=.27]{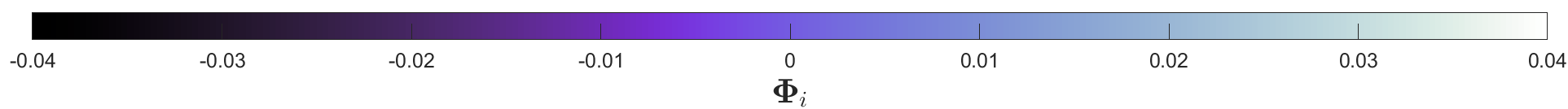}} \\
        {\includegraphics[scale=.27]{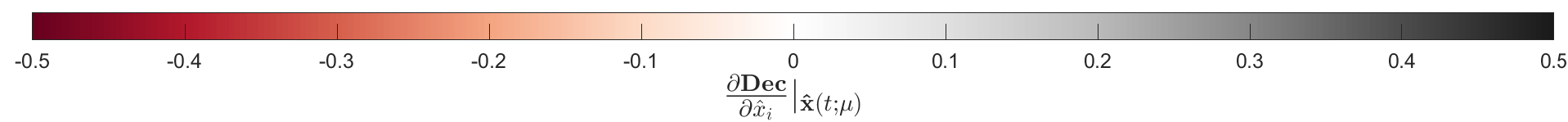}}
    \end{tabular}
    
    \captionsetup{justification=centering}
    \caption{POD modes, $\boldsymbol{\Phi}_i$, (left) and Jacobian of the decoder for the graph autoencoder of the $i^{\mathrm{th}}$ latent state variable, $\frac{\partial \mathrm{\mathbf{Dec}}}{\partial \hat{x}_i} \vert_{\mathbf{\hat{x}}(t;\boldsymbol{\mu})}$ (right) for the energy state variable ($\rho E$ in \eqref{eq:FVMeuler}) for the POD-LSPG and GD-LSPG solutions, respectively. The results are shown for the latent space dimension $M=2$ and for the test parameter $\mu_{\mathrm{in}}=1.125$. Note that the POD modes are time invariant and therefore remain fixed for all time steps, while the Jacobian of the decoder for the graph autoencoder presents time-varying mode shapes (here shown for $t=0.25$ and $t=0.75$). The Jacobian of the decoder of the graph autoencoder reveals that the graph autoencoder's latent variables primarily contain information about the bow shock that forms on the leading edge of the cylinder, offering interpretability into the GD-LSPG solution. In contrast, the POD modes used in POD-LSPG are time invariant, independent of the latent state vector, and highly diffusive. Note that the top colorbar (black/blue/white) is used to present the POD modes, while the bottom color bar (red/white/black) is used to present the Jacobian of the decoder for the graph autoencoder. (Online version in color.)}
    \label{fig:cyl_jacobian}
\end{figure}

We further assess the interpretability of the GD-LSPG solution by presenting the Jacobian of the decoder for the graph autoencoder alongside the POD modes used in the POD-LSPG solution for the latent space dimension $M=2$ and the test parameter $\mu_{\mathrm{in}} = 1.125$ in Figure \ref{fig:cyl_jacobian}. Specifically, we plot the component of the Jacobian of the decoder and the POD modes that correspond to the energy state variable (i.e., $\rho E$ in \eqref{eq:FVMeuler}) over the entire physical domain. While the POD modes are time invariant, the Jacobian of the decoder depends on the latent state vector and is shown at two distinct time instances.
At $t=0.25$, mode 1 primarily captures information about the moving shock, while mode 2 captures information about the moving shock and the high-pressure region behind it. Additionally, at $t=0.75$, mode 1 continues to represent the shock, and additionally captures part of the high-pressure region, whereas mode 2 still captures both features. On the other hand, the POD modes remain highly diffusive and time invariant, limiting their ability to represent the moving shock as well as the graph autoencoder.

\begin{figure}
    \centering
    \includegraphics[height=0.3\textwidth]{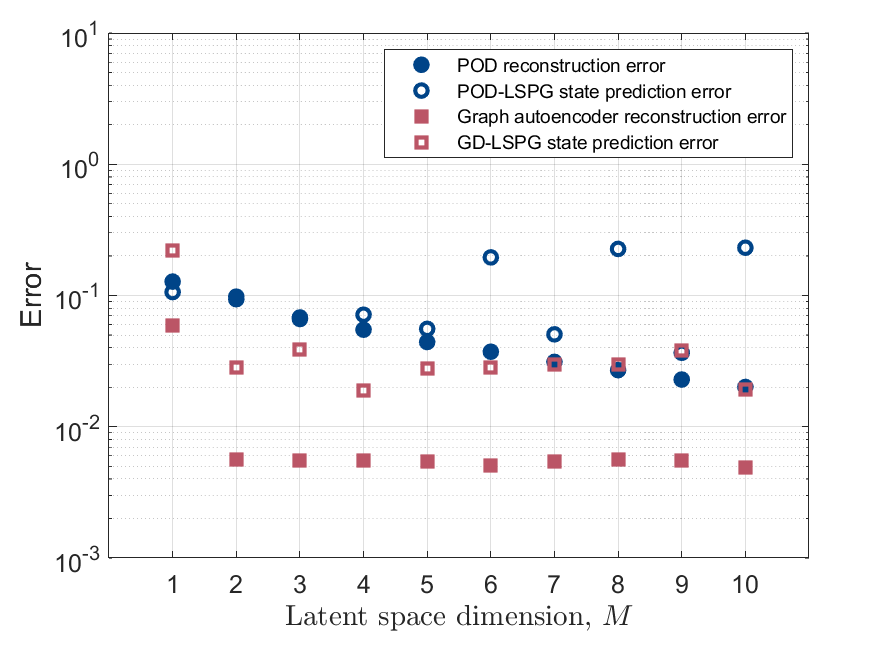}
    \captionsetup{justification=centering}
    \caption{POD reconstruction error from \eqref{eq:pod_error}, graph autoencoder reconstruction error from \eqref{eq:ae_error}, and state prediction errors from \eqref{eq:state_err} plotted with respect to the latent space dimension $M$ for 2D Euler equations for the bow shock generated by flow past a cylinder at $\mu_{\mathrm{in}}=1.125$. For latent space dimensions 2 to 10, GD-LSPG performs as well as or considerably better than POD-LSPG in terms of accuracy. (Online version in color.)}
    \label{fig:Cyl_errors}
\end{figure}

Finally, Figure \ref{fig:Cyl_errors} reports the POD and graph autoencoder reconstruction and state prediction errors. For this numerical example, we consider latent space dimensions 1 to 10, where the dimension of the FOM is $4148 \times 4 = 16592$. The graph autoencoder presents significantly lower reconstruction errors compared to POD for the latent space dimensions of choice. In addition, the GD-LSPG state prediction errors remain considerably lower than those of POD-LSPG. We note that the main reason that GD-LSPG state prediction errors for this example are not as low as those achieved in the 1D Burgers' model and 2D Riemann problem setup is the more significant phase lag error as demonstrated in Figure \ref{fig:cyl_M2_results}. Furthermore, we emphasize that identifying an appropriate error metric for advection-dominated problems is an important area of future work.

\subsection{Performance with noisy training data} \label{ssec:noisy}
\begin{figure}[t!]
    \centering
    \begingroup
    \footnotesize
    \begin{tabular}{cccc}
        $\quad$ $\sigma_{\mathrm{noise}}=0$ (no noise) & $\quad$ $\sigma_{\mathrm{noise}}=0.01$ & $\quad$ $\sigma_{\mathrm{noise}}=0.05$  & $\quad$ $\sigma_{\mathrm{noise}}=0.1$\\
        \includegraphics[width=0.17\textwidth]{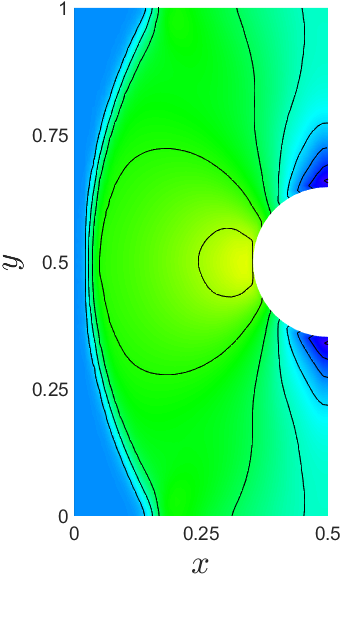} 
        & \includegraphics[width=0.17\textwidth]{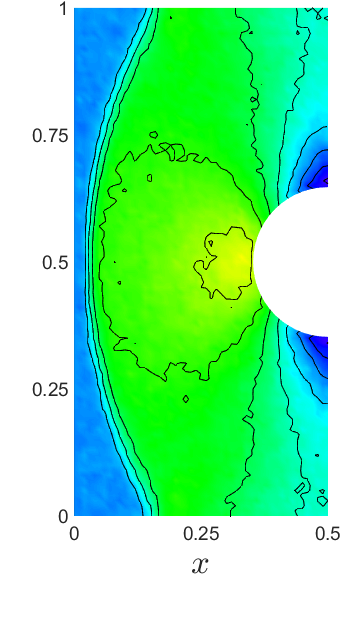}
        & \includegraphics[width=0.17\textwidth]{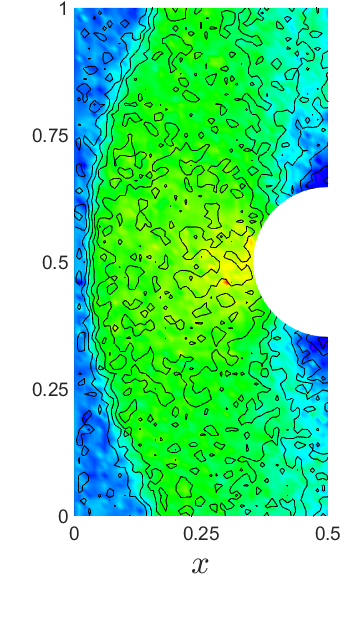} 
        & \includegraphics[width=0.17\textwidth]{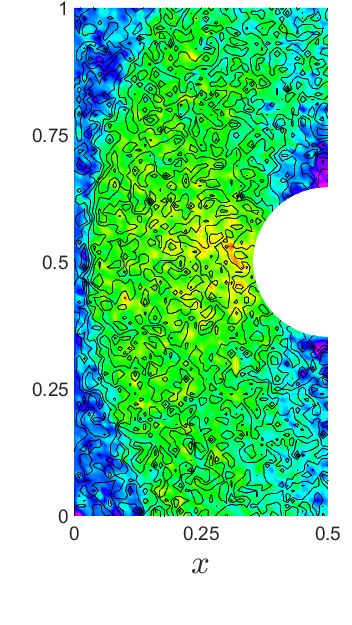}
    \end{tabular}
    \begin{tabular}{c}
        \hspace{.35cm} \includegraphics[width=0.7\textwidth]{cyl_cbar.png}
    \end{tabular}
    \endgroup
    \captionsetup{justification=centering}
    \caption{Pressure field (denoted by color) and density field (denoted by contours) are demonstrated for four different noise levels, $\sigma_{\mathrm{noise}} = 0$, $0.01$, $0.05$, and $0.1$ for the training parameter $\mu_{\mathrm{in}} = 1.1$ at $t=0.75$. Notice that for $\sigma_{\mathrm{noise}} = 0.1$, the shock features are still present, but are significantly blurred in comparison to $\sigma_{\mathrm{noise}}=0$. (Online version in color.)}
    \label{fig:cyl_noisyTrainingData}
\end{figure}

In this section, we evaluate the ability of our graph autoencoder to approximate the underlying physics when trained on noisy data. The numerical experiment is conducted on the flow past a cylinder problem, following the same training strategy outlined in \ref{section:architecture},  with the only difference being the addition of Gaussian noise to the training data. Specifically, Gaussian noise is added to each state variable ($\rho, \rho u, \rho v, \rho E$), proportional to its range within the training data,

\begin{align}
    (\,\overline{\rho}\,)_i^n &= (\rho)_i^n + \epsilon \sigma_{\mathrm{noise}}  \left((\rho)^{\mathrm{max}} - (\rho)^{\mathrm{min}}\right),\label{eq:denisty_noise} \\
    (\,\overline{\rho u}\,)_i^n &= (\rho u)_i^n + \epsilon \sigma_{\mathrm{noise}} \left((\rho u)^{\mathrm{max}} - (\rho u)^{\mathrm{min}}\right), \\
    (\,\overline{\rho v}\,)_i^n &= (\rho v)_i^n + \epsilon \sigma_{\mathrm{noise}} \left((\rho v)^{\mathrm{max}} - (\rho v)^{\mathrm{min}}\right), \\
    (\,\overline{\rho E}\,)_i^n &= (\rho E)_i^n + \epsilon \sigma_{\mathrm{noise}} \left((\rho E)^{\mathrm{max}} - (\rho E)^{\mathrm{min}}\right), \label{eq:energy_noise} 
\end{align}
where $(\,{\rho}\,)_i^n$ and 
$(\,\overline{\rho}\,)_i^n$ denote the clean and noisy density values, respectively, for the $i^{\mathrm{th}}$ cell and the $n^{\mathrm{th}}$ training snapshot, and $(\rho)^{\mathrm{max}}$ and $(\rho)^{\mathrm{min}}$ represent the maximum and minimum density values across the entire training set. The same notation applies to the momentum components, $(\,\overline{\rho u}\,)_i^n$, $(\,\overline{\rho v}\,)_i^n$, and the energy term, $(\,\overline{\rho E}\,)_i^n$). In \eqref{eq:denisty_noise}-\eqref{eq:energy_noise}, $\sigma_{\mathrm{noise}} \in \mathbb{R}_+$ denotes the noise level, and $\epsilon \sim \mathscr{N}(0,1)$ represents the added noise drawn from a standard normal distribution. In this study, we vary the noise level as $\sigma_{\mathrm{noise}}=0.0$, $0.01$, $0.05$, and $0.1$ to assess the sensitivity of the graph autoencoder to different noise levels. Figure \ref{fig:cyl_noisyTrainingData} presents examples of noisy training solutions for $\mu_{\mathrm{in}} = 1.1$ at time $t=0.75$. 

\begin{figure}[t!]
    \centering
    \begingroup
    \footnotesize
    \begin{tabular}{cccccc}
        $\quad$ Ground truth & $\quad$ $\sigma_{\mathrm{noise}}=0$ & $\quad \sigma_{\mathrm{noise}} = 0.01 $ & $\quad \sigma_{\mathrm{noise}} = 0.05$ & $\quad \sigma_{\mathrm{noise}} = 0.1$ & \\
        & $\quad$ (no noise) & & & & \\
        \raisebox{7.0em}{\multirow{2}{*}{ \includegraphics[width=0.15\textwidth]{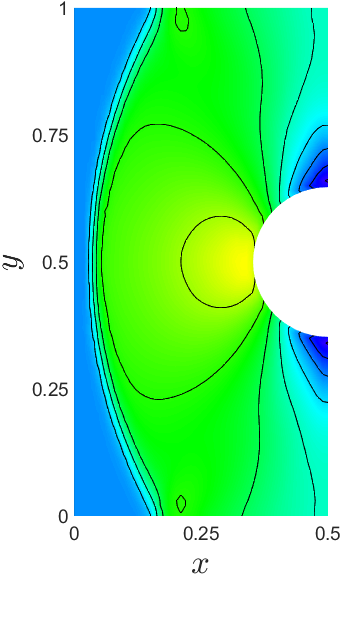}}} 
        & \includegraphics[width=0.15\textwidth]{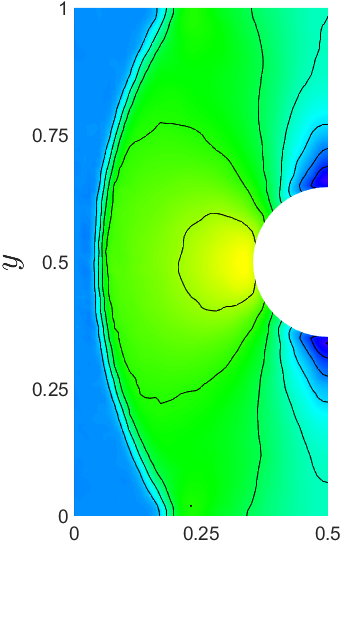}
        & \includegraphics[width=0.15\textwidth]{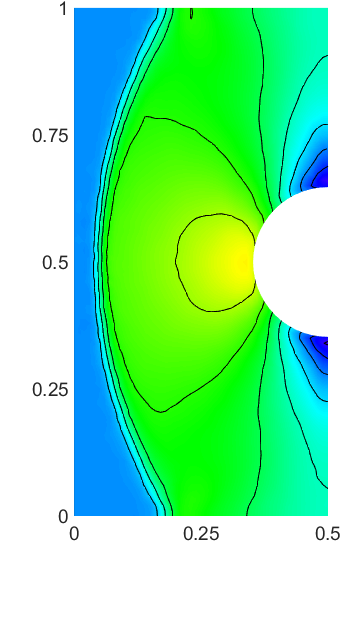} 
        & \includegraphics[width=0.15\textwidth]{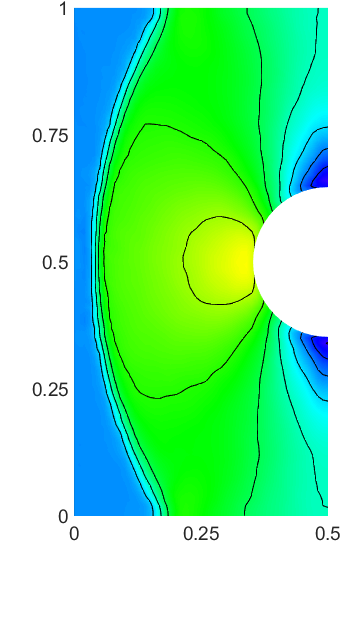} 
        & \includegraphics[width=0.15\textwidth]{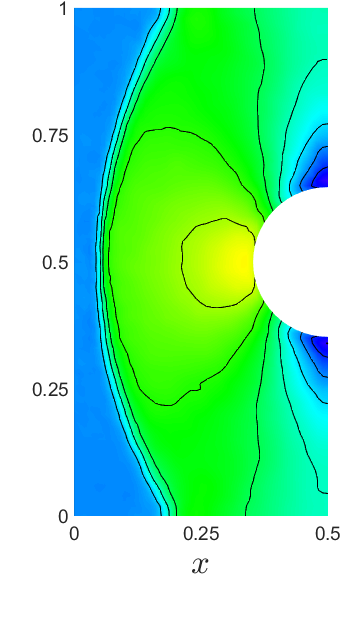}
        & \raisebox{10.5em}{\rotatebox[origin=lb]{270}{\smash{$M=2$}}}
        \\ 
        &  \includegraphics[width=0.15\textwidth]{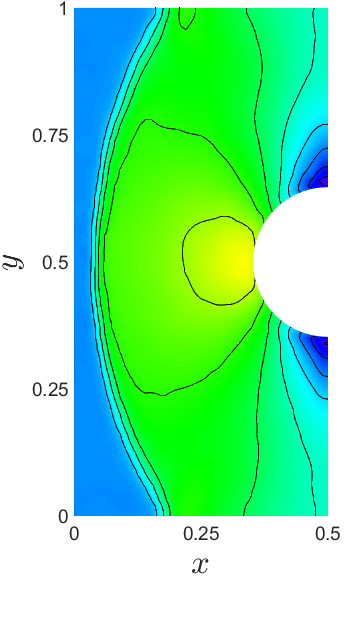} 
        & \includegraphics[width=0.15\textwidth]{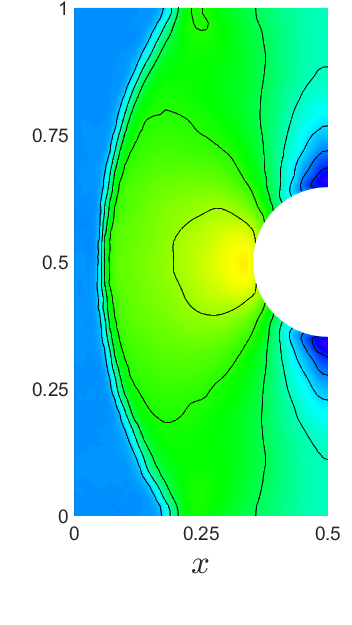} 
        & \includegraphics[width=0.15\textwidth]{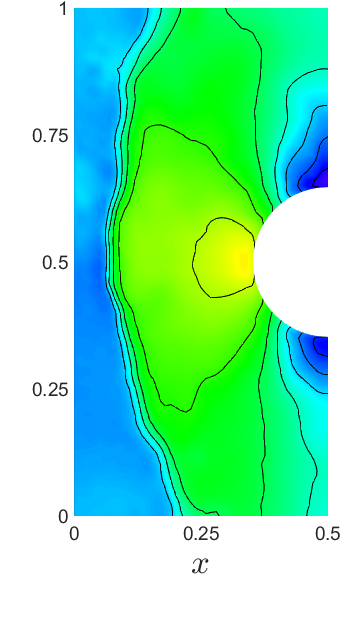}
        & \includegraphics[width=0.15\textwidth]{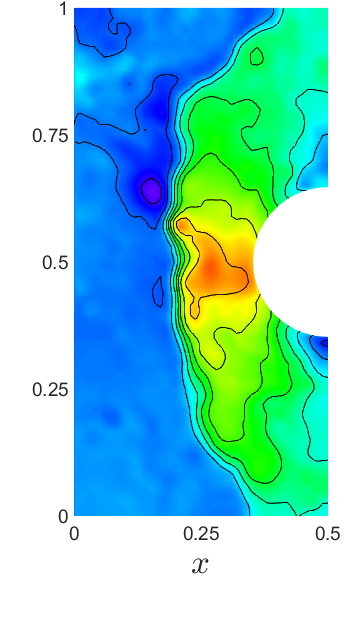}
        & \raisebox{10.5em}{\rotatebox[origin=lb]{270}{\smash{$M=10$}}}
    \end{tabular}
    \begin{tabular}{c}
        \includegraphics[scale=.35]{cyl_cbar.png}
    \end{tabular}
    \endgroup
    \captionsetup{justification=centering}
    \caption{Pressure field (denoted by color) and density field (denoted by contours) are demonstrated for the ground truth and for GD-LSPG for latent space dimensions $M=2$, and $10$ and noise levels $\sigma_{\mathrm{noise}}=0$, $0.01$, $0.05$, and $0.1$ at time $t=0.75$ for test parameter $\mu_{\mathrm{in}}=1.125$. For $M=2$, we find all solutions to be very accurate predictions of the ground truth solution. However, for $M=10$, we find that, as more noise is introduced, the GD-LSPG solution becomes more inaccurate due to noise. Still, the shock features appear to be present. (Online version in color.)}
    \label{fig:noisy_Cyl_results}
\end{figure}

Once trained over all studied noise levels, the graph autoencoder is evaluated on its ability to generate the clean (i.e., noise-free), ground truth solutions for a test parameter not seen during training. Figure \ref{fig:noisy_Cyl_results} presents the ground truth solution and GD-LSPG solutions for latent space dimensions $M=2,$ and $10$, and for test parameter $\mu_{\mathrm{in}} = 1.125$ at time $t=0.75$.
For $M=2$, the GD-LSPG solution remains consistently accurate across all studied noise levels. In contrast, for $M=10$, the GD-LSPG solution maintains accuracy for $\sigma_{\mathrm{noise}}=0$, and $0.01$, but accuracy degrades at noise levels $\sigma_{\mathrm{noise}}=0.05$, and $0.1$. While the shock features appear to still be present in the solutions, there appears to be a considerable amount of artifacts introduced into the solution, especially for noise level $\sigma_{\mathrm{noise}}=0.1$.

\begin{figure}[ht!]
    \centering
     \begin{tabular}{cccc}
        & \includegraphics[height=0.3\textwidth]{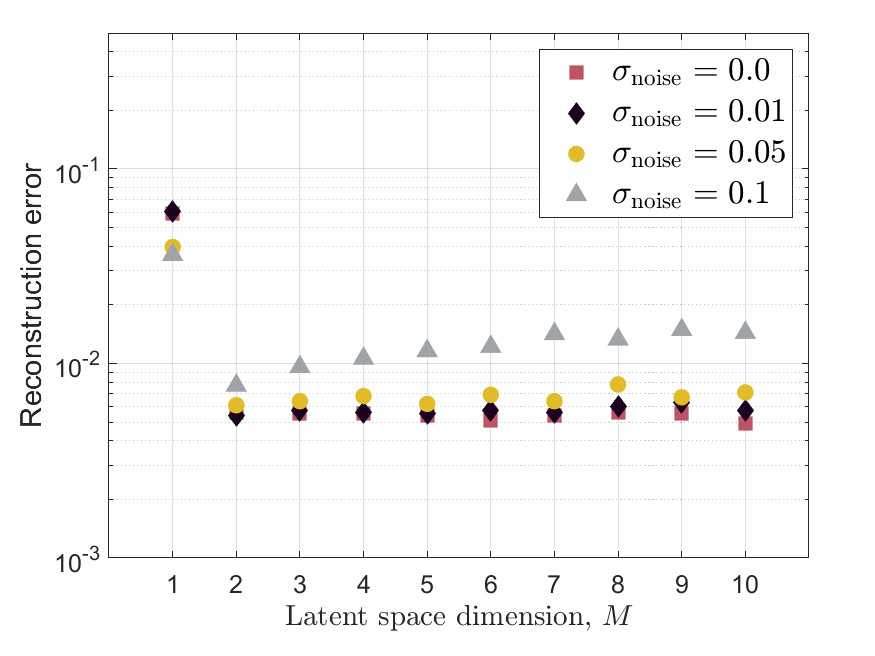}
        & \includegraphics[height=0.3\textwidth]{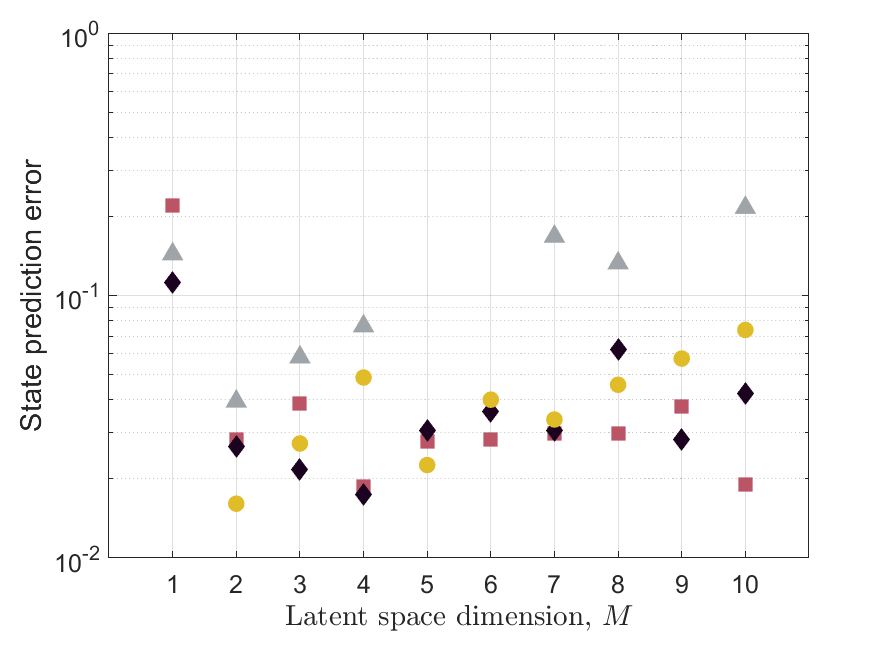}
    \end{tabular}
    \captionsetup{justification=centering}
    \caption{Graph autoencoder reconstruction errors (left) evaluated from \eqref{eq:ae_error} and GD-LSPG state prediction errors (right) evaluated from \eqref{eq:state_err} for the bow shock generated by flow past a cylinder problem when trained using noisy training data with noise levels $\sigma_{\mathrm{noise}}=0$, $0.01$, $0.05$, and $0.1$ for test parameter $\mu=1.125$. Higher noise levels in the training data lead to increased reconstructions errors, as well as higher state prediction errors. Note that GD-LSPG failed to converge due to non-physical solutions (i.e. negative pressure) for latent space dimension $M=1$ at noise level $\sigma_{\mathrm{noise}}=0.05$ and for latent space dimensions $M=5$, $6$, and $9$ at $\sigma_{\mathrm{noise}}=0.1$. (Online version in color.)}
    \label{fig:noisy_Cyl_errors}
\end{figure}

Finally, the reconstruction and state prediction errors are reported in Figure \ref{fig:noisy_Cyl_errors} for latent space dimensions $M=1$ to $10$, and noise levels $\sigma_{\mathrm{noise}}=0.0$, $0.01$, $0.05$, and $0.1$. As expected, introducing higher noise levels increased both the graph autoencoder reconstruction errors and the GD-LSPG state predictions errors. Additionally, we observe that the GD-LSPG state prediction errors for the noisy solutions increase with larger latent space dimensions. We speculate that the graph autoencoder with smaller latent space dimensions are constrained to capture only the signal itself, whereas those with larger latent space dimensions may overfit to noise. It should also be noted that several GD-LSPG solutions failed to converge due to non-physical solutions (e.g., negative pressure), primarily at higher noise levels, specifically at noise level $\sigma_{\mathrm{noise}}=0.05$ for $M=1$, and at noise level $\sigma_{\mathrm{noise}}=0.1$ for $M=5,6,9$).

\subsection{Wall clock training times}

Lastly, in addition to online computational costs, we report the times to generate the hierarchies of graphs as well as the times to train each model used in this study, which were both carried out using an NVIDIA L40S GPU. Wall-clock times to generate the hierarchy of graphs for each example are reported in Table \ref{table:hierarchy_costs} and wall-clock times to train each model are reported in Table \ref{table:training_costs}. For the second numerical experiment (Riemann problem), where we repeat the training five times, we simply report the average time for training the autoencoders for each latent space dimension. Immediately, it is worth noting that the graph autoencoder in its current form is more expensive to train than the traditional CNN-based autoencoders and considerably more expensive to train than the POD. 

The wall-clocks times reported in Table \ref{table:hierarchy_costs} are computed by repeating the hierarchical spectral clustering algorithm ten times for each model and averaging the results for each level. For the Burgers' model, we employ the $K$-means++ initialization scheme \cite{arthur2007kmeans}, as it resulted in more intuitive cluster assignments. Generating the graph hierarchy is not yet computationally efficient enough to be used on-the-fly. More efficient clustering schemes will be necessary before GD-LSPG can be effectively applied to models without a fixed mesh, such as those using adaptive meshes, Lagrangian frameworks, or arbitrary Lagrangian-Eulerian methods. 

Another noteworthy observation is that, although the solution states in the Riemann problem setup and the bow shock generated by flow past a cylinder problem have considerably larger dimensions than the Burgers' model, the cost to train the autoencoders does not appear considerably larger. There are likely several factors contributing to this. First, the larger models are likely easier to parallelize on the NVIDIA L40S GPU. Furthermore, the number of batches provided at each epoch during training is a major driving component of the training time. As seen in \ref{section:architecture}, for all autoencoders, a batch size of 20 was chosen due to random access memory limitations. However, due to the varying number of training solutions, the number of batches per epoch varies considerably depending on the problem. To elaborate, the autoencoders for the 1D Burgers' model have $36080$ solution states used for training, meaning each epoch will have 1804 batches. On the other hand, the autoencoders for the 2D Riemann problem setup have $7000$ solution states used for training, meaning each epoch will have 350 batches. Finally, the autoencoders for the 2D bow shock generated by flow past a cylinder problem have $5500$ solution states used for training, meaning each epoch will have 275 batches. 

\begin{table} [h!]
    \centering
    \caption{Average wall-clock times (in seconds) for generating the hierarchy of graphs for each example, where $\vert \mathcal{V}^i \vert$ and $\vert \mathcal{V}^{i+1} \vert$ denote the number of nodes in the $i^{\mathrm{th}}$ and $(i+1)^{\mathrm{th}}$ layers in the hierarchy of graphs, respectively. The time reported for each clustering is the average of ten repeated runs of the clustering algorithm.} 
    \begingroup
    \footnotesize
    \begin{NiceTabular}{c *{10}{c}}[colortbl-like]
    \CodeBefore
        \columncolor{gray!15}{2-4,8-10}
    \Body
        & \multicolumn{3}{c}{Burgers' model} & \multicolumn{3}{c}{Riemann problem} & \multicolumn{3}{c}{bow shock}  \\[3mm]
        \rotatebox{90}{\shortstack[l]{hierarchy \\level, $i$}} & \rotatebox{90}{$\vert \mathcal{V}^i \vert$} & \rotatebox{90}{$\vert \mathcal{V}^{i+1} \vert$} & \rotatebox{90}{\shortstack[l]{wall-clock \\time (s)}} & \rotatebox{90}{$\vert \mathcal{V}^i \vert$} & \rotatebox{90}{$\vert \mathcal{V}^{i+1} \vert$} & \rotatebox{90}{\shortstack[l]{wall-clock \\time (s)}} &  \rotatebox{90}{$\vert \mathcal{V}^i \vert$} & \rotatebox{90}{$\vert \mathcal{V}^{i+1} \vert$} & \rotatebox{90}{\shortstack[l]{wall-clock \\time (s)}} & \\
        \midrule
        0 & 316 & 64 & 15.391 & 4328 & 512 & 676.098 & 4148 & 512 & 702.034 \\
        1 & 64 & 16 & 0.029 & 512 & 64 & 1.116 & 512 & 64 & 1.095 \\
        2 & 16 & 4 & 0.037 & 64 & 8 & 0.008 & 64 & 8 & 0.008 \\
        3 & 4 & 2 & 0.002 & 8 & 2 & 0.003 & 8 & 2 & 0.002 \\
        \midrule
        Total & & & 15.425 & & & 677.224 & & & 703.140 \\
        \bottomrule
    \end{NiceTabular}
    \endgroup
    \label{table:hierarchy_costs}
\end{table}

\begin{table} [h!]
    \centering
    \caption{Wall-clock training times (in hours) for POD, CNN-based autoencoder, and graph autoencoder for all numerical examples. Note that the POD is only obtained once for all latent space dimensions where the POD basis is determined by truncating the left singular basis to the user-specified latent space dimension, $M$. Alternatively, each latent space dimension has a unique CNN-based autoencoder and graph autoencoder that must be trained individually. For the repeated training of the autoencoders for the Riemann problem, we report the average time to train each model.}
    \begingroup
    \footnotesize
    \begin{NiceTabular}{c *{8}{c}}[colortbl-like]
    \CodeBefore
        \columncolor{gray!15}{2-4,8-9}
    \Body
        \toprule
        & \multicolumn{3}{c}{Burgers' model} & \multicolumn{3}{c}{Riemann problem} & \multicolumn{2}{c}{bow shock}  \\[3mm]
        \rotatebox{90}{\shortstack[l]{latent space\\dimension, $M$}} & \rotatebox{90}{POD} & \rotatebox{90}{\shortstack[l]{CNN-based\\autoencoder}} & \rotatebox{90}{\shortstack[l]{graph\\autoencoder}} & \rotatebox{90}{POD} & \rotatebox{90}{\shortstack[l]{interpolated\\CNN-based\\autoencoder}} & \rotatebox{90}{\shortstack[l]{graph\\autoencoder}} & \rotatebox{90}{POD} & \rotatebox{90}{\shortstack[l]{graph\\autoencoder}}\\
        \midrule
        1 & \multirow{10}{*}{1.39$\times10^{-3}$} & 1.73 & 4.70 & \multirow{10}{*}{0.03} & 2.01 & 5.08 & \multirow{10}{*}{0.02} & 4.48\\
        2 & & 1.30 & 5.48 & & 1.98 & 5.01 & & 4.52\\
        3 & & 1.16 & 5.40 & & 1.90 & 5.00 & & 4.38 \\
        4 & & 1.30 & 4.64 & & 1.91 & 5.06 & & 4.41 \\
        5 & & 1.30 & 4.68 & & 1.94 & 5.00 & & 4.40\\
        6 & & 1.13 & 4.77 & & 2.23 & 4.98 & & 3.83\\
        7 & & 1.12 & 4.77 & & 2.21 & 5.03 & & 3.80\\
        8 & & 1.17 & 4.71 & & 2.20 & 5.88 & & 4.67\\
        9 & & 1.30 & 4.65 & & 1.86 & 4.95 & & 3.91 \\
        10 & & 1.28 & 4.70 & & 1.86 & 5.02 & & 4.64\\
        \bottomrule
    \end{NiceTabular}
    \endgroup
    \label{table:training_costs}
\end{table}

\section{Conclusions and future work}

In this paper, we present GD-LSPG, a PMOR method that leverages a graph autoencoder architecture to perform model reduction directly on unstructured meshes. The graph autoencoder is constructed by first generating a hierarchy of reduced graphs to emulate the compressive capabilities of CNNs. Next, message passing operations are trained for each layer in the hierarchy of reduced graphs to emulate the filtering capabilities of CNNs. In an online stage, the graph autoencoder is leveraged to perform a nonlinear manifold LSPG projection to obtain solutions to parameter sets not seen during training. We investigate the interpretability of such solutions by analyzing the Jacobian of the decoder.

In this work, we assess the performance of GD-LSPG with three different numerical experiments. First, to benchmark GD-LSPG, we apply it to a 1D Burgers' model commonly used in the literature. Specifically, this benchmarking case deploys a structured mesh and we can therefore compare the results directly to both POD-LSPG \cite{carlberg2011lspg,carlberg2017galerkinvslspg} and dLSPG \cite{lee2020deeplspg,lee2021deepconservation}. We find that GD-LSPG has accuracy that is comparable to that of dLSPG and that both methods provide considerable improvement in accuracy over POD-LSPG as they capture the moving shock behavior found in this solution more accurately in comparison to POD-LSPG. By plotting the Jacobian of the decoder, we find that the latent space of the graph autoencoder primarily captures information about the moving shock front. In the second and third numerical experiments, we apply GD-LSPG to a 2D Euler equations model that leverages an unstructured mesh in two different settings. The first setting involves a square domain that solves a Riemann problem setup. Because the domain is regular, we can interpolate the unstructured mesh onto a structured grid that can be provided to a CNN-based autoencoder and applied in dLSPG. While we find that this method can produce results that outperform GD-LSPG, it is also prone to failing to generalize well for parameter sets not seen during training. Alternatively, GD-LSPG consistently generalizes well for parameter sets not seen during training while also considerably outperforming POD-LSPG in terms of accuracy due to its ability to model the moving shock and rarefaction waves present in this problem. The second setting models a bow shock generated by flow past a cylinder with an unstructured mesh. In this setting, the modeled geometry is more naturally represented by an unstructured mesh, and therefore demonstrates the flexibility of GD-LSPG. We compare GD-LSPG to POD-LSPG and, once again, find that GD-LSPG often outperforms POD-LSPG in terms of accuracy as GD-LSPG better preserves the sharp features of the bow shock that appear on the cylinder. Alternatively, POD-LSPG smooths out the shock and introduces spurious high-pressure regions in the solution. As with the 1D Burgers' model, the Jacobian of the decoder for the graph autoencoder indicates that the latent state variables contain information about the moving bow shock. 
Finally, we also investigate the ability of the graph autoencoder to filter noise by training it on data generated by perturbing the ground truth solutions with Gaussian noise at various noise levels for the bow shock generated by flow past a cylinder problem. This setting naturally aligns with our graph autoencoder framework, as the noisy model is constrained to the same underlying governing equations as the clean data for which the residual can be computed. At smaller latent space dimensions, the graph autoencoder accurately captures the shock behavior. However, accuracy decreases as noise level and latent space dimension increase. 

As is present across an extensive amount of literature employing end-to-end training of autoencoders \cite{fries2022lasdi,lee2020deeplspg,lee2021deepconservation,barwey2023multiscale}, our graph autoencoder is limited to models with a dimension of only a few thousand. Additionally, we emphasize that the hierarchical spectral clustering scheme is limited by the scalability of the eigendecomposition of the graph Laplacian and $K-$means clustering operation which we leave as areas of future work. 
One potential future direction to achieve cost savings is to introduce a latent space dynamics model similar to operator inference \cite{peherstorfer2016data}, latent space dynamics identification \cite{fries2022lasdi}, or sparse identification of nonlinear dynamics \cite{brunton2016discovering}. An added benefit of these methods is that, unlike the LSPG schemes presented in Section \ref{sec:TimeIntegration}, they do not often require explicit knowledge of the residual. This feature makes them potentially suitable for use with experimental data or computational models where the residual is unavailable or difficult to compute.
Another possible future direction is to develop a hyper-reduction scheme to achieve cost savings. In GD-LSPG's current formulation, the high-dimensional residual must be computed and projected onto the low-dimensional latent space, forcing the operation count complexity to scale on the dimension of the FOM. A hyper-reduction scheme would alleviate this limitation by sampling and generating a sparse representation of the high-dimensional residual, thereby eliminating an operation count complexity that scales on the dimension of the FOM. To date, hyper-reduction has been achieved for dLSPG using shallow decoders \cite{kim2022masked}. However, GD-LSPG uses a deep decoder. As a result, we leave this as an open area of investigation.

\paragraph{Acknowledgements}
L. K. Magargal is supported by the Department of Defense (DoD) through the National Defense Science \& Engineering Graduate (NDSEG) Fellowship Program. P. Khodabakhshi acknowledges the support of the National Science Foundation under award 2450804.
L. K. Magargal and J. W. Jaworski acknowledge the financial support of the Department of Energy under grant DE-EE0008964. S. N. Rodriguez and J. G. Michopoulos acknowledge the support of the Office of Naval Research through U.S. Naval Research Laboratory core funding. J. W. Jaworski acknowledges the partial support of the National Science Foundation under CAREER award 1846852/2402397 and the Office of Naval Research under award N00014-24-2111. Portions of this research were conducted on Lehigh University's Research Computing infrastructure partially supported by NSF Award 2019035.

\paragraph{Data Availability Statement} The data that support the findings of this study are openly available at \url{https://doi.org/10.5281/zenodo.15269964}.

%\renewcommand\bibpreamble{By default, this template uses \texttt{bibtex} and adopts the AMS referencing style. However, the journal you’re submitting to may require a different reference style; specify the journal you're using with the class' \texttt{journal} option --- see lines 1--19 of \emph{sample.tex} for a list of options and instructions for selecting the journal.}

% If using any of the following journal options:
%   wet, dap, dce, eds, prm, flw, jdm, psy, rsm
% then use the \printbibliography line instead of:
%\bibliography{bibliography}
%\newrefcontext[sorting=nty]
%\printbibliography

\bibliographystyle{elsarticle-num} % use this one for the final paper
\biboptions{sort&compress}
\bibliography{main}

\appendix
\setcounter{table}{0}

\section{Architecture details and training}\label{section:architecture}
\setcounter{table}{0}
\setcounter{figure}{0}

All models are trained on an NVIDIA L40S GPU. We initialize all weights and biases using Xavier initialization \cite{glorot2010initialization}. No attempt was made to mitigate oversmoothing (or overdiffusion) for any of the models, which can be a concern for extremely deep GNNs. The reader is directed to \cite{rusch2023survey} for an in-depth study on quantifying and mitigating oversmoothing for GNNs. The details of the graph and CNN-based autoencoders for the test problems in Section \ref{sec:experiments} are provided in Tables \ref{table:GNN} and \ref{table:CNN}, respectively. In all test problems, the loss is evaluated using \eqref{eq:loss}. Stochastic gradient descent is performed using the Adam optimizer \cite{kingma2014adam} to update the weights and biases at each epoch. The learning rate for all problems is chosen to be $10^{-4}$, and the activation functions are taken to be ELU \cite{clevert2016elu}.
 
To train the autoencoders in the \underline{1D Burgers' model}, we first perform a training/validation split, where $4000$ solution states are stored for validation, and the remaining $36080$ solution states are used for training. We train the models for $1000$ epochs where, at each epoch, the training set is passed through the autoencoder in batches of $20$. In training the {\it graph autoencoder}, we noticed that the architecture struggles to model the solution state near the boundaries, especially when the shock approaches the boundary. We believe that this issue is related to the nonlocality of our graph autoencoder, as the boundary nodes do not receive adequate information from the space outside of the domain. This behavior is often found across the field of nonlocal modeling, including peridynamics \cite{madenci2013peridynamics} and smoothed-particle hydrodynamics \cite{monaghan1992sph}. We leave this as an open area for future investigation, but for now, we present a simple procedure for including padding nodes in a graph autoencoder. This procedure appends 30 nodes to the left side of the domain and sets their features to be the value of the left boundary condition, i.e., $\mu_1$. Along the right boundary, we find that solving Burgers' model with the finite volume solver for 30 finite volume cells to the right of the right boundary and prescribing the computed velocity values to the features of the padding nodes is appropriate. Our decoder reconstructs the solution for the nodes in the physical domain as well as those in the padding zones but only computes the loss with respect to the nodes in the physical domain of the problem. During the hierarchical spectral clustering algorithm, radius $r^i$ for layer $i$ is chosen such that \eqref{eq:radius_graph} gives roughly 7 edges for each node, i.e., $r^i = (x_{\mathrm{right}} - x_{\mathrm{left}}) \left(\frac{7}{2\vert \mathcal{V}^i \vert}\right)$, where $x_{\mathrm{right}} \in \mathbb{R}$ and $x_{\mathrm{left}} \in \mathbb{R}$ are the positions of the rightmost and leftmost padding nodes in $\mathcal{V}^0$, respectively. To train the {\it CNN-based autoencoder}, a kernel size of 25 is chosen at each layer, where half-padding is used. In the decoder, the transposed convolution layers are given an output padding of 1.

In the \underline{2D Euler equations for the Riemann problem setup}, we first perform a training/validation split, where $525$ solution states are stored for validation, and the remaining $7000$ solution states are used for training. We train the model for $5000$ epochs where, at each epoch, the training set is passed through the autoencoder in batches of $20$. To compute the hierarchy of reduced graphs in the {\it graph autoencoder}, at each layer, \eqref{eq:radius_graph} uses a radius that aims for 9 edges for each node, i.e., $r^{i} = \sqrt{\frac{9}{\pi \vert \mathcal{V}^{i} \vert}}$. We found that padding along the boundaries was unnecessary for this problem, and therefore did not include it. In the {\it CNN-based autoencoder}, a kernel size of $5\times5$ is chosen at each layer, where half-padding is used. Stride is taken as 2 for all layers. In the decoder, the transposed convolution layers are given an output padding of 1.

To train the {\it graph autoencoders} used in the \uline{2D Euler equations for the bow shock generated by flow past a cylinder problem}, we first perform a training/validation split, where $506$ solution states are stored for validation, and the remaining $5500$ solution states are used for training. We train the model for $5000$ epochs where, at each epoch, the training set is passed through the autoencoder in batches of $20$. To compute the hierarchy of reduced graphs, at each layer, \eqref{eq:radius_graph} uses a radius that aims for 9 edges for each node, i.e., $r^{i} = \sqrt{\frac{9}{\pi \vert \mathcal{V}^{i} \vert}}$. We found that padding along the boundaries was unnecessary for these models, and therefore did not include it. 

We found stacking message passing operations to be beneficial to the accuracy of the graph autoencoders. In the MPP layers, we perform message passing operations multiple times before pooling. In the UMP layers, multiple message passing operations are performed after unpooling. The number of message passing operations is represented in Table \ref{table:GNN} under the ``\# of MP operations'' column. Lastly, the final message passing operation of the decoder does not have an activation function associated with it, as the output of the last UMP layer should fall in the range $[0,1]$ because of the rescaling operations.

We additionally report the total number of tunable parameters for each layer of the autoencoders. A single SageCONV message passing operation consists of $2N_F^{i-1} N_F^i$ tunable parameters. When we stack multiple message passing operations in the same MPP layer, we ultimately have $2N_F^{i-1} N_F^i + 2 N^{\mathrm{MPP}} N_F^{i} N_F^i$ tunable parameters in the $i^{\mathrm{th}}$ layer, where $N^{\mathrm{MPP}} \in \mathbb{N}$ denotes the number of additional message passing operations in an MPP layer. Likewise, when we stack multiple message passing operations in the same UMP layer, we will have $2N_F^{i-1} N_F^i + 2 N^{\mathrm{UMP}} N_F^{i-1} N_F^{i-1}$ tunable parameters in the $i^{\mathrm{th}}$ layer, where $N^{\mathrm{UMP}} \in \mathbb{N}$ denotes the number of additional message passing operations in an UMP layer. For both CNN layers and transposed CNN layers, the total number of parameters in the $i^{\mathrm{th}}$ layer is $N_F^{i-1} N_F^i \iota + N_F^i$, where $\iota \in \mathbb{N}$ denotes the kernel size. Note that the CNN-based autoencoders use biases in the MLP/fully-connected layers, whereas the graph autoencoders do not.

\clearpage
\subsection{Details of the graph autoencoders}
\vspace{-9pt}
\begin{table}[h]
\begin{center}
\caption{Outputs of each layer of the encoder and decoder of the graph autoencoder for all test problems, where $i$ represents the layer number, $\vert \mathcal{V}^i \vert$ denotes the number of nodes in the output graph, $N_F^i$ denotes the number of features for each node in the output graph, and $M$ denotes the dimension of the latent space. The number of nodes in the output graph of the preprocessing and postprocessing layers in the 1D Burgers' model includes the 30 padding nodes on both sides of the domain. Parentheses in `\# of MP operations' column denote the number of features in the output of intermediate message passing operations.}
\begingroup
\footnotesize
\begin{NiceTabular}{*{9}c}[colortbl-like]
    \CodeBefore
        \rowcolor{gray!15}{14-25}
    \Body
    \toprule
    & & $i$ & layer type & $\vert \mathcal{V}^i \vert$ & $N_F^i$ & \# of MP operations & vector length & \# of tunable parameters\\
    \midrule
    {\Block{12-1}{\rotatebox[origin=lb]{90}{1D Burgers' model}}} & {\Block{6-1}{\rotatebox[origin=lb]{90}{encoder}}} & 0 & preprocessing & 316 & 1 & N/A & N/A & N/A\\
    & & 1 & MPP & 64 & 8 & 2 (8) & N/A & 144\\
    & & 2 & MPP & 16 & 16 & 2 (16) & N/A & 768\\
    & & 3 & MPP & 4 & 32 & 2 (32) & N/A & 3072\\
    & & 4 & MPP & 2 & 64 & 2 (64) & N/A & 12288 \\
    & & 5 & $\mathrm{MLP}_{\mathrm{enc}}$ & N/A & N/A & N/A & $M$ & 128$M$\\
    \cmidrule(l){2-9}
    & {\Block{6-1}{\rotatebox[origin=lb]{90}{decoder}}} & 0 & $\mathrm{MLP}_{\mathrm{dec}}$ & 2 & 64 & N/A & N/A & 128$M$\\
    & & 1 & UMP & 4 & 32 & 2 (64) & N/A & 12288 \\
    & & 2 & UMP & 16 & 16 & 2 (32) & N/A & 3072 \\
    & & 3 & UMP & 64 & 8 & 2 (16) & N/A & 768\\ 
    & & 4 & UMP & 316 & 1 & 2 (8) & N/A & 144 \\
    & & 5 & postprocessing & N/A & N/A & N/A & 316 & N/A \\    
    \bottomrule \vspace{1pt}
    {\Block{12-1}{\rotatebox[origin=lb]{90}{2D Riemann problem}}} & {\Block{6-1}{\rotatebox[origin=lb]{90}{encoder}}} & 0 & preprocessing & 4328 & 4 & N/A & N/A & N/A\\
    & & 1 & MPP & 512 & 16 & 2 (16) & N/A & 640 \\
    & & 2 & MPP & 64 & 64 & 2 (64) & N/A & 10240 \\
    & & 3 & MPP & 8 & 128 & 2 (128) & N/A & 49152 \\
    & & 4 & MPP & 2 & 256 & 2 (256) & N/A & 196608 \\
    & & 5 & $\mathrm{MLP}_{\mathrm{enc}}$ & N/A & N/A & N/A &  $M$ & 512$M$ \\
    \cmidrule(l){2-9}
    & {\Block{6-1}{\rotatebox[origin=lb]{90}{decoder}}} &  0 & $\mathrm{MLP}_{\mathrm{dec}}$ & 2 & 256 & N/A & N/A & 512$M$\\
    & & 1 & UMP & 8 & 128 & 2 (256) & N/A & 196608 \\
    & & 2 & UMP & 64 & 64 & 2 (128) & N/A & 49152 \\
    & & 3 & UMP & 512 & 16 & 2 (64) & N/A & 10240 \\
    & & 4 & UMP & 4328 & 4 & 2 (16) & N/A & 640 \\
    & & 5 & postprocessing & N/A & N/A & N/A & 4328 $\times$ 4 & N/A \\  
    \bottomrule \vspace{1pt}
    {\Block{12-1}{\rotatebox[origin=lb]{90}{bow shock problem}}} & {\Block{6-1}{\rotatebox[origin=lb]{90}{encoder}}} & 0 & preprocessing & 4148 & 4 & N/A & N/A & N/A\\
    & & 1 & MPP & 512 & 16 & 2 (16) & N/A & 640 \\
    & & 2 & MPP & 64 & 64 & 2 (64) & N/A & 10240 \\
    & & 3 & MPP & 8 & 128 & 2 (128) & N/A & 49152 \\
    & & 4 & MPP & 2 & 256 & 2 (256) & N/A & 196608 \\
    & & 5 & $\mathrm{MLP}_{\mathrm{enc}}$ & N/A & N/A & N/A &  $M$ & 512$M$ \\
    \cmidrule(l){2-9}
    & {\Block{6-1}{\rotatebox[origin=lb]{90}{decoder}}} & 0 & $\mathrm{MLP}_{\mathrm{dec}}$ & 2 & 256 & N/A & N/A & 512$M$\\
    & & 1 & UMP & 8 & 128 & 2 (256) & N/A & 196608 \\
    & & 2 & UMP & 64 & 64 & 2 (128) & N/A & 49152 \\
    & & 3 & UMP & 512 & 16 & 2 (64) & N/A & 10240 \\
    & & 4 & UMP & 4148 & 4 & 2 (16) & N/A & 640 \\
    & & 5 & postprocessing & N/A & N/A & N/A & 4148 $\times$ 4 & N/A \\
    \bottomrule
\end{NiceTabular}
\endgroup
\label{table:GNN}
\end{center}
\end{table}

\clearpage
\subsection{Details of the CNN-based autoencoders}
\vspace{-9pt}
\begin{table}[h]
\begin{center}
\caption{Outputs of each layer of the encoder and decoder of the CNN-based autoencoder for test problems 1 and 2 where $i$ denotes the layer number, and $M$ denotes the dimension of the latent space. Note: the $\mathrm{MLP}_{\mathrm{dec}}$ in the CNN-based autoencoder has a bias term and is followed by an ELU activation function (unlike the graph autoencoder). Empirically, we found that the inclusion of the bias term and activation function slightly improves accuracy for the CNN-based autoencoder, but not the graph autoencoder. Note that the CNN-based autoencoder for the 2D Riemann problem has $4096$ grid points in the interpolated mesh versus the $4328$ cells in the original mesh.}

\begingroup
\footnotesize
\begin{NiceTabular}{*{9}c}[colortbl-like]
    \CodeBefore
        \rowcolor{gray!15}{14-25}
    \Body
    \toprule
    & & $i$ & layer type & grid points & channels & stride & vector length & \# of tunable parameters\\
    \midrule
    {\Block{12-1}{\rotatebox[origin=lb]{90}{1D Burgers' model}}} & {\Block{6-1}{\rotatebox[origin=lb]{90}{encoder}}} & 0 & preprocessing & 256 & 1 & N/A & N/A & N/A \\
    & & 1 & 1D convolution & 128 & 8 &  2 & N/A & 208 \\
    & & 2 & 1D convolution & 32 & 16 &  4 & N/A & 3216 \\
    & & 3 & 1D convolution & 8 & 32 &  4 & N/A & 12832 \\
    & & 4 & 1D convolution & 2 & 64 &  4 & N/A & 51264 \\
    & & 5 & $\mathrm{MLP}_{\mathrm{enc}}$ & N/A & N/A & N/A & $M$ & 129$M$ \\
    \cmidrule(l){2-9}
    & {\Block{6-1}{\rotatebox[origin=lb]{90}{decoder}}} & 0 & $\mathrm{MLP}_{\mathrm{dec}}$ & 2 & 64 & N/A & N/A & 128$\left(M+1\right)$\\
    & & 1 & 1D transposed convolution & 8 & 32 &  4 & N/A  & 51232 \\
    & & 2 & 1D transposed convolution & 32 & 16 &  4 & N/A & 12816 \\
    & & 3 & 1D transposed convolution & 128 & 8 &  4 & N/A & 3208 \\
    & & 4 & 1D transposed convolution & 256 & 1 &  2 & N/A & 201 \\
    & & 5 & postprocessing & N/A & N/A & N/A & 256 & N/A \\   
    \bottomrule \vspace{1pt}
    {\Block{12-1}{\rotatebox[origin=lb]{90}{2D Riemann problem}}} & {\Block{6-1}{\rotatebox[origin=lb]{90}{encoder}}} & 0 & preprocessing & 4096 & 4 & N/A & N/A & N/A \\
    & & 1 & 2D convolution & 1024 & 8 &  2 & N/A & 208 \\
    & & 2 & 2D convolution & 256 & 16 &  2 & N/A & 3216 \\
    & & 3 & 2D convolution & 64 & 32 &  2 & N/A & 12832 \\
    & & 4 & 2D convolution & 16 & 64 &  2 & N/A & 51264 \\
    & & 5 & $\mathrm{MLP}_{\mathrm{enc}}$ & N/A & N/A & N/A & $M$ & 129$M$ \\
    \cmidrule(l){2-9}
    & {\Block{6-1}{\rotatebox[origin=lb]{90}{decoder}}} & 0 & $\mathrm{MLP}_{\mathrm{dec}}$ & 16 & 64 & N/A & N/A & 128$\left(M+1\right)$\\
    & & 1 & 2D transposed convolution & 64 & 32 &  2 & N/A  & 51232 \\
    & & 2 & 2D transposed convolution & 256 & 16 &  2 & N/A & 12816 \\
    & & 3 & 2D transposed convolution & 1024 & 8 &  2 & N/A & 3208 \\
    & & 4 & 2D transposed convolution & 4096 & 4 &  2 & N/A & 204 \\
    & & 5 & postprocessing & N/A & N/A & N/A & 4096 $\times$ 4 & N/A \\
    \bottomrule 
\end{NiceTabular}
\endgroup
\label{table:CNN}
\end{center}
\end{table}

\section{Proper orthogonal decomposition}\label{appendix:POD}
To compute the set of orthonormal POD basis vectors, we use the method of snapshots \cite{sirovich1987turbulence} in which a snapshot matrix of the time history of the FOM solutions is generated, $\mathbf{X}^{\mathrm{POD}} = \left[\mathbf{x}^1,\ldots,\mathbf{x}^{N_{\mathrm{train}}}\right] \in \mathbb{R}^{N \times N_{\mathrm{train}}}$, where $N_{\mathrm{train}}\in\mathbb{N}$ is the number of training snapshots, $\mathbf{x}^i \in \mathbb{R}^N$ is the solution state vector of snapshot $i$ with $N$ being the dimension of the state vector. Next, singular value decomposition is performed on the snapshot matrix, $\mathbf{X}^{\mathrm{POD}}$:
\begin{equation}
    \mathbf{X}^{\mathrm{POD}} = \mathbf{V} \boldsymbol{\Sigma} \mathbf{U}^T
\end{equation}
where $\mathbf{V} = \left[ \mathbf{v}_1,\ldots,\mathbf{v}_{N} \right]\in \mathbb{R}^{N \times N}$ is a matrix of $N$ orthonormal vectors which represent the POD modes in the order of decreasing singular values, $\boldsymbol{\Sigma} = \textrm{diag}\left(\sigma_1,\ldots,\sigma_{N} \right)\in \mathbb{R}^{N \times N}$ is the diagonal matrix of singular values ordered as $\sigma_1 \geq \ldots \geq \sigma_{N}$, and $\mathbf{U} = \left[\mathbf{u}_1,\ldots,\mathbf{u}_{N_{\mathrm{train}}}\right]\in \mathbb{R}^{N \times N_{\mathrm{train}}}$ provides information about the time dynamics. The POD basis is created by truncating the first $M$ left singular vectors of the snapshot matrix, i.e., $\mathbf{\Phi} = \left[\mathbf{v}_1,\ldots,\mathbf{v}_M \right]\in \mathbb{R}^{N \times M}$, which is made up of $M$ orthonormal vectors that describe the dominant mode shapes of the system. The POD basis vectors are orthonormal and optimal in the $L^2$ sense, making them a common choice in the context of PMOR \cite{taira2017modalanalysis}.
%\end{appendix}

\end{document}